\pdfoutput=1
\documentclass[12pt]{article}

\usepackage[T1]{fontenc}
\usepackage[utf8]{inputenc}
\usepackage[english]{babel}

\usepackage{amsmath}
\usepackage{amssymb}
\usepackage{amsthm}
\usepackage{mathtools}
\mathtoolsset{showonlyrefs,showmanualtags}

\allowdisplaybreaks

\usepackage[natbibapa,nodoi]{apacite}
\usepackage{authblk}

\usepackage{bbm}
\newcommand{\bbone}{\mathbbm{1}}
\usepackage{booktabs}
\usepackage[format=plain,font=small,labelfont=bf,textfont=up]{caption}
\usepackage{enumitem}
\setlist[enumerate]{itemsep=0.1ex}
\setlist[itemize]{itemsep=0.1ex}
\newcommand\expcommand\newcommand
\usepackage[letterpaper,margin=1.05in,bottom=1.45in]{geometry}
\usepackage{graphicx}
\usepackage[hidelinks]{hyperref}
\usepackage[nodisplayskipstretch]{setspace}
\usepackage{subfigure}

\usepackage{blkarray}
\usepackage[linesnumbered,ruled,vlined]{algorithm2e}
\usepackage{xcolor}

\usepackage[nohints,tight]{minitoc}

\newcommand\mainref\ref
\newcommand\suppref\ref

\DeclarePairedDelimiter\paren\lparen\rparen
\DeclarePairedDelimiter\bracket\lbrack\rbrack
\DeclarePairedDelimiter\braces\lbrace\rbrace

\DeclarePairedDelimiter\abs\lvert\rvert

\providecommand{\bbone}{\mathbf{1}}
\DeclarePairedDelimiterXPP\indicator[1]{\bbone}{\lbrack}{\rbrack}{}{#1}

\DeclarePairedDelimiterXPP\expf[1]{\exp}{\lparen}{\rparen}{}{#1}
\DeclarePairedDelimiterXPP\logf[1]{\log}{\lparen}{\rparen}{}{#1}
\DeclarePairedDelimiterXPP\maxf[1]{\max}{\lparen}{\rparen}{}{#1}
\DeclarePairedDelimiterXPP\minf[1]{\min}{\lparen}{\rparen}{}{#1}

\DeclarePairedDelimiterXPP\func[2]{#1}{\lparen}{\rparen}{}{#2}

\newcommand{\set}[1]{#1}

\DeclarePairedDelimiter\setb\lbrace\rbrace

\newcommand{\Reals}{\mathbb{R}}

\newcommand{\Naturals}{\mathbb{N}}

\DeclareMathOperator*{\argmin}{arg\,min}

\newcommand{\mat}[1]{\boldsymbol{#1}}
\renewcommand{\vec}[1]{\boldsymbol{#1}}

\newcommand{\onevec}{\vec{1}}

\newcommand{\unitM}{\mat{I}}

\newcommand{\eigvec}[1]{\vec{\eta}_{#1}}
\newcommand{\eigval}[1]{\lambda_{#1}}
\newcommand{\eigveci}{\eigvec{i}}
\newcommand{\eigvali}{\eigval{i}}
\newcommand{\eigvalmin}{\eigval{\min}}
\newcommand{\eigvalmax}{\eigval{\max}}

\newcommand{\basisv}[1]{\vec{e}_{#1}}

\makeatletter
\newcommand*{\tran}{{\mathpalette\@tran{}}}
\newcommand*{\@tran}[2]{\raisebox{\depth}{$\m@th#1\intercal$}}
\makeatother

\DeclarePairedDelimiter\iprod\langle\rangle

\DeclarePairedDelimiter\norm\lVert\rVert
\DeclarePairedDelimiterXPP\tnorm[1]{}{\lVert}{\rVert_{1}}{}{#1}
\DeclarePairedDelimiterXPP\enorm[1]{}{\lVert}{\rVert_{2}}{}{#1}
\DeclarePairedDelimiterXPP\inorm[1]{}{\lVert}{\rVert_{\infty}}{}{#1}
\DeclarePairedDelimiterXPP\pnorm[2]{}{\lVert}{\rVert_{#1}}{}{#2}

\DeclarePairedDelimiterXPP\detf[1]{\det}{\lparen}{\rparen}{}{#1}

\DeclareMathOperator{\trsym}{tr}
\DeclarePairedDelimiterXPP\tr[1]{\trsym}{\lparen}{\rparen}{}{#1}

\DeclareMathOperator{\diagsym}{diag}
\DeclarePairedDelimiterXPP\diag[1]{\diagsym}{\lparen}{\rparen}{}{#1}

\DeclareMathOperator{\ranksym}{rank}
\DeclarePairedDelimiterXPP\rank[1]{\ranksym}{\lparen}{\rparen}{}{#1}

\DeclareMathOperator{\vectorizesym}{vec}
\DeclarePairedDelimiterXPP\vectorize[1]{\vectorizesym}{\lparen}{\rparen}{}{#1}

\providecommand\given{}
\newcommand\givensymbol[1]{\nonscript\:#1\vert\allowbreak\nonscript\:\mathopen{}}

\let\Prsym\Pr
\let\Pr\relax
\DeclarePairedDelimiterXPP\Pr[1]{\Prsym}{\lparen}{\rparen}{}{%
	\renewcommand\given{\givensymbol{\delimsize}}%
	#1}

\DeclarePairedDelimiterXPP\Prsub[2]{\Prsym_{#1}}{\lparen}{\rparen}{}{%
	\renewcommand\given{\givensymbol{\delimsize}}%
	#2}

\DeclareMathOperator{\Esym}{E}
\DeclarePairedDelimiterXPP\E[1]{\Esym}{\lbrack}{\rbrack}{}{%
	\renewcommand\given{\givensymbol{\delimsize}}%
	#1}

\DeclarePairedDelimiterXPP\Esub[2]{\Esym_{#1}}{\lbrack}{\rbrack}{}{%
	\renewcommand\given{\givensymbol{\delimsize}}%
	#2}

\DeclareMathOperator{\Varsym}{Var}
\DeclarePairedDelimiterXPP\Var[1]{\Varsym}{\lparen}{\rparen}{}{%
	\renewcommand\given{\givensymbol{\delimsize}}%
	#1}

\DeclarePairedDelimiterXPP\Varsub[2]{\Varsym_{#1}}{\lparen}{\rparen}{}{%
	\renewcommand\given{\givensymbol{\delimsize}}%
	#2}

\DeclarePairedDelimiterXPP\EstVar[1]{\widehat{\Varsym}}{\lparen}{\rparen}{}{%
	\renewcommand\given{\givensymbol{\delimsize}}%
	#1}

\DeclareMathOperator{\Covsym}{Cov}
\DeclarePairedDelimiterXPP\Cov[1]{\Covsym}{\lparen}{\rparen}{}{%
	\renewcommand\given{\givensymbol{\delimsize}}%
	#1}

\DeclarePairedDelimiterXPP\Covsub[2]{\Covsym_{#1}}{\lparen}{\rparen}{}{%
	\renewcommand\given{\givensymbol{\delimsize}}%
	#2}

\DeclareMathOperator{\Corrsym}{Corr}
\DeclarePairedDelimiterXPP\Corr[1]{\Corrsym}{\lparen}{\rparen}{}{%
	\renewcommand\given{\givensymbol{\delimsize}}%
	#1}

\makeatletter
\newcommand{\indep}{\protect\mathpalette{\protect\@indep}{\perp}}
\newcommand*{\@indep}[2]{\mathrel{\rlap{$#1#2$}\mkern3mu{#1#2}}}
\makeatother

\newcommand{\bigOsym}{\mathcal{O}}
\DeclarePairedDelimiterXPP\bigO[1]{\bigOsym}{\lparen}{\rparen}{}{#1}

\newcommand{\littleOsym}{o}
\DeclarePairedDelimiterXPP\littleO[1]{\littleOsym}{\lparen}{\rparen}{}{#1}

\newcommand{\bigOpsym}{\bigOsym_p}
\DeclarePairedDelimiterXPP\bigOp[1]{\bigOpsym}{\lparen}{\rparen}{}{#1}

\newcommand{\littleOpsym}{\littleOsym_p}
\DeclarePairedDelimiterXPP\littleOp[1]{\littleOpsym}{\lparen}{\rparen}{}{#1}

\newcommand{\bigOmegasym}{\Omega}
\DeclarePairedDelimiterXPP\bigOmega[1]{\bigOmegasym}{\lparen}{\rparen}{}{#1}

\newcommand{\littleOmegasym}{\omega}
\DeclarePairedDelimiterXPP\littleOmega[1]{\littleOmegasym}{\lparen}{\rparen}{}{#1}

\newcommand{\bigThetasym}{\Theta}
\DeclarePairedDelimiterXPP\bigTheta[1]{\bigThetasym}{\lparen}{\rparen}{}{#1}

\newcommand{\sumin}{\sum_{i=1}^n}
\newcommand{\sumjn}{\sum_{j=1}^n}

\newcommand{\quadtext}[1]{\quad\text{#1}\quad}
\newcommand{\qquadtext}[1]{\qquad\text{#1}\qquad}

\newcommand{\quadand}{\quadtext{and}}
\newcommand{\qquadand}{\qquadtext{and}}

\newcommand{\quadwhere}{\quadtext{where}}
\newcommand{\qquadwhere}{\qquadtext{where}}

\theoremstyle{plain}

\newtheorem{corollary}{Corollary}
\newtheorem{lemma}{Lemma}
\newtheorem{proposition}{Proposition}
\newtheorem{theorem}{Theorem}
\newtheorem*{theorem*}{Theorem}

\newtheorem{innerreflemma}{Lemma}
\newenvironment{reflemma}[1]
{\renewcommand\theinnerreflemma{#1}\innerreflemma}
{\endinnerreflemma}

\newtheorem{innerrefproposition}{Proposition}
\newenvironment{refproposition}[1]
{\renewcommand\theinnerrefproposition{#1}\innerrefproposition}
{\endinnerrefproposition}

\newtheorem{innerreftheorem}{Theorem}
\newenvironment{reftheorem}[1]
{\renewcommand\theinnerreftheorem{#1}\innerreftheorem}
{\endinnerreftheorem}

\theoremstyle{definition}

\newtheorem{definition}{Definition}

\theoremstyle{remark}

\newcommand\resetsuppcounters{%
\setcounter{section}{0}%
\setcounter{table}{0}%
\setcounter{figure}{0}%
\setcounter{equation}{0}%
\setcounter{conjecture}{0}%
\setcounter{corollary}{0}%
\setcounter{lemma}{0}%
\setcounter{proposition}{0}%
\setcounter{theorem}{0}%
\setcounter{assumption}{0}%
\setcounter{condition}{0}%
\setcounter{definition}{0}%
\setcounter{remark}{0}%
\setcounter{example}{0}%
\setcounter{algocf}{0}%
}

\newcommand\updatesuppcounters[1]{%
\renewcommand\thesection{#1\arabic{section}}%
\renewcommand\thetable{#1\arabic{table}}%
\renewcommand\thefigure{#1\arabic{figure}}%
\renewcommand\theequation{#1\arabic{equation}}%
\renewcommand\theconjecture{#1\arabic{conjecture}}%
\renewcommand\thecorollary{#1\arabic{corollary}}%
\renewcommand\thelemma{#1\arabic{lemma}}
\renewcommand\theproposition{#1\arabic{proposition}}
\renewcommand\thetheorem{#1\arabic{theorem}}
\renewcommand\theassumption{#1\arabic{assumption}}
\renewcommand\thecondition{#1\arabic{condition}}
\renewcommand\thedefinition{#1\arabic{definition}}
\renewcommand\theremark{#1\arabic{remark}}
\renewcommand\theexample{#1\arabic{example}}
\renewcommand\thealgocf{#1\arabic{algocf}}
}

\newcommand\updatesuppanchors[1]{
\renewcommand\theHsection{#1.\arabic{section}}
\renewcommand\theHtable{#1.\arabic{table}}
\renewcommand\theHfigure{#1.\arabic{figure}}
\renewcommand\theHequation{#1.\arabic{equation}}
\renewcommand\theHconjecture{#1.\arabic{conjecture}}
\renewcommand\theHcorollary{#1.\arabic{corollary}}
\renewcommand\theHlemma{#1.\arabic{lemma}}
\renewcommand\theHproposition{#1.\arabic{proposition}}
\renewcommand\theHtheorem{#1.\arabic{theorem}}
\renewcommand\theHassumption{#1.\arabic{assumption}}
\renewcommand\theHcondition{#1.\arabic{condition}}
\renewcommand\theHdefinition{#1.\arabic{definition}}
\renewcommand\theHremark{#1.\arabic{remark}}
\renewcommand\theHexample{#1.\arabic{example}}
\renewcommand\theHalgocf{#1.\arabic{algocf}}
}

\renewcommand{\mat}[1]{\mathbb{#1}}

\newcommand{\eigveck}{\eigvec{k}}
\newcommand{\eigvalk}{\eigval{k}}

\newcommand{\ate}{\tau}

\newcommand{\est}{\widehat{\tau}}

\newcommand{\Sample}{\set{U}}

\newcommand{\z}[1]{z_{#1}}
\newcommand{\zi}{\z{i}}

\newcommand{\zv}{\vec{z}}

\newcommand{\Z}[1]{Z_{#1}}
\newcommand{\Zi}{\Z{i}}

\newcommand{\Zv}{\vec{Z}}

\newcommand{\ysym}{y}
\newcommand{\y}[2]{\ysym_{#1}\paren{#2}}
\newcommand{\yi}[1]{\y{i}{#1}}

\newcommand{\exmsym}{d}
\newcommand{\exm}[2]{\exmsym_{#1}\paren{#2}}
\newcommand{\exmi}[1]{\exm{i}{#1}}

\newcommand{\Exmsym}{D}
\newcommand{\Exm}[1]{\Exmsym_{#1}}
\newcommand{\Exmi}{\Exm{i}}

\newcommand{\exa}{e_1}
\newcommand{\exb}{e_0}
\newcommand{\exarb}{e}

\newcommand{\iex}[2]{\indicator{\Exm{#1} = #2}}
\newcommand{\iexi}[1]{\iex{i}{#1}}
\newcommand{\iexj}[1]{\iex{j}{#1}}

\newcommand{\iexia}{\iexi{\exa}}
\newcommand{\iexja}{\iexj{\exa}}

\newcommand{\iexib}{\iexi{\exb}}
\newcommand{\iexjb}{\iexj{\exb}}

\newcommand{\pex}[2]{\Pr{\Exm{#1} = #2}}
\newcommand{\pexi}[1]{\pex{i}{#1}}
\newcommand{\pexj}[1]{\pex{j}{#1}}
\newcommand{\pexia}{\pexi{\exa}}
\newcommand{\pexja}{\pexj{\exa}}
\newcommand{\pexib}{\pexi{\exb}}
\newcommand{\pexjb}{\pexj{\exb}}

\newcommand{\Y}[1]{Y_{#1}}
\newcommand{\Yi}{\Y{i}}

\newcommand{\Yv}{\vec{Y}}

\newcommand{\xve}[1]{\vec{x}_{#1}}
\newcommand{\xvi}{\xve{i}}
\newcommand{\xvj}{\xve{j}}

\newcommand{\vx}{\vec{x}}
\newcommand{\vy}{\vec{y}}

\newcommand{\xM}{\mat{X}}

\newcommand{\estcoef}[1]{W_{#1}}
\newcommand{\estcoefi}{\estcoef{i}}

\DeclareMathOperator{\Vsym}{V}
\DeclareMathOperator{\VBsym}{VB}
\DeclarePairedDelimiterXPP\varf[1]{\Vsym}{\lparen}{\rparen}{}{#1}
\DeclarePairedDelimiterXPP\varbf[1]{\VBsym}{\lparen}{\rparen}{}{#1}
\DeclarePairedDelimiterXPP\varbfest[1]{\widehat{\VBsym}}{\lparen}{\rparen}{}{#1}

\newcommand{\allexposures}{\Delta}

\newcommand{\numvar}{T}

\newcommand{\Pop}{\set{P}}
\newcommand{\Obs}{\set{S}}

\newcommand{\coef}[1]{V_{#1}}

\newcommand{\coefk}{\coef{k}}
\newcommand{\coefv}{\vec{V}}

\newcommand{\para}[1]{\theta_{#1}}

\newcommand{\parak}{\para{k}}
\newcommand{\paral}{\para{\ell}}
\newcommand{\parav}{\vec{\theta}}

\newcommand{\varM}{\mat{A}}

\newcommand{\varMe}[1]{a_{#1}}

\newcommand{\varMkl}{\varMe{k\ell}}

\newcommand{\unobspairs}{\Omega}

\newcommand{\boundM}{\mat{B}}
\newcommand{\boundMe}[1]{b_{#1}}

\newcommand{\boundMkl}{\boundMe{k\ell}}

\newcommand{\allbounds}{\mathcal{B}}

\newcommand{\slackM}{\mat{S}}
\newcommand{\slackMe}[1]{s_{#1}}

\newcommand{\slackMkl}{\slackMe{k\ell}}

\newcommand{\allslacks}{\mathcal{S}}

\newcommand{\optprocedure}{{\normalfont\textsc{OPT-VB}\xspace}}
\newcommand{\testadmissiblity}{{\normalfont\textsc{Test-Admissibility}\xspace}}

\newcommand{\objsym}{g}
\DeclarePairedDelimiterXPP\objf[1]{\objsym}{\lparen}{\rparen}{}{#1}
\newcommand{\objfS}{\objf{\slackM}}

\newcommand{\objM}{\mat{W}}

\newcommand{\otherboundM}{\tilde{\boundM}}
\newcommand{\otherslackM}{\tilde{\slackM}}

\newcommand{\inadmissiblecert}{\mat{Q}}

\newcommand{\partM}{\mat{M}}

\newcommand{\partMkl}{\partM_{k\ell}}

\newcommand{\dualM}{\mat{M}}

\title{\textbf{Optimized variance estimation under interference and complex experimental designs}}

\author[1]{Christopher Harshaw}
\author[2]{Joel Middleton}
\author[3]{Fredrik Sävje}

\affil[1]{Columbia University}
\affil[2]{Joel Middleton LLC}
\affil[3]{Uppsala University}

\date{\today}

\begin{document}

\makeatletter%
\gdef\@thefnmark{}\@footnotetext{\hspace{-1em}This paper builds on a working paper by one of the coauthors \citep{Middleton2018Unified}.
We thank P Aronow, Mateo Díaz, Peng Ding, Oscar Dowson, Jon Erickson, Avi Feller, Erin Hartman, Guido Imbens, Marek Kaluba, Miles Lubin, Cyrus Samii, Daniel Spielman and Sekhar Tatikonda for helpful comments and discussions.
Harshaw was partially supported by NSF Graduate Research Fellowship (DGE1122492) and Foundations of Data Science Institute, NSF grant DMS2023505.
Sävje was partially supported by The Bank of Sweden Tercentenary Foundation (P24-0300).
Both Harshaw and Sävje were partially supported by NSF grant MMS2316335.
The computations in this article were enabled by resources provided by the National Academic Infrastructure for Supercomputing in Sweden (NAISS), partially funded by the Swedish Research Council (2022-06725).

}%
\makeatother%

\begin{singlespace}
\maketitle

\begin{abstract}
\noindent Unbiased and consistent variance estimators generally do not exist for design-based treatment effect estimators because experimenters never observe more than one potential outcome for any unit.
The problem is exacerbated by interference and complex experimental designs.
Experimenters must accept conservative variance estimators in these settings, but they can strive to minimize the conservativeness.
In this paper, we show that the task of constructing a minimally conservative variance estimator can be interpreted as an optimization problem that aims to find the lowest estimable upper bound of the true variance given the experimenter's risk preferences and knowledge of the potential outcomes.
We characterize the set of admissible bounds in the class of quadratic forms, and we demonstrate that the optimization problem is a convex program for many natural objectives.
The resulting variance estimators are guaranteed to be conservative regardless of whether the background knowledge used to construct the bound is correct, but the estimators are less conservative if the provided information is reasonably accurate.
Numerical results show that the resulting variance estimators can be considerably less conservative than existing estimators, allowing experimenters to draw more informative inferences about treatment effects.

\vspace{0.2in}

\end{abstract}
\end{singlespace}

\doparttoc
\faketableofcontents

\clearpage

\section{Introduction}

The design-based, finite population approach to causal inference considers treatment assignment as the only source of randomness.
In this framework, the variance of treatment effect estimators depends on aspects of the joint distribution of the potential outcomes.
This poses a challenge for variance estimation because experimenters can never observe more than one potential outcome for each unit, meaning that the observed potential outcomes provide little information about their joint distribution.
Without strong assumptions, such as constant treatment effects, it is not possible to consistently estimate the variance of treatment effect estimators.

In this paper, we consider variance estimation in experiments with interference or complex experimental designs.
Interference occurs when the treatment assigned to one unit affects other units, which tend to introduce strong dependencies between units' exposures.
A complex design involves strong and intricate dependencies between units' treatments as part of the assignment mechanism itself.
The variance estimation problem is particularly difficult in these settings, because strong dependencies between units' exposures tend to make more aspects of the joint distributions inaccessible.
Conventional techniques for constructing conservative variance estimators can therefore not be used, or they produce overly conservative estimators.

We describe variance estimators that aim to minimize conservativeness.
Following the previous literature, we break up the task of constructing a variance estimator by first constructing an upper bound for the variance.
An estimator of the bound then acts as a conservative estimator of the variance.
However, unlike previous work, we consider variance estimation under arbitrary interference and arbitrary experimental designs.
We also consider a large class of linear point estimators, which includes all commonly used treatment effect estimators.

There are two main contributions of the paper.
First, in Section~\ref{sec:variance-bounds}, we describe and characterize the variance estimation problem.
We define a concept of admissibility for variance bounds, allowing us to discard a large set of poorly performing variance bounds.
The characterization allows us to understand many previously described variance bounds in a common framework, and we show that a commonly used variance bound for experiments with interference or complex designs is inadmissible relative to the class of bounds we consider.

Second, in Section~\ref{sec:finding-bounds}, we reinterpret the task of selecting a variance bound as an optimization problem.
We describe two classes of objective functions that allow experimenters to construct variance bounds based on their prior substantive knowledge and risk preferences about excessive conservativeness.
The bounds are valid and admissible no matter which class of objective functions is used and no matter if the supplied information is correct, but the resulting bound is less conservative if the information is reasonably accurate.
The underlying optimization problem is convex, meaning that it is computationally tractable.

Supplementary contributions include an investigation in Section~\ref{sec:estimate-vb} of how challenges when estimating a bound affect which bound to select.
We highlight that some bounds are easier to estimate than others, meaning that we might prefer a bound that is more conservative if we can estimate it with greater precision.
Section~\ref{sec:simulations} reports the results from a simulation exercise based on real-world data examining how the methods we describe behave in practice.

\section{Illustration and Preview of Main Results}\label{sec:illustration}

To illustrate the central question of the paper, we consider a stylized version of the study by \citet{Paluck2016Changing}, which we also use in the simulation exercise in Section~\ref{sec:simulations}.
The authors investigate whether an anticonflict intervention for students in US middle schools reduces conflict.
They were particularly interested in how the effect of the intervention spread through the students' peer networks.
They investigated this by comparing students directly exposed to the intervention with students only indirectly exposed through their peer networks.

For simplicity in this illustration, we will consider a sample of only two students.
The experiment is such that exactly one student, chosen at random with equal probability, will be directly exposed to the anticonflict intervention.
The two students are in the same peer network, so the student not directly exposed is considered to be indirectly exposed.
The estimator $\widehat{\tau}$ is the difference in observed outcomes of the two students.
Let $a_i$ denote the outcome of student $i \in \braces{1, 2}$ when directly exposed, and let $b_i$ denote outcome when indirectly exposed.
The estimator takes two values with equal probability: $a_1 - b_2$ and $a_2 - b_1$.
The variance of the estimator is therefore%
\begin{equation}
	\Var{\widehat{\tau}} = \frac{1}{4} \paren{a_1^2 + a_2^2 + b_1^2 + b_2^2}
	+ \frac{1}{2} \paren{a_1 b_1 + a_2 b_2 - a_1 a_2 - b_1 b_2 - a_1 b_2 - a_2 b_1}.
\end{equation}

At the heart of the variance estimation problem is that some terms in the variance expression are never observed.
We never observe $a_1b_1$ or $a_2b_2$, because a student cannot be assigned to both direct and indirect exposure at the same time.
Similarly, we never observe $a_1a_2$ or $b_1b_2$, because the two students are always assigned to different exposures.
The unobserved terms prevent us from constructing an unbiased, or even consistent, estimator of the variance unless we impose strong assumptions on the potential outcomes.

A common way to address this problem is to construct an upper bound for the variance.
An estimator of the bound then acts as a conservative estimator of the variance.
A simple upper bound uses the fact that $\Var{\widehat{\tau}} \leq \E{\widehat{\tau}^2}$.
Hence, the variance is upper bounded by%
\begin{equation}
	B_1
	= \E[\big]{\widehat{\tau}^2}
	= \frac{1}{2} \paren{a_1^2 + a_2^2 + b_1^2 + b_2^2} - \paren{a_1b_2 + a_2b_1}.
\end{equation}
Note that this bound holds for any values of the potential outcomes, so no additional assumptions are required for its validity.
Furthermore, the bound is estimable because all terms are observed with some positive probability.

The bound we just derived is only one of many possible bounds.
A somewhat more intricate bound uses the fact that $\paren{x^2 + y^2} / 2$ is an upper bound for the product $xy$ for any real-valued $x$ and $y$.
Applying this inequality to the problematic terms in the variance expression, we arrive at the bound%
\begin{equation}
	B_2
	= \frac{3}{4} \paren{a_1^2 + a_2^2 + b_1^2 + b_2^2} - \frac{1}{2} \paren{a_1b_2 + a_2b_1}.
\end{equation}

Both $B_1$ and $B_2$ are estimable bounds, so either can be used to construct a variance estimator that is conservative in expectation.
Indeed, there are infinitely many estimable bounds in this setting, with infinitely many corresponding conservative variance estimators.
While we do want a variance estimator that is conservative, which would be achieved by any of these bounds, we want to avoid excessive conservativeness.

The idea we explore in this paper is to use an optimization approach to choose one of these estimable bounds for the variance estimator so as to minimize conservativeness.
To make the approach tractable, we focus on bounds that are quadratic forms; the variance itself is a known quadratic form in the potential outcomes, so we find it natural to restrict attention to bounds of the same form.
We collect all estimable bounds that are quadratic forms in the set $\mathcal{B}$.
Any conservative variance estimator we consider will correspond to an element in this set.

The set $\mathcal{B}$ is always infinite, and it always contains bounds that are overly conservative.
We describe a concept of inadmissibility to characterize bounds that are unnecessarily conservative.
A bound is inadmissible if there exists another (valid) bound that is less conservative no matter what the potential outcomes might be and strictly less conservative for at least one possible set of potential outcomes.
If this is the case, we say that the second bound dominates the first.
If an inadmissible bound is used to construct a  variance estimator, we say that also the estimator inadmissible.
The bias of an inadmissible variance estimator will be larger than the bias of the variance estimator that dominates it, motivating us to use admissible estimators.

Of the two bounds considered in this section, $B_1$ dominates $B_2$, meaning that $B_2$ is inadmissible.
In particular, their difference is
$
	B_2 - B_1
	= \paren{a_1 + b_2}^2 / 4 + \paren{a_2 + b_1}^2 / 4
	\geq 0,
$
so a variance estimator based on $B_2$ will always be more biased than an estimator based on $B_1$.
Thus, $B_1$ is a better bound, because it is less conservative but still valid.

The admissibility concept allows us to discard many bounds in $\mathcal{B}$, but there will generally still be infinitely many admissible bounds, so admissibility alone does not allow us to select a bound to use in our experiments.
We suggest selecting the admissible variance bound that best conforms with one's risk preferences and any accessible background information, as encoded in an objective function $g : \mathcal{B} \to \Reals$.
The selected variance bound is the minimizer $\mathbb{B}^*$ of the objective function $g$ in $\mathcal{B}$.
The properties of the resulting variance estimator will inevitably depend on the choice of the objective function.
However, one central result of the paper is that this approach produces admissible bounds for a large class of objective functions.
This result is presented and discussed as Theorem~\ref{thm:var-bound-programs} in Section~\ref{sec:var-bound-programs} below, and it is previewed here for reference.
Importantly, the theorem holds no matter if the background information used to construct the objective function $g$ is correct.
The approach therefore allows experimenters to target the variance estimator to their setting without risk of inadvertently using an anticonservative or inadmissible variance estimator.

\begin{reftheorem}{\ref*{thm:var-bound-programs}}
If the objective function $\objsym$ is strictly monotone, then the bound based on the minimizer of $\objsym$ in $\mathcal{B}$ is conservative, estimable and admissible in finite samples.
\end{reftheorem}

\begin{figure}[t]
\centering
\includegraphics[width=0.9\textwidth]{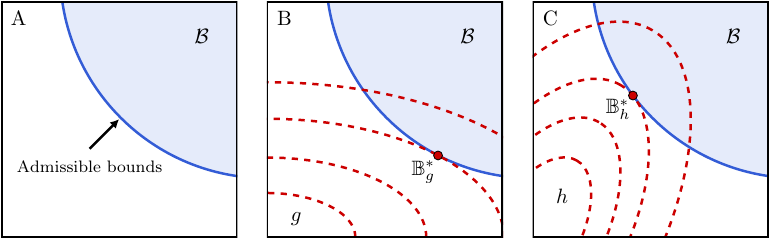}
\caption{Illustration of selecting of a variance bound using optimization.}\label{fig:opt-illustration}
\end{figure}

Figure~\ref{fig:opt-illustration} provides a graphical illustration of the approach we explore in the paper.
Each panel should be interpreted as a set of potential estimation targets.
The illustration is highly stylized, and the axes do not necessarily correspond to any particular parameterization.
In Panel A, the set $\mathcal{B}$ is plotted, where the region shaded in light blue contains all valid and estimable bounds.
The boundary of this region, marked in a stronger blue, is the set of all admissible bounds.
Panels B and C show the contour lines of two different objective functions, $g$ and $h$.
The two minimizers, $\mathbb{B}^*_g$ and $\mathbb{B}^*_h$, are different, but they are both on the boundary containing the admissible bounds.
Which bound is less conservative will depend on the potential outcomes; there are outcomes for which each bound is less conservative than the other.

\section{Related Work}\label{sec:related-lit}

\citet{Neyman1923Application} was first to recognize that the variance of a treatment effect estimator is not directly estimable.
He showed that the variance of the difference-in-means estimator under the complete randomization design depends on the covariance of unit-level potential outcomes, which cannot be estimated from the data.
Neyman applied the Cauchy--Schwarz inequality followed by the AM--GM inequality to arrive at an estimable upper bound of the variance.
He also noted that unbiased variance estimation is possible when treatment effects are constant between units, which sometimes is referred to as strict additivity.

Neyman's approach has been improved and extended in several directions.
An important line of work aims to sharpen the bound.
\citet{Robins1988Confidence} focuses on binary outcomes and derives a variance estimator that extracts all information about the joint distribution of the potential outcomes contained in the marginal distributions.
\citet{Aronow2014Sharp} use Fréchet--Hoeffding-type bounds to generalize the estimator by \citet{Robins1988Confidence} to arbitrary outcome variables.
\citet{Ding2016Potential} and \citet{Lu2019Improved} take a similar approach and use information contained in the marginal outcome distributions to improve upon Neyman's bound for binary outcomes.
\citet{Nutz2021Directional} provide further improvements under the assumption that all unit-level treatment effects are non-negative.
\citet{Imbens2021Causal} provide higher-order refinements using a bootstrap approach.
These bounds are applicable only when experimenters use the difference-in-means estimator under complete randomization; it is unclear whether and how these results generalize to more complex estimators and designs.

Another strand of the literature considers variance estimation under other experimental designs than complete randomization.
Early examples include \citet{Kempthorne1955Randomization} and \citet{Wilk1955Randomzation}, who studied variance estimation under various blocked designs.
These investigations generally impose structural assumptions on the potential outcomes, such as strict additivity, which limits their applicability.
A more recent strand of the literature has derived Neyman-type variance estimators for some types of blocked, stratified or clustered designs without such assumptions \citep[see, e.g.,][]{Gadbury2001Randomization,Abadie2008Estimation,Imai2008Variance,Higgins2015Blocking,Fogarty2018Mitigating,Pashley2021Insights}.

A related strand of the literature has derived Neyman-type variance estimators for other point estimators than the difference-in-means estimator.
\citet{Samii2012Equivalencies} investigate variance estimators for the ordinary least square regression estimator, and \citet{Aronow2013Class} do the same for the Horvitz--Thompson estimator.
\citet{Mukerjee2018Using} connect both of these strands of the literature and consider variance estimation for unbiased linear estimators of treatment effects for arbitrary experimental designs.
These authors use a formulation similar to the one in this paper to weaken the strict additivity assumption employed by \citet{Neyman1923Application} to obtain an unbiased variance estimator.

All papers mentioned so far in this section have assumed that the experimental units do not interfere with each other.
The strand of the literature closest to the current paper considers variance estimation in settings with interference under arbitrary experimental designs.
To our knowledge, the only previous result here is due to \citet{Aronow2013Conservative,Aronow2017Estimating}.
They describe a method for constructing a bound for the variance of the Horvitz--Thompson estimator when many pair-wise assignment probabilities are zero, as often is the case under interference.
In this paper, we ask whether better bounds exist in this setting.
We answer this question in the affirmative.
Indeed, as we show in Section~\ref{sec:prev-examples}, the Aronow--Samii bound is inadmissible in the class of bounds that we consider.

\section{The Variance and Variance Bounds}\label{sec:variance-bounds}

\subsection{Preliminaries}\label{sec:preliminaries}

Consider an experiment consisting of $n$ units indexed by $\Sample = \braces{1, \dotsc, n}$.
Each unit $i \in \Sample$ is assigned one of two treatment conditions $\zi \in \setb{0, 1}$.
We collect the assignments of all units into an assignment vector $\zv = \paren{\z{1}, \dotsc, \z{n}} \in \setb{0,1}^n$.
The assignments are random, and $\Zi$ denotes the random assignment for unit $i$.
Let $\Zv = \paren{\Z{1}, \dotsc, \Z{n}}$ denote the random treatment vector that collects all units' assignments.
The distribution of $\Zv$ is the \emph{design} of the experiment, which is taken to be known.

Each unit $i \in \Sample$ has an associated \emph{potential outcome} function $\ysym_i : \setb{0, 1}^n \rightarrow \Reals$ that specifies the response of unit $i$ under all possible treatment assignments.
Because the function $\ysym_i$ depends on the full assignment vector, the response of unit $i$ is allowed to depend not only on its own treatment, but potentially also on the treatments assigned to other units.
This is commonly referred to as interference.
The observed outcome of unit $i$ is $\Yi = \yi{\Zv}$, and the vector of all observed outcomes is denoted $\Yv = \paren{\Y{1}, \dotsc, \Y{n}}$.
The potential outcome functions themselves are deterministic and the randomness in the observed outcomes arises from the fact that treatment is randomly assigned.

We will use the framework described by \citet{Aronow2017Estimating} to model interference.
A related framework is described by \citet{Manski2013Identification}.
Each unit $i \in \Sample$ has an \emph{exposure mapping} $\exmsym_i: \setb{0,1}^n \rightarrow \allexposures$ that maps each assignment vector to a set of exposures $\allexposures$.
When two or more assignment vectors map to the same exposure for some unit, those assignments are considered causally equivalent with respect to that unit.
The number of exposures $\abs{\allexposures}$ is typically small compared to the number of units.

Experimenters using exposure mappings often assume that the mappings are correctly specified, and we will do the same in this paper.
The assumption states that a unit's outcome is completely determined by its exposure, in the sense that $\exmi{\zv} = \exmi{\zv'}$ implies $\yi{\zv} = \yi{\zv'}$ for all $\zv, \zv' \in \setb{0,1}^n$.
For each unit $i \in \Sample$, we define $\Exmi = \exmi{\Zv}$ to be the exposure produced by the realized treatment assignment.

The causal quantity of interest $\ate$ in this context is typically an average contrast between outcomes for two exposures.
That is, given two exposures $\exa, \exb \in \allexposures$, experimenters aim to estimate
$
	\ate = n^{-1} \sum_{i=1}^n \bracket[\big]{ \yi{\exa} - \yi{\exb} },
$
where we have overloaded the notation by writing $\yi{\exarb}$ to denote the outcome of unit $i$ under exposure $\exarb \in \allexposures$.
Many commonly studied estimands, including total, direct and indirect treatment effects, are of this form \citep{Hudgens2008Causal}.
The conventional average treatment effect under a no-interference assumption also takes this form.
While experimenters almost exclusively consider estimands that are unweighted averages of contrasts of potential outcomes, all results in this paper generalize to estimands that are arbitrary linear functions of the potential outcomes.

We consider the class of \emph{linear estimators} of the effect $\ate$.
An estimator in this class can be written as a random linear combination of the observed outcomes:
\begin{equation}
	\est = \frac{1}{n} \sumin \estcoefi \Yi,
	\label{eq:linear-est-def}
\end{equation}
where the coefficients $\estcoefi$ may depend arbitrarily on the treatment assignments $\Zv$ and characteristics of the units, but they cannot depend on the observed outcomes $\Yv$.
Thus, the coefficients can, and typically will, be random.
In Section~\suppref{sec:example-estimators} of the supplement, we show that the class of linear estimators includes all estimators commonly used by experimenters to estimate treatment and exposure effects.
This includes Horvitz--Thompson (IPW), difference-in-means, H\'ajek, OLS and linear AIPW estimators.

\subsection{The Variance of Linear Estimators}

Any estimator in the class of linear estimators can be written as
$
	\est
	= n^{-1} \sum_{k \in \Pop} \coefk \parak
	= n^{-1} \coefv^\tran \parav,
$
where $\Pop = [K]$ is a set of indices, $\coefv = \paren{\coef{1}, \dotsc, \coef{K}}$ is a vector of known random variables, and $\parav = \paren{\para{1}, \dotsc, \para{K}}$ is a vector of unknown, non-random potential outcomes.
In many cases, the estimator depends only on two types of potential outcomes, in which case $K = 2n$, and the elements of $\coefv$ will take the form $\estcoefi \indicator{\Exmi = \exarb}$.
In Section~\suppref{sec:reform-lin-est} of the supplement, we show how to go from the estimator written as in Equation~\eqref{eq:linear-est-def} to the current form in a general setting.

The advantage of writing the estimator in the current form is that all randomness is isolated in the coefficient vector $\coefv$.
This makes the derivation of the variance of the estimator straightforward, as shown in the following lemma.
All proofs appear in Section~\suppref{sec:proofs} of the supplement.

\expcommand{\lemlinearestvariance}{%
The variance of a linear estimator $\est = n^{-1} \coefv^\tran \parav$ is $\Var{\est} = n^{-2} \parav^\tran \varM \parav$, where $\varM = \Cov{\coefv}$ is the covariance matrix of the coefficient vector $\coefv$.
}

\begin{lemma}\label{lem:linear-est-variance}
	\lemlinearestvariance{}
\end{lemma}

The lemma is useful because $\varM = \Cov{\coefv}$ does not depend on the potential outcomes, so it is known.
Furthermore, because $\varM$ is a covariance matrix, it is positive semidefinite.
The variance is thus a known positive semidefinite quadratic form of the potential outcome vector, which makes it conducive to analysis.

A possible complication is that the covariance matrix may be difficult to derive analytically for some estimators and designs.
If that turns out to be the case, experimenters can use numerical methods to compute the matrix \citep{Fattorini2006Applying}.
This generally does not cause troubles because experimenters can run the Monte Carlo simulation until the matrix is known to desired precision.
However, to avoid distractions from the main ideas and insights we explore in the paper, we will proceed under the assumption that $\varM$ is known.

\subsection{The Variance Is Not Estimable}

The preceding subsection reduced the task of estimating the variance of a linear estimator to a task of estimating a (known) quadratic form in the (unknown) potential outcome vector $\parav$.
For our purposes, the central problem is that some quadratic forms cannot be estimated well.
In particular, some pairs of potential outcomes may never, or only very rarely, be observed at the same time, and this makes it difficult or impossible to estimate the quadratic form.

Let $\Obs$ be a random subset of $\Pop$ that collects the indices $k \in \Pop$ of potential outcomes $\parav = \paren{\para{1}, \dotsc, \para{K}}$ that are observed under the realized treatments $\Zv$.
If $\Pr{k, \ell \in \Obs} = 0$ for some pair $k, \ell \in \Pop$, then the corresponding product $\parak \paral$ is never observed.
These unobservable products will be central to our discussion, so we collect them in a set:
$
	\unobspairs = \setb[\big]{ (k, \ell) \in \Pop \times \Pop : \Pr{k, \ell \in \Obs} = 0 }.
$
As formalized in the following definition and proposition, estimable quadratic forms are those that are compatible with this pattern of observability.

\begin{definition}\label{def:design-compatible}
A quadratic form $\parav^\tran \mat{Q} \parav$ is \emph{design compatible} if the probability of simultaneously observing $\parak$ and $\paral$ is zero only when the corresponding element in $\mat{Q}$ is zero:
\begin{equation}
	\forall k, \ell \in \Pop,\; (k, \ell) \in \unobspairs \implies q_{k\ell} = 0,
\end{equation}
where $q_{k\ell}$ is the element in the $k$th row and $\ell$th column of $\mat{Q}$.
\end{definition}

\expcommand{\propunbiaseddesigncompatible}{%
An unbiased estimator exists for a quadratic form if and only if it is design compatible.
}

\begin{proposition}\label{prop:unbiased-design-compatible}
	\propunbiaseddesigncompatible{}
\end{proposition}

Because it is impossible to simultaneously observe two potential outcome of the same unit, a quadratic form representing variances will always be design incompatible, no matter the design.
The pattern of design incompatibility is such that the bias is large also in large samples, so consistent variance estimation is also impossible.
Furthermore, the sign of the bias will generally not be known, so inferences based on such a biased variance estimator could be anti-conservative.

The variance estimation problem is exacerbated by interference and complex experimental designs.
When units interfere, the structure of the exposure mappings often prevents certain combinations of exposures to be simultaneously realizable.
For example, when units interact with each other in a network, all neighbors of a unit that is treated will necessarily be indirectly exposed to treatment, meaning that they cannot be in a pure control condition if any of their neighbors are treated.
A similar problem occurs with complex experimental designs, which often introduce strong dependencies between the treatment assignments of different units.
This could either be in an effort to improve precision, such as with the matched-pair design, or because the design is forced on the experimenter by external factors, such as with the cluster-randomized design.

\subsection{Conservative Variance Bounds}\label{sec:conservative-var-bounds}

A variance estimator with bias of unknown sign could lead to misleading conclusions.
To address this, experimenters tend to use conservative variance estimators that systematically overestimate the variance, providing a pessimistic assessment of the precision of the point estimator.
Confidence intervals based on conservative variance estimators err on the side of caution, in the sense that they support firm conclusions only under disproportionately strong evidence.

We can understand conservative variance estimators as estimators of an upper bound of the variance.
A \emph{variance bound} is a function $\VBsym : \Reals^{K} \to \Reals$ that satisfies $\varbf{\parav} \geq \Varsub{\parav}{\est}$ for all potential outcomes $\parav \in \Reals^{K}$.
If a variance bound is also estimable, then we can use it to construct a conservative variance estimator, because the estimator will estimate a quantity that is guaranteed to be larger than the variance.
For variance bounds that are quadratic forms, design compatibility is our concept of estimability.

The focus in much of the previous literature has implicitly been on upper bounds that themselves are positive semidefinite quadratic forms.
We do the same in this paper.
That is, we consider bounds of the form
$
	\varbf{\parav} = n^{-2} \parav^\tran \boundM \parav = n^{-2} \sum_{k \in \Pop} \sum_{\ell \in \Pop} \boundMkl \parak \paral,
$
where $\boundM$ is a $K$-by-$K$ positive semidefinite matrix, and $\boundMkl$ is the element in the $k$th row and $\ell$th column of $\boundM$.
Throughout the remainder of the paper, we will use $\boundM$ to refer to both the coefficient matrix and the variance bound function $\varbf{\parav}$.

To serve its role as the basis for a conservative estimator, we require the variance bounds to be both conservative and design compatible.
This imposes two types of constraints on the coefficient matrix $\boundM$.
To satisfy design conservativeness, $\boundM$ must be larger than $\varM$, in the sense that $\parav^\tran \varM \parav \leq \parav^\tran \boundM \parav$ for all vectors $\parav$.
This is precisely the \emph{Loewner partial order} on symmetric matrices, where $\varM \preceq \boundM$ denotes that $\boundM - \varM$ is positive semidefinite.
To satisfy design compatibility, $\boundM$ must be such that $\boundMkl = 0$ for all pairs $(k, \ell) \in \unobspairs$.
We refer to symmetric matrices that satisfy these two conditions as valid variance bounds.

\begin{definition}\label{def:valid-bound}
	A symmetric matrix $\boundM$ is a \emph{valid} variance bound for $\varM$ if it is larger than $\varM$ in the Loewner order and design compatible under the current design.
	Let $\allbounds$ collect all valid variance bounds: $\allbounds = \setb[\big]{ \boundM : \varM \preceq \boundM \text{ and } \boundMkl = 0 \text{	for all } (k, \ell) \in \unobspairs }$.
\end{definition}

An alternative, but equivalent, way to characterize the set of variance bounds is to use a slack matrix $\slackM$.
A variance bound is constructed by adding the slack matrix to the variance matrix: $\boundM = \varM + \slackM$.
The resulting variance bound is conservative if and only if $\slackM$ is positive semidefinite.
Thus, the slack captures what we are adding to the variance matrix in order to achieve design compatibility.
The set of slack matrices that produces valid variance bounds is $\allslacks = \setb[big]{ \slackM : 0 \preceq \slackM \text{ and } \slackMkl = - \varMkl \text{ for all } (k, \ell) \in \unobspairs }$,
where $\slackMkl$ is the element in the $k$th row and $\ell$th column of $\slackM$.
While the two representations are equivalent, it is often more convenient to work with slack matrices.

\subsection{Admissibility}\label{sec:admissibility}

Some variance bounds will introduce slack beyond what is required for design compatibility, making them unnecessarily conservative.
Experimenters will typically want to use a variance bound that introduces as little conservativeness, or slack, as possible.
The amount of slack will depend on the potential outcomes, so there is no universal ordering of the bounds with respect to conservativeness.
But, even if there exists no universally best bound, some bounds can be ruled out because they introduce more slack than some other bound no matter what the potential outcomes might be.
To formalize this idea, we define a concept of admissibility for variance bounds.
In the following definition, let $\mathcal{F}$ be a set of valid variance bounds, i.e. functions $\VBsym : \Reals^{K} \to \Reals$ such that $\varbf{\parav} \geq \Varsub{\parav}{\est}$ for all potential outcomes $\parav \in \Reals^{K}$.

\begin{definition}\label{def:general-admissible}
	A variance bound $\VBsym : \Reals^{K} \to \Reals$ is \emph{inadmissible} with respect to a set of candidate variance bounds $\mathcal{F}$ if there exists some bound $\VBsym' \in \mathcal{F}$ such that $\VBsym'(\parav) \leq  \varbf{\parav}$ for all $\parav \in \Reals^{K}$, and $\VBsym'(\parav) <  \varbf{\parav}$ for at least one $\parav \in \Reals^{K}$.
	A variance bound is \emph{admissible} with respect to $\mathcal{F}$ if it is not inadmissible.
\end{definition}

If a variance bound $\VBsym$ is admissible with respect to some class $\mathcal{F}$, then there exists no variance bound in $\mathcal{F}$ that is strictly less conservative than $\VBsym$.
Thus, the admissibility concept is similar to Pareto optimality; an admissible bound is not optimal in some universal sense -- universally optimal bounds typically do not exist -- but it is optimal in the sense that it cannot be uniformly improved.
We can in principle consider admissibility with respect to any function class $\mathcal{F}$, but as we are considering quadratic variance bounds in this paper, we take the function class to be the set of such bounds: $\mathcal{F} = \allbounds$.

\begin{proposition}\label{prop:quadratic-admissible}
A quadratic variance bound $\boundM \in \allbounds$ is inadmissible with respect to $\allbounds$ if there exists a bound $\boundM' \in \allbounds$ such that $\boundM' \neq \boundM$ and $\boundM' \preceq \boundM$.
\end{proposition}

Proposition~\ref{prop:quadratic-admissible} demonstrates that the set of admissible quadratic bounds consists exactly of the minimal elements of $\allbounds$ with respect to the Loewner order.
Because this is a partial order, there will be many minimal elements, mirroring the fact that there exists no universally best bound.
Moreover, there will generally be infinitely many admissible quadratic bounds.

We say that a procedure for generating variance bounds is admissible if it produces admissible bounds for all input instances.
The procedures for deriving variance bounds that we describe in this paper are admissible by construction.
However, if a bound is constructed in some other way, it could be inadmissible.
In Section~\suppref{sec:testing} of the supplement, we describe a procedure for testing whether an arbitrary quadratic bound is admissible.
Experimenters can use this procedure to confirm that the variance estimator they are using is admissible.

Admissibility of variance bounds has not previously been considered in the literature.
The primary focus has instead been whether a bound is sharp.
The concept of sharpness has, however, rarely been formally defined, and statisticians have used at least two distinct sharpness concepts without clearly differentiating them.
The most common use of the term ``sharp'' is to describe a variance bound that is equal to the true variance for at least one set of potential outcomes.
The following proposition shows that, for quadratic variance bounds, admissibility is a stronger concept than this conventional definition of sharpness.

\begin{definition}\label{def:sharp}
A variance bound $\VBsym : \Reals^{K} \to \Reals$ is \emph{sharp} if there exists a nonzero vector of potential outcomes $\parav \in \Reals^{K}$ such that $\varbf{\parav} = \Varsub{\parav}{\est}$.
\end{definition}

\expcommand{\admissibleimpliessharp}{%
	If a quadratic variance bound $\boundM \in \allbounds$ is admissible with respect to $\allbounds$, then it is also sharp.
	However, there exist sharp quadratic variance bounds that are not admissible.
}
\begin{proposition} \label{prop:admissible-implies-sharp}
	\admissibleimpliessharp
\end{proposition}

The alternative sharpness concept, which we refer to as ``uniform marginal sharpness,'' is stronger than the sharpness concept in Definition~\ref{def:sharp}, but it has a narrower application and cannot be used in the settings we consider.
Uniform marginal sharpness has exclusively been used in the literature on Fréchet--Hoeffding-type variance bounds \citep[see, e.g.,][]{Aronow2014Sharp}.
Although not formally defined in this literature, sharpness is here understood to mean that a bound uses all available information in the marginal distributions of potential outcomes, meaning that the bound holds with equality for some potential outcomes for every possible set of marginal distributions.
While marginal distributions are useful as summaries of the potential outcomes under complete randomization, they are not apprioriate summaries in experiments with interference or complex designs.
Our focus in the main paper will therefore exclusively be on the sharpness concept in Definition~\ref{def:sharp}.
We provide a detailed discussion about the alternative sharpness concept in Section~\suppref{sec:fh-bounds} in the supplement.

\subsection{Examples of Variance Bounds}\label{sec:prev-examples}

We can use our formalization of the variance estimation problem to understand existing variance estimators.
Our first example is the variance estimator described by \citet{Neyman1923Application}.
In the setting of complete randomization with two equally sized treatment groups, Neyman showed that the variance of the difference-in-means estimator is $\Var{\est} = \paren{ \sigma^2_1 + \sigma^2_0 + 2 \rho } / \paren{n - 1}$, where $\sigma^2_1$ and $\sigma^2_0$ are the population variances of the potential outcomes under treatment and control among all units in the experiment, and $\rho$ is the covariance between the two potential outcomes.
The covariance is not estimable and must be bounded.
Neyman's solution was to use the Cauchy--Schwartz inequality followed by the AM-GM inequality on $\rho$ to obtain the upper bound $\Var{\est} \leq 2 \paren{ \sigma^2_1 + \sigma^2_0 } / \paren{n - 1}$.
This upper bound can be estimated using sample variances.%
\footnote{%
	\citet{Neyman1923Application} also considers the variance bound attained by applying only the Cauchy--Schwartz inequality, which is typically less conservative.
	This bound is not unbiasedly estimatable, but consistent estimators can be constructed under mild regularity conditions.
	The bound attained by also applying the AM-GM inequality is, however, more often used in practice \citep{Imbens2015Causal}.
}

The Neyman bound can be expressed in our framework.
As above, the variance of the estimator can be written $n^{-2} \parav^\tran \varM \parav$, where the covariance matrix is
\begin{equation}
	\varM = \Cov{\coefv} = \frac{n}{n-1}
	\begin{bmatrix}
		 \mat{H} & \mat{H} \\
		 \mat{H} & \mat{H}
	\end{bmatrix}
	\qquadtext{and}
	\mat{H} = \unitM - \onevec \onevec^\tran / n,
\end{equation}
and the potential outcome vector is $\parav = \paren[\big]{\y{1}{1}, \dotsc, \y{n}{1}, \y{1}{0}, \dotsc, \y{n}{0}}$.

The matrix $\varM$ is not design compatible because the diagonal elements in the off-diagonal blocks are nonzero, but the corresponding pairs of potential outcomes are never simultaneously observed.
For example, $(1, n + 1) \in \unobspairs$, so the product $\para{1}\para{n+1} = \y{1}{1} \y{1}{0}$ is never observed, but entry in row $1$ and column $n + 1$ of $\varM$ is one: $\varMe{1,n+1} = 1$.
To address this, the Neyman variance estimator implicitly uses the slack matrix
\begin{equation}
	\slackM = \frac{n}{n-1}
	\begin{bmatrix}
		\phantom{-}\mat{H} & -\mat{H} \\ -\mat{H} & \phantom{-}\mat{H}
	\end{bmatrix},
	\quad
	\text{yielding the variance bound}
	\quad
	\boundM =
	\frac{2n}{n-1}
	\begin{bmatrix}
		\mat{H} & 0 \\
		0 &  \mat{H}
	\end{bmatrix}.
\end{equation}
This bound is valid because $\slackM$ is positive semidefinite and $\slackMkl = - \varMkl$ for all $(k, \ell) \in \unobspairs$.

\expcommand{\Neymanisadmissible}{%
	Under complete randomization with two equally sized treatment groups and no interference, the Neyman bound is admissible with respect to the class of quadratic bounds $\allbounds$.
}
\begin{proposition}\label{prop:neyman-is-admissible}
	\Neymanisadmissible
\end{proposition}

Our second example is the class of variance estimators described by \citet{Aronow2013Conservative,Aronow2017Estimating}.
The authors consider variance estimation for the Horvitz--Thompson point estimator under arbitrary experimental designs, and they describe a bound based on Young's inequality for products.
The most straightforward version of Young's inequality states that $2xy \leq x^2 + y^2$ for any two real numbers $x$ and $y$.
Recall that the central problem here is that the variance contains terms $\varMkl \parak \paral$ such that $\parak \paral$ is unobservable and $\varMkl$ is not zero.
To address this, the Aronow--Samii bound apply Young's inequality separately on each of these problematic terms: $\varMkl \parak \paral \leq \abs{\varMkl} \paren[\big]{\parak^2 + \paral^2} / 2$.

We can use our quadratic form representation to write the Aronow--Samii bound as a slack matrix.
Let $\partMkl$ be a $K \times K$ matrix with zeros entries except in the $(k, \ell)$th block, which instead is given by $\abs{\varMkl}$ in the diagonal entries $(k,k)$ and $(\ell,\ell)$, and $-\varMkl$ in the off-diagonal entries $(k,\ell)$ and $(\ell,k)$.
The slack matrix corresponding to the Aronow--Samii bound is $\slackM = \sum_{(k, \ell) \in \unobspairs} \partMkl / 2$.
This bound is design compatible because $\slackMkl = - \varMkl$ by construction for all $(k, \ell) \in \unobspairs$.
Furthermore, because all matrices $\partMkl$ are positive semidefinite, their sum $\slackM$ will also be positive semidefinite.
Hence, the bound is conservative.
However, as the following proposition shows, it is not admissible.

\expcommand{\ASinadmissible}{%
The Aronow--Samii bounding procedure is inadmissible in the class of quadratic bounds $\allbounds$.
}

\begin{proposition}\label{prop:AS-inadmissible}
	\ASinadmissible
\end{proposition}

The bound $B_2$ in the illustration in Section~\ref{sec:illustration} is the Aronow--Samii bound, so the fact that bound $B_1$ dominates $B_2$ is an instance of Proposition~\ref{prop:AS-inadmissible}.
It is possible to construct similar examples in more involved settings, including with larger samples and more intricate designs, but we omit these examples in the interest of space.%
\footnote{%
\citet{Aronow2013Class} provide a refinement to the Aronow--Samii bound in the case of a cluster-randomized design, suggesting that it was known that the original bound was not admissible at least in that case.
However, this refinement can be shown to also be inadmissible in the class of quadratic bounds.
}

We provide a third example in Section~\suppref{sec:imai-bounds} in the supplement, where we consider Imai's variance bound for experiments using the matched-pair design in settings without interference.
We show that this bound is admissible in the class of quadratic bounds.
The bounds by \citet{Fogarty2018Mitigating} and \citet{Pashley2021Insights} are similar to Neyman's and Imai's bounds, suggesting that they also are admissible in their respective settings, but it is beyond the scope of the current paper to formally investigate this.

\section{Constructing Variance Bounds}\label{sec:finding-bounds}

\subsection{Variance Bound Programs}\label{sec:var-bound-programs}

There is currently no method that allows experimenters to construct variance estimators that minimize conservativeness.
Indeed, there exists no method to even construct admissible variance bounds for general exposure mappings, designs and estimators.
To address this, we describe a computational approach that selects a variance bound from $\allbounds$ using an optimization formulation.
For some real-valued function $\objsym$ on symmetric matrices, we aim to find a slack matrix $\slackM \in \allslacks$ that minimizes $\objsym$.
We refer to this procedure as \optprocedure{}, which is the following mathematical program:
\[
\tag{\optprocedure{}}
\slackM^* \in \argmin_{\slackM \in \allslacks} \objfS.
\]
The properties of a variance bound constructed in this way, and the associated variance estimator, will depend on the choice of objective function $\objsym$.
The ideal objective function is $\objfS = \parav^\tran \slackM \parav$, where $\parav$ refers to the true potential outcomes, because then the objective captures the actual conservativeness in the current experiment.
But such an objective is infeasible, as it would require exact knowledge of the potential outcomes.
Instead, the approach we explore in this paper is to encode in $\objsym$ the experimenter's preferences concerning risk trade-offs and any background knowledge they might have about the potential outcomes.

Unless otherwise noted, all objective functions discussed in this paper are such that they strictly penalize matrices that are weakly larger in the Loewner order, which we refer to as strict monotonicity.
Strict monotonicity ensures that the variance bound produced by \optprocedure{} using the objective is admissible.
This result was previewed in Section~\ref{sec:illustration}, and the following definition and proposition present the formal result in full.

\begin{definition}\label{def:strictly-monotone}
A real-valued function $g$ on symmetric matrices is \emph{strictly monotone} if $g(\mat{Q}) < g(\mat{P})$ whenever $\mat{Q} \neq \mat{P}$ and $\mat{Q} \preceq \mat{P}$.
\end{definition}

\expcommand{\varboundprograms}{%
If the input objective function $\objsym$ is strictly monotone, then \optprocedure{} returns a variance bound that is conservative, design compatible and admissible in finite samples.
}

\begin{theorem}\label{thm:var-bound-programs}
	\varboundprograms
\end{theorem}

Definition~\ref{def:strictly-monotone} differs from the conventional definition of strict monotonicity based on the strict Loewner order.
The conventional definition states that a strictly monotone function $f$ satisfies $f(\mat{Q}) < f(\mat{P})$ whenever $\mat{Q} \prec \mat{P}$.
This definition, though well-motivated in many applications, does not align with our notion of admissibility, necessitating us to extend it slightly.

The \optprocedure{} program is generally computationally tractable.
As a rule of thumb, an optimization program is tractable if it is convex \citep{Rockafellar1993Lagrange,Boyd2004Convex}.
The set of slack matrices $\allslacks$ is convex, implying that \optprocedure{} is a convex program if $\objsym$ is a convex function.
All objective functions considered in this paper are convex, so they admit efficient algorithms for finding optimal solutions, up to desired tolerances.
However, the implementation of these algorithms might still be practically challenging.

\subsection{Norm Objectives}\label{sec:norm-objectives}

We will first consider when an experimenter has little or no background knowledge about the potential outcomes.
Our goal here is to select a variance bound that is not excessively conservative for most potential outcomes.
This corresponds to selecting a quadratic form of small magnitude.
We will use the family of Schatten $p$-norms of the bounds' coefficient matrices to make this idea precise.

There is an implicity trade-off between average performance and worst-case performance when selecting a norm for the coefficient matrix, corresponding to the experimenter's risk preference concerning excessive conservativeness.
To understand how the Schatten norms capture this risk trade-off, consider the spectral decomposition of the coefficient matrix of a bound $\boundM \in \allbounds$.
Because the matrix is symmetric, we can write $\boundM = \sum_{k=1}^{K} \eigvalk \eigveck \eigveck^\tran$, where $\eigveck$ is the $k$th eigenvector of $\boundM$ and $\eigvalk$ is the corresponding $k$th eigenvalue.
This allows us to write any quadratic bound as
\begin{equation}
	\varbf{\parav}
	= \frac{1}{n^2} \parav^\tran \boundM \parav
	= \frac{\norm{\parav}^2}{n^2} \sum_{k=1}^{K} w_k \eigvalk,
\end{equation}
where $w_k = \iprod{\parav, \eigveck}^2 / \norm{\parav}^2$ captures the alignment of the potential outcome vector to the $k$th eigenvector $\eigveck$.
Because $\boundM$ is positive semidefinite, all eigenvalues are non-negative.
By construction, the coefficients $w_k$ are non-negative and sum to one, so they act as weights in a convex combination of the eigenvalues.
The conservativeness of the variance bound is therefore determined by the eigenvalues and the alignment of the potential outcomes to the eigenvectors of $\boundM$.
If we make the eigenvalues of the variance bound matrix $\boundM$ small, we ensure that the bound is not excessively conservative.

The Schatten norms are different ways of measuring the magnitude of these eigenvalues.
Formally, a Schatten $p$-norm of $\boundM$ is the usual $p$-norm applied to the vector of singular values of $\boundM$, which in our case coincide with the eigenvalues:
$
	\pnorm{p}{\boundM} = \paren[\big]{ \sum_{k=1}^{K} \abs{\eigvalk}^p }^{1/p}.
$
When $p$ is small, the norm tolerates a few large eigenvalues if it means that many other eigenvalues are small.
When $p$ is large, the norm is disproportionately affected by large eigenvalues, diminishing the influence of smaller eigenvalues.
Therefore, the choice of $p$ is directly related to the experimenter's risk preferences.
A risk averse experimenter would want to avoid a situation with a few large eigenvalues even if most eigenvalues are small, as there is then a risk that the bound is excessively conservative if the potential outcomes happen to align with an eigenvector corresponding to one of the few large eigenvalues.
To avoid such situations, the experimenter would use a Schatten $p$-norm with a large $p$.
A risk tolerant experimenter on the other hand would prefer a Schatten $p$-norm with a small $p$, as they tolerate bounds that occasionally are excessively conservative provided that the eigenvalues are smaller overall, leading to less conservativeness in general.

The following proposition shows that we achieve admissibility no matter the choice of $p$.

\expcommand{\thmSchattenIsMonotone}{
For all $p \in [1, \infty)$, the Schatten $p$-norm objective $\objfS = \pnorm{p}{\varM + \slackM}$ is strictly monotone, ensuring that the variance bound produced by \optprocedure{} using $\objsym$ is admissible.
}
\begin{theorem}\label{thm:schatten-is-monotone}
\thmSchattenIsMonotone
\end{theorem}

The Schatten $p$-norm coincides with some more familiar matrix norms for particular values of $p$.
If we set $p = 1$, the Schatten $p$-norm is simply the sum of the absolute values of the eigenvalues.
This is the nuclear norm, which also is called the trace norm.
Using this norm produces a bound with the best average performance, in the sense that it puts uniform weight on all eigenvalues no matter their magnitude.

At the other extreme, when we let $p \to \infty$, we obtain the operator norm induced by the $2$-norm, which in our setting coincides with the maximum eigenvalue of $\boundM$.
An experimenter who is maximally risk adverse would use this norm, as it would trade-off any amount of average conservativeness for even a minute reduction is worst-case conservativeness.
The operator norm is not strictly monotone according to Definition~\ref{def:strictly-monotone}, so an arbitrary minimizer of an objective function using this norm is not guaranteed to be admissible.
This can be addressed by using a large but not infinite $p$-norm, which will behave like the operator norm for practical purposes.
Alternatively, we describe a regularization procedure of the operator norm in Section~\suppref{sec:reg-of-operator-norm} of the supplement that ensures admissibility.

Finally, we recover the Frobenius norm when $p = 2$.
This norm provides an intermediate point in the risk trade-off.
It disproportionately penalizes large eigenvalues, making sure that no eigenvalue gets very large, but it does not ignore the smaller eigenvalues completely.

\subsection{Targeted Linear Objectives}\label{sec:linear-obj}

The norm objectives in the previous subsection cannot encode background knowledge experimenters might have about the potential outcomes.
We describe a class of targeted objective functions to fill this role.
The prior knowledge the experimenter encodes in the objective function need not be correct, not even approximately, to ensure the validity and admissibility of the resulting variance bound.
But if they are able to provide reasonably accurate information, the bound will be less conservative.
This idea is related to the model-assisted tradition that originated in the literature on design-based survey sampling (see, e.g., \citealp{Saerndal1992Model}, and \citealp{Basse2018Model}).

The class of \emph{targeted linear objectives} takes the form $\objfS = \iprod{\slackM , \objM}$,
where $\slackM$ is a slack matrix, $\objM$ is a positive semidefinite targeting matrix of the same dimensions, and $\iprod{ \cdot, \cdot}$ denotes the trace inner product on matrices: $\iprod{\slackM , \objM} = \tr{\slackM \objM}$.
The purpose of $\objM$ is to target particular potential outcomes, motivated by prior substantive knowledge.
All objective functions in this class are linear in the coefficients of the slack matrix, so the optimization problem underlying \optprocedure{} becomes a semidefinite program, ensuring computational tractability.

By construction, the bound returned by \optprocedure{} using a targeted linear objective will be valid.
What makes the class of targeted linear objectives stand out compared to the norm objectives is a type of completeness result.
Namely, the class of targeted linear objectives completely characterizes the set of all admissible variance bounds.

\expcommand{\thmsdpprocedureadmissibility}{%
A bound $\boundM$ is admissible if and only if it can be obtained from \optprocedure{} using the objective function $\objfS = \iprod{\slackM , \objM}$ for some positive definite targeting matrix $\objM$.
}

\begin{theorem}\label{thm:sdp-procedure-admissibility}
	\thmsdpprocedureadmissibility
\end{theorem}

The proof that every bound returned by \optprocedure{} using a positive definite targeting matrix is admissible proceeds by showing that every targeted linear objective is strictly monotone and then appeals to Theorem~\ref{thm:var-bound-programs}.
The proof of the opposite direction, that every admissible bound can be obtained as a solution to \optprocedure{} using some targeted linear objective, is more involved and appeals to the separating hyperplane theorem from convex analysis.
The proof is provided in the supplement.

Theorem~\ref{thm:sdp-procedure-admissibility} shows that we always obtain an admissible bound when we use a targeted linear objective with a positive definite targeting matrix.
Furthermore, due to the one-to-one correspondence between admissible bounds and targeted linear objectives, the theorem allows us to re-interpret other procedures for constructing variance bounds by showing what matrix they implicitly target, which by extension shows what potential outcomes they implicitly target.

\subsection{Example on the Construction of Targeting Matrices}\label{sec:targeting-matrix}

\newcommand{\arbv}{\widehat{\parav}}

Recall that the variance bound using coefficients $\boundM = \varM + \slackM$ is
\begin{equation}
	n^2 \varbf{\parav}
	= \parav^\tran \boundM \parav
	= \parav^\tran \varM \parav
	+ \parav^\tran \slackM \parav.
\end{equation}
If the true potential outcomes were known, the experimenter would use the targeting matrix $\objM = \parav \parav^\tran$, because it directly targets the actual conservativeness of the bound: $\iprod{\slackM, \objM} = \parav^\tran \slackM \parav$.
Of course, the challenge here is that the potential outcomes are unknown.

As an example of how the targeting matrix can be constructed, suppose the experimenter has some prior, partial knowledge about the potential outcomes, and they encode that knowledge in a generative model.
That is, we would consider $\parav$ as a random variable drawn from some known distribution.
Seen from this perspective, the value of the variance bound is random, because the randomness of $\parav$ is passed on to $\varbf{\parav}$.
A natural target would then be to minimize the expectation of the variance bound $\varbf{\parav}$ with respect to the stipulated generative model.

We use a subscripted expectation operator $\Esub{\parav}{\cdot}$ to denote the expectation with respect to the imagined distribution of $\parav$, rather than the true randomization distribution induced by the experimental design, as in the rest of the paper.
The expected value of a variance bound given by $\boundM = \varM + \slackM$ is then
\begin{equation}
	n^2 \Esub{\parav}{\varbf{\parav}}
	= \iprod[\big]{\boundM , \Esub{\parav}{\parav \parav^\tran}}
	= \iprod{\varM , \Esub{\parav}{\parav \parav^\tran}}
	+ \iprod{\slackM , \Esub{\parav}{\parav \parav^\tran}}.
\end{equation}
To take advantage of their prior knowledge about the potential outcomes, the experimenter would use a targeted linear objective with matrix $\objM = \Esub{\parav}{\parav \parav^\tran}$, because this directly minimizes the expected variance bound under the stipulated generative model.

It should be emphasized that the interpretation of $\parav$ as a random variable here is simply a convenient way to express prior (partial) knowledge about the potential outcomes.
It is not assumed nor required for any of our results that the stipulated generative model accurately reflects how the potential outcomes actually were generated.
The resulting bound is valid no matter what distribution one uses for $\parav$, and the resulting bound is admissible as long as $\Esub{\parav}{\parav \parav^\tran}$ is positive definite.
However, the bound will be less conservative if the distribution is a good approximation of the true potential outcomes.

Experimenters following this approach should take care to ensure that $\objM = \Esub{\parav}{\parav \parav^\tran}$ indeed is positive definite.
Targeting matrices that are not positive definite disregard some dimensions of the potential outcome vector space, meaning that they do not penalize excessive conservativeness in those dimensions.
A simple way to ensure that a targeting matrix is positive definite is to include independent noise in the generative model.

To make this example more concrete, consider when the experimenter knows, or presumes to know, that the potential outcomes can be well-approximated by a linear function of some set of covariates.
For simplicity, we consider when there are two exposures, $\exa$ and $\exb$, so the potential outcome vector is $\parav = \paren[\big]{\y{1}{\exa}, \dotsc, \y{n}{\exb}}$.
Letting $\xM$ be a $n$-by-$m$ matrix collecting $m$ covariates for the $n$ units, a possible generative model for the potential outcomes could be written as $\parav = ( \xM \vec{\beta}_{\exa} + \vec{\varepsilon}_{\exa}, \xM \vec{\beta}_{\exb} + \vec{\varepsilon}_{\exb} )$, where $\vec{\beta}_{\exa}$ and $\vec{\beta}_{\exb}$ are coefficient vectors describing how the covariates relate to the potential outcomes, and $\vec{\varepsilon}_{\exa}$ and $\vec{\varepsilon}_{\exb}$ describe aspects of the potential outcomes not captured by the covariates.

The covariate matrix $\xM$ is observed and fixed, but we might not have a good sense of $\paren{\vec{\beta}_{\exa}, \vec{\beta}_{\exb}, \vec{\varepsilon}_{\exa}, \vec{\varepsilon}_{\exb}}$.
We can express our ignorance about these vectors as a distribution.
For simplicity, we consider when the coefficients $\vec{\beta}_{\exa}$ and $\vec{\beta}_{\exb}$ are fixed, so they are non-random vectors, and the coordinates of $\paren{\vec{\varepsilon}_{\exa}, \vec{\varepsilon}_{\exb}}$ are independent and follow a standard normal distribution.
With this generative model, the targeting matrix becomes
\begin{equation}
	\objM = \Esub{\parav}{\parav \parav^\tran} =
	\begin{bmatrix}
		\xM & 0 \\
		0 & \xM
	\end{bmatrix}
	\begin{bmatrix}
		\vec{\beta}_{\exa} \\
		\vec{\beta}_{\exb}
	\end{bmatrix}
	\begin{bmatrix}
		\vec{\beta}_{\exa}^\tran & \vec{\beta}_{\exb}^\tran
	\end{bmatrix}
	\begin{bmatrix}
		\xM^\tran & 0 \\
		0 & \xM^\tran
	\end{bmatrix}
	 + \unitM.
\end{equation}
More intricate generative working models yield more elaborate targeting matrices.

While this type of generative working model can be constructed in any way that reflect the experimenters prior knowledge about the potential outcomes, experimenters should be careful using data from their current experiment in this exercise.
Using data from a pilot study or baseline outcome measurements is unproblematic, but constructing the model based on outcome data collected after randomization could make the bound anticonservative.
Such an approach could potentially work well in large samples under suitable regularity conditions and provided that the objective function is sufficiently regularized.
For example, the act of estimating the coefficient vectors $\vec{\beta}_{\exa}$ and $\vec{\beta}_{\exb}$ in a model similar to the one above from the experimental data will typically not introduce instability in the variance estimators in large samples due to the addition of the noise terms, thereby maintaining validity and admissibility.
This follows from standard results in the literature on sensitivity analysis in optimization \citep[see, e.g.,][]{Fiacco1990Sensitivity}, which is unrelated to the literature on sensitivity analysis in causal inference.
It is, however, beyond the scope of the current paper to explore this approach.

\subsection{Composite Objectives}\label{sec:composite-obj}

There are situations where experimenters want a combination of properties offered by different objective functions.
Using the fact that monotonicity is maintained under positive combinations, the following proposition shows that a combination of elementary objectives can be used with \optprocedure{}.

\expcommand{\compositeobjective}{%
If $g$ is strictly monotone and $h$ is monotone, then the function $g + \gamma h$ is strictly monotone for any $\gamma \geq 0$.
}

\begin{proposition}\label{prop:composite-objective}
	\compositeobjective
\end{proposition}

One situation in which a composite objective is useful is when an experimenter wants to regularize a targeted linear objective, perhaps because they are not very confident in the information encoded in the targeting matrix.
They can then use a composite objective that includes one of the norm objectives discussed in Section~\ref{sec:norm-objectives}.
For some Schatten $p$-norm and coefficient $\gamma > 0$, deciding the relative focus on the two objectives, the composite objective function is
\begin{equation}
	\objfS = \iprod{\slackM, \objM} + \gamma \pnorm{p}{\varM + \slackM}.
\end{equation}
If $p \in [1, \infty)$, this composite objective is strictly monotone even if $\objM$ is only positive semidefinite, so the composite objective always yields a bound that is conservative, estimable and admissible.

\section{Estimating Variance Bounds}\label{sec:estimate-vb}

\subsection{Precision of Variance Bound Estimator}

A quadratic form can be reinterpreted as a linear function of the elements $\parak \paral$ of the outer product $\parav \parav^\tran$.
This means that we can use any estimator in the class of linear estimators to estimate a quadratic variance bound itself once it has been derived, yielding a conservative variance estimator.
For example, experimenters could take advantage of background information about the units and use a covariate-adjusted version of the estimator, similar to the GREG or AIPW estimators.
It is, however, beyond the scope of this paper to investigate which of these estimators is best suited for estimation of quadratic forms.
In this section, we will instead consider how the choice of the variance bound influences how challenging the estimation task becomes.

We restrict our focus to the Horvitz--Thompson estimator of the bounds.
This estimator is sufficiently simple so as to not distract from the main ideas and insights we aim to explore.
For a variance bound $\boundM$, the corresponding Horvitz--Thompson estimator is
\begin{equation}
	\varbfest{\parav}
	=
	\frac{1}{n^2} \sum_{k \in \Obs} \sum_{\ell \in \Obs} \frac{\boundMkl \parak \paral}{\Pr{k, \ell \in \Obs}},
\end{equation}
where, as above, $\Obs$ is the random subset that collects indices of the observed potential outcomes and $\Pr{i,j \in \Obs}$ is the probability of simultaneously observing potential outcomes $\parak$ and $\paral$.

The Horvitz--Thompson estimator is unbiased whenever the variance bound is design compatible.
However, unbiasedness does not ensure that the estimator is precise.
While the precision of the estimator critically depends on the potential outcomes and the experimental design, the experimenter's choice of bound also plays a part.
To explore this, we define
\begin{equation}
	R_{k\ell} = \frac{\indicator{k, \ell \in S} \times \indicator{\boundMkl \neq 0}}{\Pr{k, \ell \in S}}
\end{equation}
to be a random variable for each pair $(k, \ell) \in \Pop \times \Pop$, capturing the inverse propensity weighting done by the variance bound estimator.
We define the ratio of zero and zero to be zero, meaning that $R_{k\ell} = 0$ if potential outcomes $k, \ell \in \Pop$ are never observed simultaneously.
Collecting the $K^2$ variables $R_{k\ell}$ in a vector $\vec{R}$, we use the covariance matrix $\Cov{\vec{R}}$ to characterize the precision of the variance bound estimator.

\expcommand{\finitesampleMSEbound}{%
If a variance bound $\boundM$ is design compatible, the normalized mean squared error of the Horvitz--Thompson estimator of the bound is bounded as
\begin{equation}
	\E[\Big]{ \paren[\big]{ n \varbfest{\parav} - n \varbf{\parav} }^2 }
	\leq
	\frac{1}{n^2}
	\pnorm{\infty}{\Cov{\vec{R}}} \times
	\pnorm{\infty}{\parav}^2 \times
	\pnorm{2}{\boundM}^2,
\end{equation}
where $\pnorm{\infty}{\Cov{\vec{R}}}$ is the operator norm of the covariance matrix of the inverse propensity variables, $\pnorm{\infty}{\parav} = \max_{k \in \Pop} \abs{\parak}$ is the largest magnitude of the potential outcomes, and $\pnorm{2}{\boundM}^2$ is the squared Frobenius norm of the coefficient matrix of the bound.
}

\begin{proposition}\label{prop:finite-sample-MSE-bound}
	\finitesampleMSEbound
\end{proposition}

The proposition allows us to consider the design, potential outcomes and variance bound separately when building understanding of the behavior of variance bound estimators.
The factor $\pnorm{\infty}{\Cov{\vec{R}}}$ captures aspects of the experimental design.
This norm will be small for designs that do not induce too much dependence between the exposures.
A design that induces highly correlated exposures might make precise estimation impossible even in large samples, which would then be reflected in a large $\pnorm{\infty}{\Cov{\vec{R}}}$.

Importantly, while properties of the design that facilitate precise point estimation roughly coincide with those that facilitate precise variance estimation, they do not coincide exactly.
In particular, the performance of the point estimator is governed by first-order exposure probabilities, but the construction of $\vec{R}$ uses second-order probabilities.
A design can have well-behaved first-order probabilities but still have many second-order probabilities being close to zero.
In such cases, experimenters should consider extending Definition~\ref{def:design-compatible} so that design compatibility requires $\Pr{k, \ell \in S} \geq c$ for some constant $c > 0$, rather than just not being zero.
This will make the bound more conservative, but one would ensure that $\pnorm{\infty}{\Cov{\vec{R}}}$ is well-controlled.
For the purpose of this section, we will proceed under the presumption that experimenters have taken the steps necessary to ensure that $\pnorm{\infty}{\Cov{\vec{R}}}$ is well-controlled.

The factor $\pnorm{\infty}{\parav}$ captures the scale of the potential outcomes.
If the potential outcomes are large in magnitude, the estimator will naturally be less precise in absolute terms.
For simplicity, we use the uniform norm to measure the scale of the potential outcomes.
If the potential outcomes are known to be in some interval, this norm is asymptotically bounded by construction.
However, the uniform norm can paint an overly pessimistic picture, and we show in Section~\suppref{app:consistent-var-est} of the supplement that it is possible to replace the uniform norm with other moments of the potential outcomes, at the cost of making the mean square error bound more sensitive to outliers among the coefficients in the variance bound $\boundM$.

The final factor $\pnorm{2}{\boundM}^2$ measures the magnitude of the coefficients in the variance bound.
If $\pnorm{2}{\boundM}^2$ is large relative to $\pnorm{2}{\varM}^2$, then the bound is achieving design compatibility by overweighting a subset of the potential outcome products, making the variance bound estimators disproportionally sensitive to estimation errors in those terms.
The following corollary, which follows directly from Proposition~\ref{prop:finite-sample-MSE-bound}, states that control over $\pnorm{2}{\boundM}^2$ ensure that the error of the variance bound estimator is small with high probability in large samples.

\begin{corollary}\label{cor:consistency-var-bound}
	If $\pnorm{\infty}{\Cov{\vec{R}}}$ and $\pnorm{\infty}{\parav}$ are asymptotically bounded and $\pnorm{2}{\boundM}^2$ is dominated by $n^2$, then the variance bound estimator is consistent: $n \varbfest{\parav} - n \varbf{\parav} = \littleOp{1}$.
\end{corollary}

The corollary suggests that a useful heuristic to improve precision of the variance bound estimator is to make $\pnorm{2}{\boundM}^2$ small.
A way to achieve this is to use the Frobenius norm as the objective in \optprocedure{}, as discussed in Section~\ref{sec:norm-objectives}.
This will not ensure consistency, as there may be no valid bound with sufficiently small norm.
In cases where the potential outcomes are not bounded, experimenters should consider using a Schatten $p$-norm for some $p > 2$ to account for the fact that $\pnorm{\infty}{\parav}$ might not be well-controlled.

\subsection{Accuracy With Respect to the True Variance}\label{sec:accuracy-wrt-true-var}

The previous subsection considered the precision of the variance bound estimator with respect to the variance bound itself.
This does not account for the fact that the variance bound potentially could be very conservative, in which case the variance bound estimator would give a misleading picture of the precision of the point estimator even if it is itself precise.
We can address this using a composite objective, as discussed in Section~\ref{sec:composite-obj}.

Consider the normalized mean square error of the variance bound estimator $\varbfest{\parav}$ with respect to the true variance $\Var{\est}$.
Using the usual bias--variance decomposition, we can write the overall error with respect to the variance as
\begin{equation}
	\E[\Big]{ \paren[\big]{ n \varbfest{\parav} - n \Var{\est} }^2 }
	= \paren[\big]{ n \varbf{\parav} - n \Var{\est} }^2
	+ \E[\Big]{ \paren[\big]{ n \varbfest{\parav} - n \varbf{\parav} }^2 }.
\end{equation}
The first term is the slack that is introduced to ensure that the variance bound design compatible.
This was the focus of the investigation in Section~\ref{sec:finding-bounds}.
For example, if following the model-assisted approach described in Section~\ref{sec:targeting-matrix}, we would use the inner product of $\objM = \Esub{\parav}{\parav \parav^\tran}$ and $\boundM$ as a proxy for this first term.
The second term is the precision of the variance bound estimator with respect the bound itself, which was the focus of the previous subsection.
We could use the bound from Proposition~\ref{prop:finite-sample-MSE-bound} as proxy for this second term.
This leads to the following composite objective as a heuristic for the mean square error of the variance estimator:
\begin{equation}
	\objfS = \iprod{\slackM, \objM} + \gamma \pnorm{2}{\varM + \slackM}^2,
	\qquadwhere
	\gamma = \pnorm{\infty}{\Cov{\vec{R}}} \times \pnorm{\infty}{\parav}.
\end{equation}

It is not generally obvious what the appropriate relative weighting of the two terms is in this objective, and $\gamma$ effectively functions as a tuning parameter for the composite objective.
However, the bound will be valid and admissible as long as $\gamma > 0$.

\section{Numerical Illustration}\label{sec:simulations}

In our simulation exercise, we use data collected by \citet{Paluck2016Changing} from a randomized network experiment involving 24,183 students in 56 public middle schools in New Jersey.
The study investigated the effectiveness of an intervention that aimed to reduce conflict and bullying among adolescents.
The authors were particularly interested in measuring the spillover effects of the intervention on perceived social norms and behavior in the students' peer networks.
In addition to treatment status and outcome data, the authors recorded the self-reported social networks among students as well as numerous covariates.
Our goal here is not to perform a re-analysis of this study, but rather to use the original data to construct an empirical setting that reasonably reflects a real-world study.

The experimental design used by \citet{Paluck2016Changing} was a two stage randomization process.
A set of ``seed students,'' who were well-connected in the social network, was selected in each school.
The 56 schools were grouped into 14 blocks and half of the schools in each block were selected at random to receive treatment.
Among the schools selected to receive treatment, half of the seed students were selected at random to receive the intervention.
The exposure mapping is defined by a tuple of binary variables $(s_i, z_i, a_i)$, where $s_i$ indicates whether the school was selected for treatment, $z_i$ indicates whether the student received the intervention, and $a_i$ indicates whether at least one of the student's peers in the network received the intervention.
In our simulations, we focus on estimating the direct effect of the intervention, corresponding to the contrast between exposures $e_1 = (1,1,0)$ and $e_0 = (0,0,0)$.
We focus on these exposures because they have the most second-order positivity violations in the experiment, making the problem of finding a variance bound particularly challenging.
We consider other exposures in the online supplement.
Because only seed students can receive the exposure $e_1 = (1,1,0)$ under this design, the effect cannot be estimated for all students.
Instead, we focus to the subset of 2,906 seed students who receive each exposure with a probability of at least $0.5\%$.

We consider two types of potential outcomes.
In the first setting, the outcomes are generated by a linear function of a set of observed covariates $\parav = ( \xM \vec{\beta}_{\exa} + \vec{\varepsilon}_{\exa}, \xM \vec{\beta}_{\exb} + \vec{\varepsilon}_{\exb} )$, in line with the generative model discussed in Section~\ref{sec:targeting-matrix}.
The included covariates are age, height, weight, gender, and grade, normalized to have zero mean and unit standard deviation.
We use $\vec{\beta}_{\exa} = 11/10 \times \mathbf{1}$ and $\vec{\beta}_{\exb} = 9/10 \times \mathbf{1}$ for the coefficients, and $\vec{\varepsilon}_{\exa}$ and $\vec{\varepsilon}_{\exb}$ are independent standard normal.
In the second setting, we use two reported outcomes from the original study: adoption of an anti-bullying wristband and school-reported disciplinary actions.
The potential outcome under exposure $e_1 = (1,1,0)$ is the wristband outcome and the potential outcome under exposure $e_0 = (0,0,0)$ is the disciplinary action outcome.
Because we are mixing different types of outcomes, this exercise does not capture a real-world causal effect, but the approach allows us to use the reported data unaltered, and it retains any peculiarities of the outcome distributions.

We use the Horvitz--Thompson estimator to estimate the effect.
We construct the variance matrix $\varM$ using a Monte Carlo with 5 million replicates from the experimental design.
The set of unobservable products $\unobspairs$ consists of all unit-exposure pairs that are realized with probability of less than $0.2\%$.
We examine variance bounds produced by several objective functions described in the paper: (i) the trace norm, (ii) the Frobenius norm, (iii) a targeted linear objective, and (iv) a composite objective.
We also examine the Aronow--Samii bound, which is the only existing bound that can be used in this setting.
The matrix used in the targeted linear objective is constructed as described in Section~\ref{sec:targeting-matrix} with $\vec{\beta}_{\exa} = \vec{\beta}_{\exb} = \mathbf{1}$ and standard normal noise, using the same covariates as above.
The targeted linear objective is therefore nearly correctly specified for the synthetic outcomes based on a linear model, but likely not particularly informative for the outcomes based on real-world data.
The first setting can be seen as representing a best-case scenario for the approach we describe in this paper, while the second setting can be seen as a more typical scenario.
The composite objective is the targeted linear objective, without noise, with a Frobenius penalty as described in Section~\ref{sec:accuracy-wrt-true-var} with $\gamma = 1$.

In order to solve the \optprocedure{} program, we use the JuMP modeling software \citep{Dunning2017} and the SCS solver \citep{scs}.
We use the Horvitz--Thompson estimator for quadratic forms to estimate the bounds as described in Section~\ref{sec:estimate-vb}.
We construct Wald-type 95\% confidence intervals using the square root of the variance estimator as an estimate for the standard error of the point estimator.

\begin{table}[ht]
\centering
\caption{Simulation results}\label{tab:sim-res}
\resizebox{.99\textwidth}{!}{%
	\begin{tabular}{lrrrrcrrrr}
	\toprule
	& \multicolumn{4}{c}{Panel A: Synthetic outcomes} & & \multicolumn{4}{c}{Panel B: Real data outcomes} \\ \cmidrule{2-5} \cmidrule{7-10}
	& Bias & Precision & Coverage & Width & & Bias & Precision & Coverage & Width \\
	\midrule
	Aronow-Samii & 4.0461 & 0.771 & 1.000 & 1.000 &   & 2.781 & 0.707 & 1.000 & 1.000 \\
Trace & 0.1144 & 0.417 & 0.968 & 0.465 &   & 0.770 & 0.419 & 0.992 & 0.682 \\
Frobenius & 0.0947 & 0.391 & 0.961 & 0.461 &   & 0.764 & 0.392 & 0.992 & 0.682 \\
Targeted & 0.0452 & 0.370 & 0.960 & 0.450 &   & 0.743 & 0.436 & 0.991 & 0.677 \\
Composite & 0.0301 & 0.348 & 0.955 & 0.448 &   & 0.764 & 0.391 & 0.992 & 0.682\\ \bottomrule
	\multicolumn{10}{l}{\footnotesize\emph{Note:} Based on 5 million simulation replicates.}
\end{tabular}
}
\end{table}

The simulation results are presented in Table~\ref{tab:sim-res}, with one panel for each of the two outcomes.
The first column in each panel presents the bias of each variance estimator relative to the true variance: $\braces{\E{ \varbfest{\parav} } - \Var{\est}} / \Var{\est}$.
We find that the Aronow--Samii variance estimator introduces bias of $405\%$ for the synthetic outcomes and $278\%$ for the outcomes based on real data.
The bias for the estimators based on the optimized variance bound are consistently smaller, being less than $12\%$ for the synthetic outcomes and around $76\%$ for the real data outcomes.
The bias is therefore between $3.6$ and $134$ times larger for estimators based on the Aronow--Samii bound compared to the optimized bounds.
The targeted linear objectives (with and without penalty) have the smallest bias for the synthetic outcomes, reflecting the fact that the covariates are highly informative in that setting.
The four optimized variance estimators have similar biases for the real data outcomes, indicating that the covariates are less informative here.

The second column presents the precision of the variance estimators, as measured by their standard errors relative to the true variance: $\mathrm{SD}\bracket{ \varbfest{\parav} } / \Var{\est}$.
We find that all variance estimators based on optimized bounds are more precise than the Aronow--Samii estimator, showing that we do not achieve less bias at the cost of imprecision when using the optimized bounds.
The third column presents coverage rates for confidence intervals at the $95\%$ nominal level.
The variance estimators are conservative by construction, so all intervals overcover, as expected.
Conditions ensuring asymptotic validity of Wald-type intervals for this point estimator when paired with a conservative, consistent variance estimator are provided by \citet{Aronow2017Estimating}.
However, confidence intervals based on the Aronow--Samii variance estimator stand out with $100\%$ coverage for both outcomes.
The last column presents average width of confidence intervals constructed based on each variance estimator relative to the confidence intervals based on the Aronow--Samii estimator.
As expected from the reduction in bias, we find that intervals based on optimized bounds are narrower, making the confidence intervals more informative.
The intervals are less than half as wide for the synthetic outcomes and about $32\%$ narrower for the outcomes based on real data.

\begin{figure}
	\centering
	\subfigure[Synthetic Outcomes]{%
		\label{fig:first}%
		\includegraphics[width=0.47\textwidth]{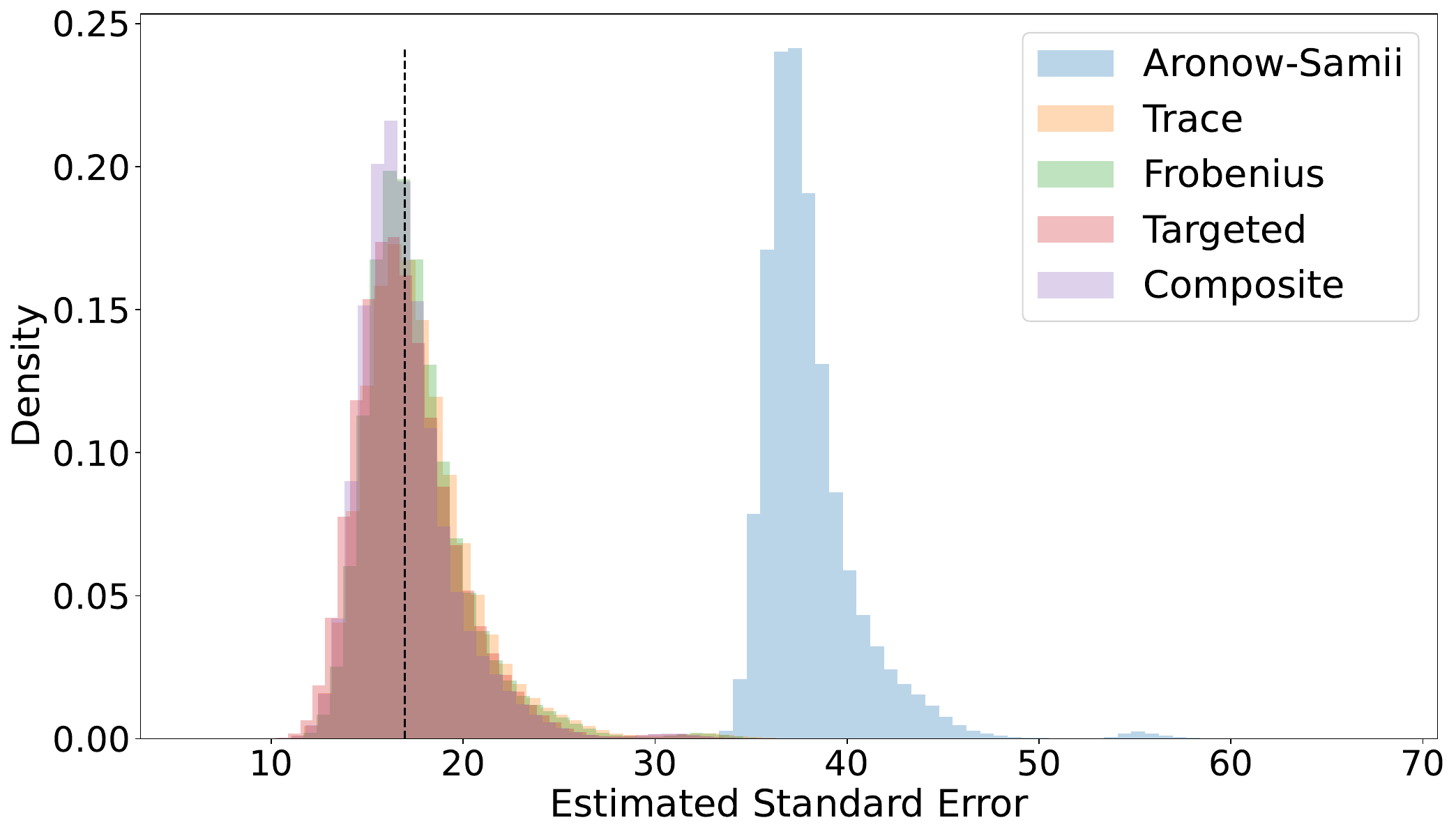}}%
	\qquad
	\subfigure[Real Data Outcomes]{%
		\label{fig:second}%
		\includegraphics[width=0.47\textwidth]{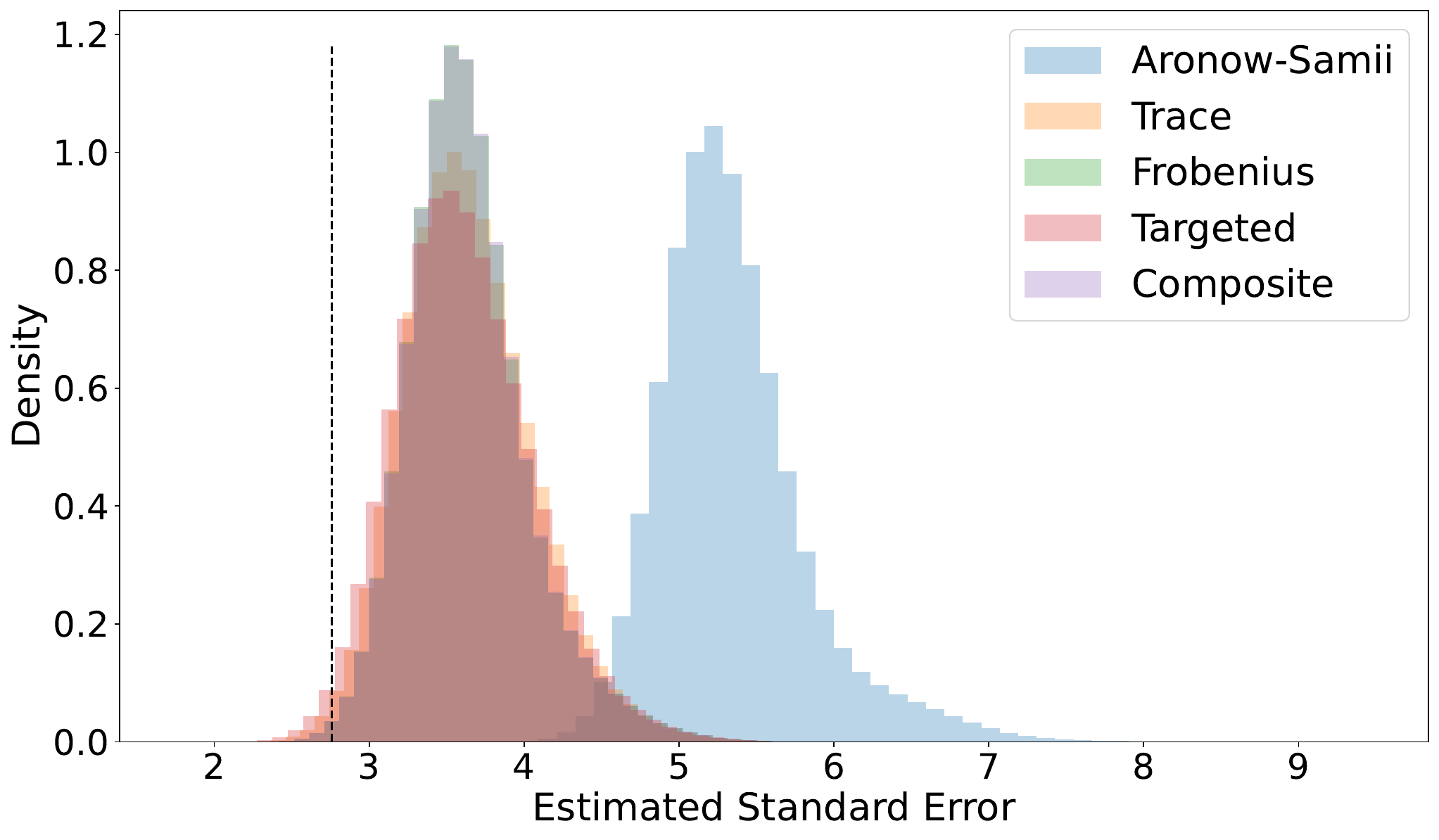}}%
	\caption{Sampling Distributions of Standard Error Estimators}\label{fig:std-est-hist}
\end{figure}

Figure~\ref{fig:std-est-hist} presents the sampling distributions of the five standard error estimators, obtained by taking the square root of the variance estimators.
The dotted black line is the true standard error of the point estimator.
For both outcomes, we see that the estimator based on the Aronow--Samii bound is much further to the right than the other estimators.
This shows that the standard error estimates from the optimized bounds will be smaller than those derived from the Aronow--Samii bound with a high probability.
The differences between the four optimized bounds are small compared to the Aronow--Samii bound.

Section~\suppref{sec:additional-sims} of the supplement provides additional simulation results.
We investigate additional outcomes, vary the cut-off parameter which determines the set of unobservable products, consider other exposures with different level of dependence, and consider settings with misspecified exposures.
These additional results are in line with those reported here.
The variance estimators appear to be reasonably robust to misspecification, remaining conservative also under relatively severe misspecification.
However, the confidence intervals sometimes undercover when the exposures are misspecified, primarily driven by bias in the point estimator.

\section{Concluding Remarks}

Unbiased and consistent variance estimators generally do not exist for point estimators of treatment effects.
Experimenters therefore opt for conservative estimators to avoid misleading inferences, but they want to avoid excessive conservativeness.
The methods we have described in this paper allow experimenters to construct valid variance estimators that aim to minimize conservativeness.
Experimenters can take advantage of background information about the potential outcome to reduce conservativeness by using a targeted linear objective.
In case no such information is available, experimenters can use a norm objective to reduce conservativeness for most potential outcomes.
No matter the approach, the resulting estimator is guaranteed to be conservative and admissible, even if the experimenters perceived knowledge of the potential outcomes happens to be incorrect.

Experimenters can explore various variance bounds and pick the one that is best suited for their particular settings.
However, this choice should not be based on the estimated variances of the different bounds.
The estimated variances will contain sampling variability, meaning that selection on the realized estimates will introduce negative bias relative to the variance bound, potentially making the estimation procedure anticonservative.
A simple way to make sure this does not happen is to select the variance bound before accessing the outcome data.
For example, experimenters can use outcome data from pilot studies or baseline measurements when selecting the bound without the risk of introducing negative bias.

There are several extensions and open questions that are yet to be explored.
We have considered the class of linear point estimators in this paper, and it remains an open question whether our results and methods can be extended to a larger class of estimators.
While the class of linear estimators includes almost all treatment effect estimators, recently developed estimators based on machine learning techniques do not fall in this class \citep[see, e.g.,][]{Aronow2013Class,Wager2016High,Chernozhukov2018Double,Wu2018LOOP}.
The key challenge is that the variance of these estimators are not quadratic forms in the potential outcome vector, so bounds that themselves are quadratic forms will generally not be valid.
A possible way forward is to linearize the point estimators, in which case one could construct quadratic bounds that are asymptotically valid.
However, extending the finite-sample results in the current paper to this larger class of treatment effect estimators appears to currently be beyond reach.

Our discussion about how to select a variance bound to minimize the mean squared error of the variance estimator in Section~\ref{sec:estimate-vb} was based on a bound on the precision of the estimator.
This bound will occasionally be loose, so the approach we describe in this paper is best seen as a heuristic.
While we believe this heuristic is useful and appropriate in most circumstances, it remains an open question whether one can select the variance bound so as to directly minimize mean squared error.

Finally, our investigation relies on the assumption that the exposures are correctly specified.
The interference literature has recently considered estimation of exposure effects when the exposures are misspecified or unrestricted \citep[see, e.g.,][]{Auerbach2023Local,Leung2022Causal,Li2022Random,Savje2024Causal}.
While our extended simulation study in the supplement indicates that the variance estimators are reasonably robust to misspecification, it is an open question to what degree our theoretical results extend to settings with misspecification.

\begin{singlespace}
\bibliography{ms}
\end{singlespace}

\clearpage
\newcommand{\suppletter}{S}
\newcommand{\supptitle}{Supplement}
\newcommand{\supplabel}{supp:proofs}

\renewcommand{\numberline}[1]{#1\ \ \ }

\mtcaddpart[Supplement]

\resetsuppcounters
\updatesuppanchors{\suppletter}
\updatesuppcounters{\suppletter}

\makeatletter
\immediate\write\@auxout{\string\newlabel{\supplabel}{{\suppletter}{\thepage}{\supptitle}{supplement.\suppletter}{}}}
\makeatother
\hypertarget{supplement.\suppletter}{}

\begin{center}
	\LARGE \textbf{Supplement}
\end{center}
\bigskip

\parttoc

\section{Examples of Linear Estimators}\label{sec:example-estimators}

We here provide examples of linear estimators, according to the definition in Section~\mainref{sec:preliminaries} in the main paper.
The class of linear estimators contains many more members than these examples.

\begin{enumerate}
	\item The Horvitz--Thompson estimator \citep{Horvitz1952} uses inverse probability weighting to account for non-uniform assignment probabilities.
	We can write this estimator in the form of Eq.\ (\mainref{eq:linear-est-def}) in the main paper by using the coefficients
	\begin{equation}
		\estcoefi = \frac{\iexia}{\pexia} - \frac{\iexib}{\pexib}.
	\end{equation}

	\item The difference-in-means estimator \citep{Imbens2015Causal} contrasts the sample means between the two groups that received the exposures of interest.
	We can write this estimator in the linear form by using the coefficients
	\begin{equation}
		\estcoefi = \frac{\iexia}{n^{-1} \sumjn\iexja} - \frac{\iexib}{n^{-1} \sumjn\iexjb}.
	\end{equation}

	\item The H\'ajek estimator \citep{Hajek1971} is a generalization of the difference-in-means estimator that accommodates non-uniform assignment probabilities.
	We can write this estimator in the linear form by using the coefficients
	\begin{equation}
		\estcoefi =
		\paren[\Bigg]{ \frac{\iexia}{\pexia} \bigg/ \frac{1}{n} \sumjn \frac{\iexja}{\pexja} }
		-
		\paren[\Bigg]{ \frac{\iexib}{\pexib} \bigg/ \frac{1}{n} \sumjn \frac{\iexjb}{\pexjb} }.
	\end{equation}

	\item The conventional OLS regression estimator of the average treatment effect \citep[see, e.g.,][]{Duflo2007Using} is obtained by using the coefficients
	\begin{equation}
		\estcoefi = n \basisv{2}^\tran \paren[\big]{\mat{Q}^\tran \mat{Q}}^{-1} \mat{Q}^\tran \basisv{i},
	\end{equation}
	where $\basisv{i}$ is the $i$th standard basis vector of appropriate dimension, and $\mat{Q} = \begin{bmatrix}\onevec, \Zv, \xM\end{bmatrix}$.
	This estimator has been shown to perform poorly in some situations \citep{Freedman2008Regression}.
	\citet{Lin2013Agnostic} describes a modified OLS regression estimator that addresses the issue.
	The modified estimator has the same form as the original but with the matrix $\mat{Q} = \begin{bmatrix}\onevec, \Zv, \xM_{\textsc{dm}}, \xM_{\textsc{int}}\end{bmatrix}$, where the matrix $\xM_{\textsc{dm}} = \xM - n^{-1} \onevec \onevec^\tran \xM$ is the demeaned covariate matrix, and $\xM_{\textsc{int}} = \Zv \onevec^\tran \circ \xM_{\textsc{dm}}$ is the demeaned covariate matrix interacted column-wise with the treatment vector.

	\newcommand{\betaa}{\boldsymbol{\beta}_1}
	\newcommand{\betab}{\boldsymbol{\beta}_0}
	\newcommand{\betad}{\boldsymbol{\beta}_\exarb}
	\newcommand{\estbetaa}{\widehat{\boldsymbol{\beta}}_1}
	\newcommand{\estbetab}{\widehat{\boldsymbol{\beta}}_0}
	\newcommand{\estbetad}{\widehat{\boldsymbol{\beta}}_\exarb}

	\item The Generalized Regression Estimator \citep{Cassel1976Some}, which is sometimes called the Augmented Inverse Propensity Weighted Estimator, allows for both covariate adjustment and non-uniform assignment probabilities.
	When using linear covariate adjustment, the estimator is written as
	\begin{equation}
		\est
		= \frac{1}{n} \sumin \bracket[\Bigg]{ \xvi^\tran \paren[\big]{\estbetaa - \estbetab}
		+ \frac{ \iexia \paren[\big]{\Yi - \xvi^\tran \estbetaa} }{\pexia}
		- \frac{ \iexjb \paren[\big]{\Yi - \xvi^\tran \estbetab} }{\pexjb} },
	\end{equation}
	where the linear functions $\estbetad$ are chosen to minimize $\sumin \iexi{\exarb} \paren{\Yi - \xvi^\tran \betad }^2$.
	We can write $\estbetad$ in closed form as $\paren[\big]{\xM_\exarb^\tran \xM_\exarb}^{-1} \xM_\exarb^\tran \Yv$, where the $i$th row of $\xM_\exarb$ is equal to $\xvi$ if $\Exmi = \exarb$ and otherwise equal to zero.
	This means that we can write the estimator in the linear form by using the coefficients
	\begin{equation}
		\estcoefi = \frac{\iexia}{\pexia} - \frac{\iexib}{\pexib} + \sumjn \vec{Q}_j^\tran \basisv{i},
	\end{equation}
	where $\basisv{i}$ is the $i$th standard basis vector of dimension $n$, and
	\begin{equation}
		\vec{Q}_j^\tran = \paren[\bigg]{1 - \frac{\iexja}{\pexja}} \xvj^\tran \paren[\big]{\xM_1^\tran \xM_1}^{-1} \xM_1^\tran
		- \paren[\bigg]{1 - \frac{\iexjb}{\pexjb}} \xvj^\tran \paren[\big]{\xM_0^\tran \xM_0}^{-1} \xM_0^\tran.
	\end{equation}

\end{enumerate}

\section{Reformulation of Linear Estimators}\label{sec:reform-lin-est}

Recall from Section~\mainref{sec:preliminaries} in the main paper that our focus is linear estimators that take the form
\begin{equation}
	\est = \frac{1}{n} \sumin \estcoefi \Yi,
\end{equation}
where the coefficients $\estcoefi$ may depend arbitrarily on the treatment assignments $\Zv$ and characteristics of the units, but they cannot depend on the observed outcomes $\Yv$.
Further recall that the exposure mapping is some unit-specific function $\exmsym_i: \setb{0,1}^n \rightarrow \allexposures$ that maps the treatment assignment to a set of exposures $\allexposures$, and the realized exposure is $\Exmi = \exmi{\Zv}$.

When the exposure mapping is correctly specified, we can write the observed outcome for each unit as
\begin{equation}
	\Yi = \sum_{\exarb \in \allexposures} \indicator{\Exmi = \exarb} \yi{\exarb}.
\end{equation}
Hence, any linear estimator can be written as
\begin{equation}
	\est = \frac{1}{n} \sumin \sum_{\exarb \in \allexposures} \estcoefi \indicator{\Exmi = \exarb} \yi{\exarb}.
\end{equation}

For each unit-exposure pair $(i,\exarb) \in \Sample \times \allexposures$, define variables
\begin{equation}
	\coef{i,\exarb} = \estcoefi \indicator{\Exmi = \exarb}
	\qquadand
	\para{i,\exarb} = \yi{\exarb}.
\end{equation}
Using these variables, the estimator can be written
\begin{equation}
	\est
	= \frac{1}{n} \sumin \sum_{\exarb \in \allexposures} \coef{i,\exarb} \para{i,\exarb}
	= n^{-1} \coefv^\tran \parav,
\end{equation}
where $\coefv$ and $\parav$ are column vectors obtained by collecting the variables $\coef{i,\exarb}$ and $\para{i,\exarb}$, respectively.
As noted in the main paper, all randomness is collected in the vector $\coefv$, which has a known distribution, and the unknown vector $\parav$ is non-random.

\section{Testing Admissibility of Variance Bounds}\label{sec:testing}

In the main paper, we introduced the notion of admissibility of variance bounds and proposed several methods for computing such bounds.
In this section, we present a semidefinite program for testing admissibility of a given quadratic variance bound.
This allows experimenters to check whether a given variance bound, possibly obtained by different methods than those described in this paper, is admissible before running an experiment.

The procedure \testadmissiblity{} decides whether a variance bound is admissible by testing whether the optimal value of a particular semidefinite program is positive.
Recall that a variance bound $\boundM = \varM + \slackM$ is inadmissible if there exists another variance bound $\otherboundM = \varM + \otherslackM$ such that
\begin{equation}
	\boundM - \otherboundM = \slackM - \otherslackM
\end{equation}
is nonzero and positive semidefinite.
If we can certify that no such matrices $\otherboundM$ and $\otherslackM$ exist, then the variance bound $\boundM$ is admissible.

To this end, the procedure \testadmissiblity{} searches over all slack matrices $\otherslackM \in \allslacks$ with the extra constraint that $\slackM - \otherslackM$ is positive semidefinite.
What remains to be shown is whether there exists a feasible solution such that this difference is nonzero.
To determine this, \testadmissiblity{} maximizes the trace of the difference, $\tr{\slackM - \otherslackM}$.
If the optimal value is positive, then the difference is nonzero and the original variance bound is inadmissible; otherwise, the optimal value is zero and the variance bound is admissible.
The test for admissibility is given formally below in Algorithm~\ref{alg:test-admissibility}.

\begin{algorithm}
	\caption{\testadmissiblity}\label{alg:test-admissibility}
	\SetKwInOut{Input}{Input}
	\Input{Variance bound slack matrix $\slackM$ and unobservable pairs $\unobspairs$.}
	Solve the following semidefinite program
	\begin{equation} \label{eq:admissible-program}
		\tag{Admissible-SDP}
			\begin{aligned}
				\alpha \gets& \underset{\otherslackM}{\text{maximize}}
				& & \tr{\slackM - \otherslackM} \\
				& \text{subject to}
				& & \tilde{s}_{k\ell} = \slackMkl \; \text{ for all } (k,\ell) \in \unobspairs, \\
				&&& 0 \preceq \otherslackM \preceq \slackM.
			\end{aligned}
	\end{equation} \\
	\Return{ {\normalfont\texttt{False}} if optimal value $\alpha > 0$ and {\normalfont\texttt{True}} otherwise.}
\end{algorithm}

Note that $\otherslackM = \slackM$ is always a feasible solution to the optimization underlying \testadmissiblity{}, but this yields a objective value of zero.
The following theorem guarantees correctness of the \testadmissiblity{} procedure.

\begin{theorem}
	\testadmissiblity{} returns {\normalfont\texttt{True}} if and only if the variance bound is admissible.
\end{theorem}
\begin{proof}
	Suppose that the variance bound $\boundM = \varM + \slackM$ is admissible.
	Then, there does not exist a matrix $\otherslackM \in \allslacks$ such that $\slackM - \otherslackM$ is positive semidefinite and nonzero.
	Thus, the only feasible solution to \eqref{eq:admissible-program} is $\slackM$, which yields an objective value of $0$.
	In this case, \testadmissiblity{} returns \texttt{True}, which is the correct answer.

	Suppose that the variance bound is inadmissible.
	Then, there exists a matrix $\otherslackM \in \allslacks$ such that $\slackM - \otherslackM$ is positive semidefinite and nonzero.
	If $\slackM - \otherslackM$ is positive semidefinite and nonzero, then all of its eigenvalues are non-negative and at least one is positive, meaning that the sum of the eigenvalues is positive.
	Because the trace of a matrix is equal to the sum of the eigenvalues, the feasible matrix $\otherslackM$  yields a positive objective value: $\tr{\slackM - \otherslackM} > 0$.
	In this case, \testadmissiblity{} returns \texttt{False}, which is the correct answer.
\end{proof}

There are some numerical considerations when implementing \testadmissiblity{}.
Semidefinite programs can only be solved up to some desired accuracy.
This means that testing whether the optimal objective is exactly zero is generally not possible using finite precision arithmetic, except in certain restricted cases.
For this reason, the main practical use case of \testadmissiblity{} will be to certify that a variance bound is sufficiently admissible, rather than certifying exact admissibility.
This numerical issue should not be of great concern, as an experimenter can use \testadmissiblity{} to certify that a variance bound is approximately admissible (up to an arbitrary desired tolerance) which is generally sufficient for practical purposes.

In order to decide that the input variance bound is inadmissible, \testadmissiblity{} needs only to produce a feasible solution where $\tr{\slackM - \otherslackM} > 0$.
If the input variance bound is inadmissible, this may require significantly less computation than solving the underlying optimization program to optimality.
Therefore, early stopping may be used in \testadmissiblity{} to increase computational efficiency.

\section{Regularization of Operator Norm Objective}\label{sec:reg-of-operator-norm}

We here describe how to obtain an admissible bound when using the operator norm objective, which minimizes worst-case conservativeness, as discussed in Section~\mainref{sec:norm-objectives} in the main paper.
The operator norm is monotone, but not strictly monotone, so it could yield inadmissible bounds if used on its own.
We address this by regularizing the operator norm with the Frobenius norm in a composite objective, as discussed in Section~\mainref{sec:composite-obj} in the main paper.
That is, for some small $\gamma > 0$, the experimenter would use the objective
\begin{equation}
	\objfS = \pnorm{\infty}{\varM + \slackM} + \gamma \pnorm{2}{\varM + \slackM}^2.
\end{equation}
Proposition~\mainref{prop:composite-objective} applies to this objective because the operator norm is monotone and the Frobenius norm is strictly monotone, meaning that the composite objective is strictly monotone.
Therefore, the resulting bound is conservative, design compatible and admissible.
Experimenters should set $\gamma$ as small as possible here, as this ensures that the worst-case conservativeness is minimized as well as possible.
The following proposition formalizes this intuition using a limiting argument.
The proof of the proposition is given in Section~\ref{sec:proof-formal-limiting-argument} of the supplement.

\expcommand{\propformallimitingargument}{%
Let $\setb{ \gamma_k }_{k=1}^\infty$ be a sequence of positive regularization parameters converging to zero.
Let $ \setb{ \slackM_k }_{k=1}^\infty$ be a sequence of slack matrices obtained as solutions to the regularized programs
\begin{equation}
	\slackM_k = \argmin_{\slackM \in \allslacks} \paren[\Big]{ \pnorm{\infty}{\varM + \slackM} + \gamma_k \pnorm{2}{\varM + \slackM}^2 }.
\end{equation}
The sequence of slack matrices converges to $\slackM_k \rightarrow \slackM^*$, where $\slackM^*$ is the slack matrix minimizing the squared Frobenius norm among all minimizers of the operator norm:
\begin{equation}
	\begin{aligned}
		\slackM^* = \argmin_{\slackM \in \allslacks} \quad & \pnorm{2}{\varM + \slackM}^2,\\
		\textrm{s.t.} \quad & \pnorm{\infty}{\varM + \slackM} = \inf \braces[\big]{ \pnorm{\infty}{\varM + \mat{Q}} : \mat{Q} \in \allslacks }.
	\end{aligned}
\end{equation}
}

\begin{proposition}\label{prop:formal-limiting-argument}
	\propformallimitingargument
\end{proposition}

\section{Admissibility of Imai's variance bound}\label{sec:imai-bounds}

We here consider the variance bound introduced by \citet{Imai2008Variance} for experiments using the matched pair design.
We will show that the bound is admissible.

In an experiment using the  matched pair design, there are $n = 2m$ units that are grouped into pairs.
Without loss of generality, we assume that units are consecutively paired so that units 1 and 2 are paired, 3 and 4 are paired, and in general $(2k-1, 2k)$ are paired for $k=1 \dots m$.
Let $\mathcal{P}$ be the set of all pairs, such that $P = \setb{i,j} \in \mathcal{P}$ if $i$ and $j$ are paired.

The distribution of treatment assignments $(Z_1 \dots Z_n)$ in this design is such that two paired units $P = \setb{i,j}$ are assigned to opposite treatments, each with equal probability:
\[
\Pr{ Z_i = 1, Z_j = 0} = \Pr{ Z_i = 0, Z_j = 1 } = 1/2 \enspace,
\]
and treatment assignment is independent between non-paired units.
\citet{Imai2008Variance} considers the usual Horvitz--Thompson estimator
\[
\widehat{\tau} = \frac{1}{n} \sum_{i=1}^n Y_i \paren[\Big]{ \frac{\indicator{Z_i = 1}}{\Pr{Z_i = 1}} - \frac{\indicator{Z_i = 0}}{\Pr{Z_i = 0}} }
\enspace.
\]
The corresponding variance can be written as a quadratic form $n^2 \Varsub{\parav}{\widehat{\tau}} = \parav^\tran \varM \parav$ where
\[
\varM =
\begin{bmatrix}
	\mat{H} & \mat{H} \\
	\mat{H} & \mat{H}
\end{bmatrix}
\qquadtext{where}
\mat{H}(i,j) = \left\{
\begin{array}{lr}
	1 & \text{ if } i = j\\
	-1 & \text{ if $i$ and $j$ are paired} \\
	0 & \text{otherwise}.
\end{array}
\right.
\]
As before, $\parav = (y_1(0), \dotsc, y_n(0),y_1(1), \dotsc, y_n(1))$.
Note that $\mat{H}$ is a block diagonal matrix where the 2-by-2 blocks correspond to the paired units and are given by
\[
\begin{bmatrix}
	1 & -1 \\
	-1 & 1
\end{bmatrix}
\enspace.
\]
Those pairs of potential outcomes which cannot be observed are (1) different outcomes for the same unit, such as $y_i(1) y_i(0)$, and (2) the same outcome for paired units, such as $y_i(0) y_j(0)$.
Formally, we can write
\[
\unobspairs = \setb[\Big]{ (i,i+n), (i+n,i) : \ i \in [n] } \cup \setb[\Big]{ (i,j), (i+n,j+n) : i,j \in P \in \mathcal{P} }
\enspace.
\]
Imai's variance estimator takes the following form:
\[
\widehat{\textrm{VB}} = \frac{4}{n(n-2)} \sum_{\substack{P \in \mathcal{P} \\ P = \setb{i,j}}} \paren[\Bigg]{ \frac{1}{2} \paren[\big]{  \widehat{\tau}_i + \widehat{\tau}_j } - \widehat{\tau} }^2
\enspace.
\]

We take his variance bound to be the expectation of his variance estimator, i.e. $\textrm{VB}(\parav) = \Esub{\parav}{ \widehat{\textrm{VB} }}$.
With some tedious (but straightforward) calculations, we can verify that the variance bound is the quadratic form $n^2 \textrm{VB} = \parav^\tran \boundM \parav$, where
\[
\boundM =
\begin{bmatrix}
	\mat{Q} & - \mat{Q} \\
	-\mat{Q} & \mat{Q}
\end{bmatrix}
\quadwhere
\mat{Q}(i,j) =  \left\{
\begin{array}{lr}
	2 & \text{ if } i = j\\
	0 & \text{ if $i$ and $j$ are paired} \\
	\frac{2}{n-2} & \text{otherwise}
\end{array}
\right.
\]
Thus, the slack matrix $\slackM$ corresponding to Imai's bound is given by
\[
\slackM =
\begin{bmatrix}
	\mat{Q} - \mat{H} & - \paren[\big]{ \mat{Q} - \mat{H} } \\
	- \paren[\big]{ \mat{Q} - \mat{H} } & \mat{Q} - \mat{H}
\end{bmatrix}
\enspace.
\]

By Theorem~\mainref{thm:sdp-procedure-admissibility}, in order to show that Imai's bound is admissiblle, it suffices to show that Imai's bound may be obtained through \optprocedure{} with a positive definite linear objective.
Our proof is derived from the duality theory of semi-definite programming.
Indeed, the matrix $\dualM$ in the proof below may be understood as a matrix-valued dual variable; however, no knowledge of duality theory is required to verify the proof.
While we find these proofs to be useful and relatively short, we remark that they are somewhat challenging to produce, due to the ``guess-and-check'' nature involved with identifying a positive definite linear objective and dual certificates simultaneously.

\expcommand{\Imaiisadmissible}{%
	In the setting of matched pair design, Imai's bound is admissible with respect to $\allbounds$.
}

\begin{proposition}\label{prop:imai-is-admissible}
	\Imaiisadmissible
\end{proposition}

\begin{proof}
	To show that Imai's variance bound is admissible, it suffices to show (by Theorem~\mainref{thm:sdp-procedure-admissibility}) that the slack matrix $\slackM$ is a solution to the \optprocedure{} program for a linear objective $\objfS = \iprod{\slackM , \objM}$ where $\objM$ is positive definite.
	We claim that $\objM = \unitM$ is one such objective where Imai's bound is the optimal solution.

	To begin, we claim that the slack matrix $\slackM$ corresponding to Imai's bound attains the objective value $\iprod{\slackM , \objM} = 2n$.
	This is verified by the following calculation:
	\begin{align*}
		\iprod{\slackM , \objM}
		&= \iprod{\slackM}
			&\text{($\objM = \unitM$)} \\
		&= \tr[\Bigg]{
			\begin{bmatrix}
				\mat{Q} - \mat{H} & - \paren[\big]{ \mat{Q} - \mat{H} } \\
				- \paren[\big]{ \mat{Q} - \mat{H} } & \mat{Q} - \mat{H}
			\end{bmatrix}
		}
			&\text{(def of Imai's $\slackM$)} \\
		&= 2  \tr{ \mat{Q} - \mat{H} }
			&\text{(linearity of $\trsym$)} \\
		&= 2  \sum_{i=1}^n \mat{Q}(i,i) - \mat{H}(i,i)
			&\text{(definition of $\trsym$)} \\
		&= 2 \sum_{i=1}^n 1
			&\text{(definition of $\mat{H}$, $\mat{Q}$)} \\
		&= 2n \enspace.
	\end{align*}

	Next, we will show that any feasible slack matrix $\slackM \in \allslacks$ has an objective value which is bounded below as $\iprod{\slackM , \objM} \geq 2n$.
	This would immediately imply optimality of Imai's bound for this linear objective and render the proof complete.
	To this end, we define a new $2n$-by-$2n$ matrix $\dualM$ which is given by
	\[
	\dualM =
	\begin{bmatrix}
		\frac{1}{2} \mat{R} & -\frac{1}{2} \unitM\\
		-\frac{1}{2} \unitM & \frac{1}{2} \mat{R}
	\end{bmatrix}
	\]
	where $\mat{R}$ is a block matrix of 2-by-2 blocks corresponding to the paired units given by
	\[
	\begin{bmatrix}
		0 & 1 \\
		1 & 0
	\end{bmatrix}
	\]
	We claim that the largest eigenvalue of $\dualM$ is at most 1.
	We will show this using a few steps.
	First, we will begin by using the Courant-Fischer variational characterization of eigenvalues.
	Let $\parav = ( \vec{x}, \vec{y}) \in \Reals^{2n}$ be given and observe that
	\begin{align*}
		\parav^\tran \dualM \parav
		&= \begin{bmatrix}
			\vec{x} \\
			\vec{y}
		\end{bmatrix}^\tran
		\begin{bmatrix}
			\frac{1}{2} \mat{R} & -\frac{1}{2} \unitM\\
			-\frac{1}{2} \unitM & \frac{1}{2} \mat{R}
		\end{bmatrix}
		\begin{bmatrix}
			\vec{x} \\
			\vec{y}
		\end{bmatrix} \\
		&= \frac{1}{2} \vec{x}^\tran \mat{R} \vec{y} + \frac{1}{2} \vec{y}^\tran \mat{R} \vec{y}
		+ \vec{x}^\tran \vec{y} \\
		&\leq \frac{1}{2} \vec{x}^\tran \mat{R} \vec{y} + \frac{1}{2} \vec{y}^\tran \mat{R} \vec{y}
		+ \frac{1}{2} \paren[\Big]{ \norm{\vec{x}}^2 + \norm{\vec{y}}^2 }
			&\text{(Cauchy-Schwarz and AM-GM)} \\
		&\leq \frac{1}{2}  \eigvalmax(\mat{R}) \paren[\Big]{ \norm{\vec{x}}^2 + \norm{\vec{y}}^2 }
		+ \frac{1}{2} \paren[\Big]{ \norm{\vec{x}}^2 + \norm{\vec{y}}^2 }
			&\text{(Courant-Fischer)} \\
		&= \frac{1}{2} \paren{ \eigvalmax(\mat{R}) + 1 } \cdot  \paren[\Big]{ \norm{\vec{x}}^2 + \norm{\vec{y}}^2 } \\
		&=  \frac{1}{2} \paren{ \eigvalmax(\mat{R}) + 1 } \cdot \norm{\parav}^2 \enspace,
	\end{align*}
	which implies that $\eigvalmax(\dualM) \leq  \frac{1}{2} \paren{ \eigvalmax(\mat{R}) + 1 }$.
	Next, observe that $\mat{R}$ is a block-diagonal matrix so that its largest eigenvalue is equal to the largest eigenvalue of one of its blocks.
	Fortunately, all of its blocks have the same structure and it is simple to verify that
	\[
	\eigvalmax \paren[\Big]{
		\begin{bmatrix}
			0 & 1 \\
			1 & 0
		\end{bmatrix}
 	}
 	= 1 \enspace.
	\]
	Thus, $\eigvalmax(\mat{R}) = 1$.
	Plugging this into the above yields that $\eigvalmax(\dualM) \leq 1$, as desired.
	Thus, we have that $\dualM \preceq \unitM = \objM$.

	Let $\slackM \in \allslacks$ be a valid variance bound.
	Given the pattern of unobservability $\unobspairs$ associated with the matched pair design, this means that the valid slack matrix $\slackM$ must satisfy that for all $i \in [n]$,
	\begin{equation} \label{eq:slack-constrained-imai-singles}
		\slackM(i,i+n)
		= - \varM(i,i+n)
		= -1
	\end{equation}
	and also for all pairs $p = \setb{i,j}$ that
	\begin{equation} \label{eq:slack-constrained-imai-pairs}
		\slackM(i,j+n)
		= - \varM(j+n,i)
		= 1
		\enspace.
	\end{equation}
	This implies that the objective value of any feasible solution may be bounded below as
	\begin{align*}
		\iprod{\slackM , \objM}
		&\geq \iprod{ \slackM, \dualM }
		&\text{($\dualM \preceq \unitM$ and $\mathbf{0} \preceq \slackM$)} \\
		&= \tr[\Big]{\slackM
			\begin{bmatrix}
				\frac{1}{2} \mat{R} & -\frac{1}{2} \unitM\\
				-\frac{1}{2} \unitM & \frac{1}{2} \mat{R}
			\end{bmatrix}
		}
		&\text{(def of $\dualM$)}
		\intertext{Observe that the non-zero entries in $\mat{R}$ exactly correspond to the missing patterns of paired units' outcomes under the same treatment. Similarly, the off-diagonal matrix $\unitM$ corresponds to the missing patterns of the single units' two outcomes under different treatments. Thus, we can write the trace as}
		&= \underbrace{2 \sum_{i \in [n]} \slackM(i,i+n) \cdot \paren[\Big]{ - \frac{1}{2} }}_{\textrm{single units}} + \underbrace{2 \sum_{ \substack{P \in\mathcal{P} \\ P = \setb{i,j}} } 2 S(i,j+n) \cdot \frac{1}{2} R(i,j)}_{\textrm{paired units}} \\
		&= 2 \cdot n \cdot \paren[\Big]{ \frac{1}{2} } + 2 m \paren[\big]{ 2 \cdot 1 \cdot \frac{1}{2} }
			&\text{(\eqref{eq:slack-constrained-imai-singles} and \eqref{eq:slack-constrained-imai-pairs})}\\
		&= n + 2m \\
		&= 2n \enspace.
	\end{align*}
	Thus, we have shown that Imai's variance bound is optimal for this positive definite linear objective.
	Therefore, it is admissible, as desired.
\end{proof}

\section{Frechet-Hoeffding-type bounds}\label{sec:fh-bounds}

In this section, we review the Frechet-Hoeffding style variance bound of \citet{Aronow2014Sharp} and draw comparisons to \optprocedure{} presented in this paper.
A summary of our conclusions is as follows: while the Frechet--Hoeffding approach provides variance bounds which satisfy a stronger notion of sharpness, both this type of bounds and this concept of sharpness are only applicable to simple and symmetric experimental settings.
In particular, the Frechet--Hoeffding method for constructing variance bounds{\textemdash}at least as it is currently conceived{\textemdash}cannot produce minimally conservative variance bounds even for designs that are only minor departures from the conventional complete randomization design, let alone experiments featuring interference and complex designs.

\subsection{Review of Aronow et al. (2014)}

The variance of the difference-in-means estimator in an experiment with no interference under the complete randomized design is
\begin{align*}
\frac{1}{(n-1)}  &\underbrace{\frac{1}{n} \sum_{i=1}^n \paren[\Big]{  y_i(1) - \frac{1}{n} \sum_{j=1}^n y_j(1) }^2}_{\triangleq \ \sigma_1^2}
+ \frac{1}{(n-1)} \underbrace{ \frac{1}{n} \sum_{i=1}^n \paren[\Big]{  y_i(0) - \frac{1}{n} \sum_{j=1}^n y_j(0) }^2}_{\triangleq \ \sigma_0^2} \\
&\qquad + \frac{1}{(n-1)} 2 \cdot \underbrace{ \frac{1}{n} \sum_{i=1}^n \paren[\Big]{  y_i(1) - \frac{1}{n} \sum_{j=1}^n y_j(1) } \cdot \paren[\Big]{  y_i(0) - \frac{1}{n} \sum_{j=1}^n y_j(0) }}_{\triangleq \ \rho} \enspace,
\end{align*}
where the first two terms $\sigma_1^2$ and $\sigma_0^2$ are the variances of the potential outcomes under treatment and control and the third term is the correlation between treatment and control outcomes.

The preliminary insight into the variance estimator of \citet{Aronow2014Sharp} is to recognize that, in this particular experimental scenario, that the variance is a function of the joint cumulative distribution function (CDF) of the potential outcomes.
In the design-based context, the potential outcomes are non-random, and so the CDF is viewed merely as a formal way to describe the collection of joint potential outcomes.
Formally, we may define the joint CDF $\mathbb{F} : \Reals^2 \to [0,1 ]$ as the function
\[
\mathbb{F}(s,t) = \frac{1}{n} \sum_{i=1}^n \indicator{y_i(1) \leq s, y_i(0) \leq t}
\enspace.
\]
The relevant quantities in the variance (i.e. $\sigma_1^2$, $\sigma_0^2$, and $\rho$) are all functions of the joint CDF $\mathbb{F}$.
In other words, the joint CDF contains all the relevant information necessary to determine the variance of the effect estimator.
For this reason, we will use the shorthand $\Varsub{\mathbb{F}}{\widehat{\tau}}$, $\sigma_1^2(\mathbb{F})$, $\sigma_0^2(\mathbb{F})$, and $\rho(\mathbb{F})$ to illustrate that these quantities are indeed functions of the joint CDF $\mathbb{F}$.

Unfortunately, the joint CDF cannot be directly estimated because the experimenter will never simultaneously observe both $y_i(1)$ and $y_i(0)$ for any unit.
This is the fundamental problem of causal inference.
However, the marginal CDFs of the potential outcomes can be estimated with good precision in large samples.
Indeed, one could argue that the marginal CDFs contain all possible information that can be ascertained about the potential outcomes in experiments with simple, symmetric designs without interference or covariate information.
Motivated by these insights, \citet{Aronow2014Sharp} seek to determine the information about these relevant quantities (i.e. $\sigma_1^2$, $\sigma_0^2$, $\rho$) that are contained in the marginal CDFs.
For each treatment $k \in \setb{0,1}$, the marginal cumulative distribution function $\mathbb{F}_{(k)}: \Reals \to [0,1]$ of the corresponding potential outcome is
\[
\mathbb{F}_{(k)}(t) = \sum_{i=1}^n \indicator{ y_i(k) \leq t } \enspace.
\]

These marginal CDFs can typically be estimated to high precision, but for simplicity we will take them as known in this discussion.
For the purposes of deriving a variance bound, the question then becomes:  how much information about the relevant quantities $\sigma_1^2$, $\sigma_0^2$, and $\rho$ can be learned from the two marginal CDFs $\mathbb{F}_{(1)}$ and $\mathbb{F}_{(0)}$?

The quantities $\sigma_1^2$ and $\sigma_0^2$ are exactly determined by $\mathbb{F}_{(1)}$ and $\mathbb{F}_{(0)}$, respectively.
For this reason, we write $\sigma_1^2(\mathbb{F}_{(1)})$ and $\sigma_0^2(\mathbb{F}_{(0)})$ to illustrate that these quantities are indeed functions of the marginal CDFs.
The correlation parameter $\rho$, however, is not fully determined by the marginal CDFs.
To understand how much we can learn about $\rho$ from the marginal CDFs, we need to introduce the key ingredients for the Frechet--Hoeffding bound.

To this end, let us provide a few definitions.
Let $\mathcal{H}_{(2)}$ be the set of all possible cumulative distribution functions on $\Reals^2$, i.e.
\[
\mathcal{H}_{(2)} =
	\left\{
	\begin{array}{lr}
		F : \Reals^2 \to [0,1]
		\textrm{ such that }
		F \textrm{ is monotone non-decreasing,} \\
		\lim_{(t,s) \to (-\infty, -\infty)} F(t,s) = 0, \text{ and }
		\lim_{(t,s) \to (\infty, \infty)} F(t,s) = 1
	\end{array}
	\right\}
\enspace.
\]
Likewise, define $\mathcal{H}$ be the set of all possible cumulative distribution functions on $\Reals$:
\[
\mathcal{H} = \setb[\Big]{
	F : \Reals \to [0,1]
	:
	F \textrm{ is monotone non-decreasing},
	\lim_{t \to -\infty} F(t) = 0,
	\lim_{t \to \infty} F(t) = 1
}.
\]
Let $Q : \mathcal{H}_{(2)} \to \mathcal{H}$ be the marginal map which send the joint CDF to its two marginal CDFs, i.e. $Q(\mathbb{F}) = (\mathbb{F}_{(1)}, \mathbb{F}_{(0)})$.
Given two marginal CDFs $\mathbb{F}_{(1)}$ and $\mathbb{F}_{(0)}$, we define the \emph{worst-case correlation} $\rho_{\max}(\mathbb{F}_{(1)}, \mathbb{F}_{(0)})$ to be the largest value that the correlation $\rho$ may take for any joint CDF having the specified marginals, i.e.
\[
\rho_{\max}(\mathbb{F}_{(1)}, \mathbb{F}_{(0)})
\triangleq \sup_{\substack{ \mathbb{F} \in \mathcal{H}_{(2)} \\ Q(\mathbb{F}) = (\mathbb{F}_{(1)}, \mathbb{F}_{(0)}) }}
\rho(\mathbb{F})
\enspace.
\]
The Frechet-Hoeffding bound provides two insights.
First, it shows that the supremum is actually attained by a specific joint CDF.
Second, it derives a closed form expression for this worst case correlation quantity $\rho_{\max}(\mathbb{F}_{(1)}, \mathbb{F}_{(0)})$.
For the purposes of our summary and comparison to \citet{Aronow2014Sharp}, the actual closed form expression does not matter.

To summarize, \citet{Aronow2014Sharp} describe a variance bound based on the following two insights:
\begin{enumerate}
	\item The first two terms in the variance, (i.e. $\sigma_1^2$ and $\sigma_0^2$), are exactly determined by the marginals $\mathbb{F}_{(1)}$ and $\mathbb{F}_{(0)}$.
	\item The Frechet-Hoeffding bound exactly describes the largest value $\rho_{\max}(\mathbb{F}_{(1)}, \mathbb{F}_{(0)})$ that the third term can take, provided only information about the two marginals $\mathbb{F}_{(1)}$ and $\mathbb{F}_{(0)}$.
\end{enumerate}
Together, this yields the Aronow, Green and Lee (2014) (AGL) variance bound,
\[
\textrm{VB}_{\textrm{AGL}} = \frac{1}{n-1} \paren[\Big]{ \sigma_1^2(\mathbb{F}_{(1)}) + \sigma_2^2(\mathbb{F}_{(0)}) + 2 \rho_{\max}(\mathbb{F}_{(1)}, \mathbb{F}_{(0)}) }.
\]
While the marginals $\mathbb{F}_{(1)}$ and $\mathbb{F}_{(0)}$ cannot be known exactly, they can be estimated to high precision in large samples.
In this way, one can obtain plug-in estimates for $\sigma_1^2(\mathbb{F}_{(1)})$, $\sigma_0^2(\mathbb{F}_{(0)})$, and $\rho_{\max}(\mathbb{F}_{(1)}, \mathbb{F}_{(0)})$, which in turn can be used to construct a consistent estimator for the AGL variance bound.

\subsection{Admissibility and Sharpness of the AGL Bound}

We need to introduce additional notation to discuss admissibility of the AGL variance bound and various notions of sharpness.
Recall that $\mathcal{H}$ is the set of all possible cumulative distribution functions on $\Reals$:
\[
\mathcal{H} = \setb[\Big]{
	F : \Reals \to [0,1]
	:
	F \textrm{ is monotone non-decreasing},
	\lim_{t \to -\infty} F(t) = 0,
	\lim_{t \to \infty} F(t) = 1
 }.
\]
We introduce the class of \emph{marginal CDF bounds} $\mathcal{F}_{\textrm{CDF}}$ to be those variance bounds that may be obtained by first extracting the marginal information from the potential outcomes.
We formally define this class of variance bounds in the following way: for all $\textrm{VB} \in \mathcal{F}_{\textrm{CDF}}$, we have that $\textrm{VB} = h \circ g$, where $g: \Reals^{2n} \to \mathcal{H} \times \mathcal{H}$ is the function that maps potential outcomes to their respective marginal cumulative distribution functions $\mathbb{F}_{(1)}$ and $\mathbb{F}_{(0)}$, and $h : \mathcal{H} \times \mathcal{H} \to \Reals$ is any function that maps two marginal CDFs to the reals.
Note that the AGL variance bound $\textrm{VB}_{\textrm{AGL}}$ belongs in the class of marginal CDF bounds $\mathcal{F}_{\textrm{CDF}}$, where the corresponding $h$ function is the mapping
\[
h_{\textrm{AGL}}(\mathbb{F}_{(1)}, \mathbb{F}_{(0)}) \mapsto \frac{1}{n-1} \paren[\Big]{
	\sigma_1^2(\mathbb{F}_{(1)})
	+ \sigma_0^2(\mathbb{F}_{0})
	+  2\rho_{\max}(\mathbb{F}_{(1)} , \mathbb{F}_{(0)})
}
\enspace.
\]
The following proposition shows that this bound is admissible within this class.

\begin{proposition}
	The \citet{Aronow2014Sharp} variance bound $\textrm{VB}_{AGL}$ is admissible with respect to the class of marginal CDF bounds $\mathcal{F}_{\textrm{CDF}}$.
\end{proposition}
\begin{proof}
	Suppose for sake of contradiction that the \citet{Aronow2014Sharp} variance bound $\textrm{VB}_{\textrm{AGL}}$ is not admissible in the class of marginal CDF bounds $\mathcal{F}_{\textrm{CDF}}$.
	This means that there exists a valid variance bound $\textrm{VB}^* \in \mathcal{F}_{\textrm{CDF}}$ which dominates $\textrm{VB}_{\textrm{AGL}}$.
	In particular, there exists a potential outcome vector $\parav \in \Reals^{2n}$ such that $\textrm{VB}^*(\parav) < \textrm{VB}_{\textrm{AGL}}(\parav)$.
	We shall show that this implies a contradicition, namely that $\textrm{VB}^*$ is not a variance bound.

	Let $\mathbb{F}_{(1)}$ and $\mathbb{F}_{(0)}$ be the marginal CDFs corresponding to potential outcomes $\parav$, i.e. $g(\parav) = (\mathbb{F}_{(1)}, \mathbb{F}_{(0)})$.
	Because both variance bounds are in the class of marginal CDF bounds $\mathcal{F}_{\textrm{CDF}}$, we have that they take the form
	\[
	\textrm{VB}^* = h^* \circ g
	\quadand
	\textrm{VB}_{\textrm{AGL}} = h_{\textrm{AGL}} \circ g \enspace,
	\]
	where $g$ is the function that maps potential outcomes to marginal CDFs and $h^* : \mathcal{H} \times \mathcal{H} \to \Reals$ is used in the dominating variance bound.
	Using the fact that $\textrm{VB}^*(\parav) < \textrm{VB}_{\textrm{AGL}}(\parav)$ and unpacking these definitions above, we have that
	\begin{align*}
	h^*( \mathbb{F}_{(1)} , \mathbb{F}_{(0)} )
	&<  h_{\textrm{AGL}}( \mathbb{F}_{(1)} , \mathbb{F}_{(0)} ) \\
	&= \frac{1}{n-1} \paren[\Big]{
		\sigma_1^2(\mathbb{F}_{(1)})
		+ \sigma_0^2(\mathbb{F}_{(0)})
		+  2\rho_{\max}(\mathbb{F}_{(1)} , \mathbb{F}_{(0)})
	} \\
	&= \frac{1}{n-1} \paren[\Big]{
		\sigma_1^2(\mathbb{F}_{(1)})
		+ \sigma_0^2(\mathbb{F}_{(0)})
		+  2 \cdot \sup_{\substack{ \mathbb{F} \in \mathcal{H}_{(2)} \\ Q(\mathbb{F}) = (\mathbb{F}_{(1)}, \mathbb{F}_{(0)}) }}
		\rho(\mathbb{F})
	} \enspace,
\intertext{
	where we have used the definition of the AGL bound as well as the definition of the worst-case correlation term.
	Next, we bring the supremum outside to include the first two terms:
}
	&= \sup_{\substack{ \mathbb{F} \in \mathcal{H}_{(2)} \\ Q(\mathbb{F}) = (\mathbb{F}_{(1)}, \mathbb{F}_{(0)}) }}
		\frac{1}{n-1} \paren[\Big]{
			\sigma_1^2(\mathbb{F}_{(1)})
			+ \sigma_0^2(\mathbb{F}_{(0)})
			+ 2 \cdot \rho(\mathbb{F})
		}
	\intertext{At this point, we can recognize that the variance quantities are not only functions of the individual marginal CDFs, but also of the joint CDF. Thus, we can replace this in the notation below:}
	&= \sup_{\substack{ \mathbb{F} \in \mathcal{H}_{(2)} \\ Q(\mathbb{F}) = (\mathbb{F}_{(1)}, \mathbb{F}_{(0)}) }}
	\frac{1}{n-1} \paren[\Big]{
		\sigma_1^2(\mathbb{F})
		+ \sigma_0^2(\mathbb{F})
		+ 2 \cdot \rho(\mathbb{F})
	}  \\
	&= \sup_{\substack{ \mathbb{F} \in \mathcal{H}_{(2)} \\ Q(\mathbb{F}) = (\mathbb{F}_{(1)}, \mathbb{F}_{(0)}) }} \Varsub{\mathbb{F}}{\widehat{\tau}}
	\end{align*}
	This means that there exists a joint CDF $\mathbb{F} \in \mathcal{H}_{(2)}$ with exactly the marginals $\mathbb{F}_{(1)}$ and $\mathbb{F}_{(0)}$ such that
	\[
	h^*( \mathbb{F}_{(1)} , \mathbb{F}_{(0)} ) < \Varsub{\mathbb{F}}{\widehat{\tau}}
	\enspace.
	\]
	Let $\parav'$ be a set of potential outcomes corresponding to this joint CDF $\mathbb{F}$.
	Because variance bounds in the class of marginal CDF bounds $\mathcal{F}_{\textrm{CDF}}$ factor through the marginals, we have that
	\[
	\textrm{VB}^*(\parav') < \Varsub{\parav'}{\widehat{\tau}}
	\enspace,
	\]
	so that $\textrm{VB}^*$ is not a valid variance bound.
	Thus, we have arrived at a contradiction and our proof is complete.
\end{proof}

Recall from Definition~\mainref{def:sharp} in Section~\mainref{sec:admissibility} that a variance bound $\textrm{VB} : \Reals^{2n} \to \Reals$ is said to be \emph{sharp} if there exists potential outcomes for which the variance bound and the true variance are equal.
The following proposition shows that the AGL variance bound satisfies this definition of sharpness.

\begin{proposition}
	The \citet{Aronow2014Sharp} variance bound $\textrm{VB}_{AGL}$ is sharp.
\end{proposition}
\begin{proof}
	To show that the AGL bound is sharp, it suffices to show that there is a joint CDF $\mathbb{F}$ for which the actual correlation term is equal to the worst-case correlation term, i.e. $\rho(\mathbb{F}) = \rho_{\max}(\mathbb{F}_{(1)}, \mathbb{F}_{(0)})$.
	This follows from the Frechet-Hoeffding theorem which shows that the supremum is actually attained.
\end{proof}

There is, however, a stronger notion of sharpness which may be employed when considering variance bounds in the class of marginal CDF bounds.
Because marginal CDF variance bounds  $\textrm{VB} \in \mathcal{F}_{\textrm{CDF}}$ factorize through the marginal CDFs, we can slightly abuse notation and write them as $\textrm{VB}(\mathbb{F}_{(1)}, \mathbb{F}_{(0)})$.

\begin{definition}
	A variance bound $\textrm{VB} \in \mathcal{F}_{\textrm{CDF}}$ satisfies \emph{uniform marginal sharpness} if for every pair of marginal CDFs $\mathbb{F}_{(1)}$ and $\mathbb{F}_{(0)}$, there exists joint CDF $\mathbb{F}$ which matches the marginals, i.e.
	$
	Q(\mathbb{F}) = (\mathbb{F}_{(1)}, \mathbb{F}_{0})
	$,
	for which the variance bound is equal to the variance: $\textrm{VB}(\mathbb{F}_{(1)}, \mathbb{F}_{(0)}) = \Varsub{\mathbb{F}}{\widehat{\tau}}$.
\end{definition}

The definition of uniform marginal sharpness is stronger than the usual definition of sharpness, in the sense that uniform marginal sharpness implies sharpness.
It says that for every pair of marginal CDFs, there are potential outcomes for which the variance bound is equal to the variance.
Because we never observe more than one potential outcome for any unit, the marginal CDFs can be understood as containing all information about the potential outcomes in this setting.
Uniform marginal sharpness can therefore be understood to say that we extract as much information as possible regarding the joint CDF of the potential outcomes from their marginal CDFs.
The following proposition shows that the AGL bound satisfies uniform marginal sharpness.

\begin{proposition}
	The \citet{Aronow2014Sharp} variance bound $\textrm{VB}_{AGL}$ satisfies uniform \\ marginal sharpness.
\end{proposition}
\begin{proof}
	The proof follows directly from the construction of $\rho_{\max}(\mathbb{F}_{(1)}, \mathbb{F}_{(0)})$.
	In particular, the Frechet-Hoeffding theorem shows that for every pair of marginal CDFs, there exists a joint CDF matching the marginals which achieves the supremum.
	This is precisely the definition of uniform marginal sharpness.
\end{proof}

\citet{Aronow2014Sharp} never formally define their sharpness concept, but it can implicitly be understood from their discussions that they refer to uniform marginal sharpness.

\subsection{Comparison to \optprocedure{}}

We now review key differences between the variance bounds produced by \optprocedure{} and the AGL variance bound.

\paragraph{Admissibility}
First, the relevant notions of admissibility for \optprocedure{} and AGL are different and ultimately incomparable.
The variance bounds considered in this paper are quadratic functions.
As such, the relevant notion of admissibility is with respect to the set of valid quadratic variance bounds, $\mathcal{F}= \allbounds$.
On the other hand, the Frechet--Hoeffding variance bound considered by \citet{Aronow2014Sharp} is not a quadratic function; rather, it belongs to the class of marginal CDF bounds, $\mathcal{F}_{\textrm{CDF}}$.
Because the two bounds are in two different function classes, their relevant notions of admissibility are incomparable.

\paragraph{Sharpness}
As we have discussed above, the AGL bound satisfies the notion of ``uniform marginal sharpness'', which makes its use attractive.
However, there is no clear generalization of this notion beyond the class of marginal CDF bounds.
In particular, the marginal CDFs do not contain sufficient information about the units and the design to construct valid and minimally conservative variance bound under interference and complex designs, so any generalized concept of uniform sharpness must consider some intermediate structure other than the only the marginal CDFs.
While there exists such intermediate structures that contain sufficient information about the variance (e.g. the entire vector of potential outcomes $\parav \in \Reals^{K}$), they are generally not estimable,and thus that they cannot be the basis for an estimator.
It is unclear whether an intermediate structure similar to the marginal CDFs can be constructed to extend the concept of uniform sharpness to a general setting.
This is related to the discussion that immediately follows.

\paragraph{Applicability to Interference and Complex Design}
The major distinction between \optprocedure{} and the bound of \citet{Aronow2014Sharp} is that the latter cannot be applied in general experimental settings.
Indeed, as currently constructed, the use of the Frechet--Hoeffding bound is appropriate only for simple and symmetric designs.
As an example of a setting where the use of Frechet--Hoeffding-type bounds is not appropriate, consider the Horvitz--Thompson estimator under a Bernoulli design in the no-interference settings, where each unit is assigned to treatment with \emph{non-uniform probability} $p_i \in (0,1)$.
Its scaled variance  $n \cdot \Var{\widehat{\tau}}$ will be
\[
\underbrace{ \frac{1}{n} \sum_{i=1}^n y_i(1)^2 \cdot \paren[\Big]{\frac{1}{p_i} - 1} }_{\triangleq s_1^2}
+ \underbrace{ \frac{1}{n} \sum_{i=1}^n y_i(0)^2 \cdot \paren[\Big]{\frac{1}{1-p_i} - 1} }_{\triangleq s_1^2}
+ 2\underbrace{ \frac{1}{n} \sum_{i=1}^n y_i(1) y_i(0)  }_{\triangleq \gamma}
\enspace.
\]
When the treatment assignment probabilities are non-uniform (i.e. $p_i \neq p_j$) the marginal CDFs $\mathbb{F}_{(1)}$ and $\mathbb{F}_{(0)}$ do not provide sufficient information to determine the quantities $s_1^2$ and $s_2^2$.
The reason is that the marginal CDFs ``anonymize'' the potential outcomes, so that any marginal CDF bound cannot map a unit $i \in [n]$ to its treatment assignment probability $p_i$.
In this case, even the terms $s_1^2$ and $s_0^2$ cannot be bounded tightly if only the class of marginal CDF bounds are considered.
The underlying issue here is that the variance itself is no longer a function of the joint CDF $\mathbb{F}$ and so the marginals CDFs{\textemdash}which provide even less information{\textemdash}are not sufficient for obtaining minimally conservative bounds on the variance.
This means that the approach of \citet{Aronow2014Sharp} will not only be inapplicable for general experiments, but also that the relevant function class may be inappropriate for producing desirable variance bounds.

In the setting of binary outcomes, \citet{Ding2016Potential} and \citet{Lu2019Improved} construct variance bounds in a manner which is similar to \citet{Aronow2014Sharp}.
Namely, they identify the inestimable part of the variance and use optimization techniques to determine the maximal value it can take given the marginal CDFs.
Because the outcomes are assumed to be binary, the space of possible CDFs is smaller and so the resulting bound is tighter.
They also claim sharpness of their bounds, which we understand to be essentially the definition of uniform marginal sharpness presented above.

\section{Proofs}\label{sec:proofs}

\subsection{Theorem~\mainref{thm:var-bound-programs}: Monotonicity Implies Admissibility}

\begin{reftheorem}{\mainref{thm:var-bound-programs}}
	\varboundprograms
\end{reftheorem}

\begin{proof}
Let $\boundM^*$ be the bound returned by \optprocedure{}, and let $\slackM^* = \boundM^* - \varM$ be the corresponding slack matrix.
By definition of the program, $\slackM^*$ is a minimizer of $g$ in $\allslacks$.
Because $\slackM^* \in \allslacks$, we know that $\boundM^*$ is conservative and design compatible.
For sake of contradiction, assume that $\boundM^*$ is not admissible.
Then, there exists a conservative and design compatible bound $\boundM \in \allbounds$ with corresponding slack matrix $\slackM \in \allslacks$ such that $\boundM + \inadmissiblecert = \boundM^*$ for some nonzero positive semidefinite matrix $\inadmissiblecert$.
By subtracting $\varM$ from both sides, we can write this equality in terms of slack matrices, $\slackM + \inadmissiblecert = \slackM^*$.
By assumption, the objective $g$ is strictly monotone and $\inadmissiblecert$ is nonzero positive semidefinite.
This implies that
\[
	g(\slackM)
	< g(\slackM + \inadmissiblecert)
	= g(\slackM^*) .
\]
However, $\slackM^*$ is a minimizer of $g$ over $\allslacks$, so we have arrived at a contradiction.
\end{proof}

\subsection{Theorem~\mainref{thm:schatten-is-monotone}: Schatten Norms Are Admissible}

\begin{reftheorem}{\mainref{thm:schatten-is-monotone}}
	\thmSchattenIsMonotone
\end{reftheorem}

\begin{proof}
	Let $\varM$ be an $K$-by-$K$ positive semidefinite matrix and let $\slackM$ be an $K$-by-$K$ positive semidefinite matrix that is not zero.
	Let the eigenvalues of $\varM + \slackM$ be denoted $\mu_1, \mu_2, \dotsc, \mu_{K}$ and let the eigenvalues of $\varM$ be denoted $\eigval{1}, \eigval{2}, \dotsc, \eigval{K}$.

	Because $\slackM$ is positive semidefinite, it follows that $\mu_{\ell} \geq \eigval{\ell}$ for each $1 \leq \ell \leq K$.
	We now show that at least one of these inequalities is strict.
	Recall that the trace of a matrix is the sum of the eigenvalues so that
	\[
	\sum_{\ell =1}^{K} \eigval{\ell}
	= \tr{\varM}
	< \tr{\varM} + \tr{\slackM}
	= \tr{\varM + \slackM}
	= \sum_{\ell =1}^{K} \mu_{\ell} \enspace,
	\]
	where the strict inequality follows from the fact that $\slackM$ is nonzero and positive semidefinite.
	Thus, the inequality is strict for at least one $1 \leq \ell \leq K$.

	The strict monotonicity of $g(\slackM)$ is established using the result above and observing that the function $x \mapsto x^p$ is strictly monotone on the real line.
\end{proof}

\subsection{Theorem~\mainref{thm:sdp-procedure-admissibility}: Admissibility and Linear Objectives}

\begin{reftheorem}{\mainref{thm:sdp-procedure-admissibility}}
	\thmsdpprocedureadmissibility{}
\end{reftheorem}

\begin{proof}
	To show that every variance bound obtained using the objective $g(\slackM) = \iprod{\objM, \slackM}$ is admissible, we show that $g$ is strictly monotone and appeal to Theorem~\mainref{thm:var-bound-programs}.
	Let $\inadmissiblecert$ be a nonzero positive semidefinite matrix.
	Then we may write
	\[
	g(\slackM + \inadmissiblecert) - g(\slackM)
	= \iprod{\objM, \slackM + \inadmissiblecert} - \iprod{\objM, \slackM}
	= \iprod{\objM , \inadmissiblecert} ,
	\]
	where the last equality follows by linearity of the inner product.
	Let the eigendecomposition of $\inadmissiblecert$ be given as $\inadmissiblecert =  \sum_{i=1}^{K} \eigvali \eigveci \eigveci^\tran$.
	Then, the inner product may be rewritten as
	\[
	\iprod{\objM , \inadmissiblecert}
	= \iprod{\objM , \sum_{i=1}^{K} \eigvali \eigveci \eigveci^\tran}
	= \sum_{i=1}^{K} \eigvali \eigveci^\tran \objM \eigveci
	.
	\]
	Because $\objM$ is positive definite, each of the $\eigveci^\tran \objM \eigveci$ terms are positive.
	Likewise, because $\inadmissiblecert$ is positive semidefinite and nonzero, there exists at least one positive eigenvalue $\eigvali > 0$.
	This establishes that $g(\slackM + \inadmissiblecert) > g(\slackM)$ so that $g$ is strictly monotone.
	Thus, by Theorem~\mainref{thm:var-bound-programs}, the resulting variance bound is admissible.

	Now, suppose that $\boundM$ is an admissible variance bound and write the corresponding slack matrix as $\slackM = \boundM - \varM$.
	Define the set
	\[
	\mathcal{F}_\boundM
	= \setb[\big]{
		\otherboundM = \boundM - \inadmissiblecert
		:  \inadmissiblecert \text{ is nonzero and positive semidefinite}
	}
	\enspace.
	\]
	Because $\boundM$ is admissible, there does not exist another variance bound $\otherboundM \in \allbounds$ which is in the set $\mathcal{F}_\boundM$.
	In other words, the intersection of $\mathcal{F}_\boundM$ and  $\allbounds$ is empty.
	Because the two sets $\mathcal{F}_\boundM$ and  $\allbounds$ are disjoint and convex, there exists a separating hyperplane between them.
	That is, there exists a matrix $\objM$ and a scalar $\alpha$ so that
	\begin{align}
	\iprod{ \objM , \boundM} \geq \alpha &\text{ for all } \boundM \in \allbounds \\
	\iprod{ \objM , \otherboundM} < \alpha &\text{ for all } \otherboundM \in \mathcal{F}_\boundM
	\end{align}

	Let us first establish that $\iprod{\objM , \boundM} = \alpha$.
	For sake of contradiction, suppose that $\iprod{\objM , \boundM} = \alpha + \epsilon$ for some $\epsilon > 0$.
	Consider the matrix $\mat{H} = \boundM - \beta \cdot \unitM$, where $\beta = \frac{\epsilon}{2 \tr{\objM}}$.
	It follows that $\mat{H}$ is in the set $\mathcal{F}_\boundM$.
	However, we can compute
	\[
	\iprod{\objM , \mat{H}}
	= \iprod{\objM, \boundM - \beta \cdot \unitM}
	= \iprod{\objM, \boundM} - \beta \iprod{\objM, \unitM}
	\geq \alpha + \epsilon / 2
	\enspace,
	\]
	which is a contradiction of the separating hyperplane.
	Thus, $\iprod{\objM , \boundM} = \alpha$.

	Let us next establish that $\objM$ is positive definite.
	Let the eigenvalue decomposition of $\objM$ be given as $\objM = \sum_{i=1}^{\numvar} \eigvali \eigveci \eigveci^\tran$.
	For sake of contradiction, suppose that one of the eigenvalues $\eigval{k}$ is non-positive.
	Let $\eigvec{k}$ be the corresponding eigenvector.
	Consider the matrix $\mat{H} = \boundM - \eigvec{k} \eigvec{k}^\tran$, which is in the set $\mathcal{F}_\boundM$.
	However, we can obtain that
	\[
	\iprod{\objM, \mat{H}}
	= \iprod{\objM, \boundM - \eigvec{k} \eigvec{k}^\tran}
	= \iprod{\objM, \boundM} - \eigvec{k}^\tran \objM \eigvec{k}
	= \alpha - \eigval{k}
	\geq \alpha \enspace,
	\]
	which is a contradiction of the separating hyperplane.
	Thus, $\objM$ is positive definite.

	Finally, we show that $\boundM$ may be obtained by using the objective $g(\slackM) = \iprod{\objM, \slackM}$.
	First, observe that the corresponding slack matrix takes value
	\[
	g(\slackM) = \iprod{\objM, \slackM}
	= \iprod{\objM, \varM + \slackM} - \iprod{\objM, \varM}
	= \iprod{\objM, \boundM} - \iprod{\objM, \varM}
	= \alpha - \iprod{\objM, \varM} \enspace.
	\]
	By the separating hyperplane, any other slack matrix $\otherslackM \in \allslacks$ (with corresponding bound matrix $\otherboundM \in \allbounds$) has objective value at most
	\[
	g(\otherslackM) = \iprod{\objM, \otherslackM}
	= \iprod{\objM, \varM + \otherslackM} - \iprod{\objM, \varM}
	= \iprod{\objM, \otherboundM} - \iprod{\objM, \varM}
	\geq \alpha - \iprod{\objM, \varM} \enspace.
	\]
	Thus, $\slackM$ is a minimizer of $g$ over $\allslacks$.
\end{proof}

\subsection{Proposition~\mainref{prop:unbiased-design-compatible}: Design Compatibility and Unbiasedness}

\begin{refproposition}{\mainref{prop:unbiased-design-compatible}}
	\propunbiaseddesigncompatible{}
\end{refproposition}

\begin{proof}
To show the first direction, that an unbiased estimator exists if a quadratic form is design compatible, we prove that the Horvitz--Thompson estimator described in Section~\mainref{sec:estimate-vb} is unbiased for a quadratic form $\parav^\tran \boundM \parav$.
After defining $0/0$ to be zero, the estimator can be written as
\begin{equation}
	\frac{1}{n^2} \sum_{k = 1}^{K} \sum_{\ell = 1}^{K} \frac{\boundMkl \parak \paral \indicator{k, \ell \in \Obs}}{\Pr{k, \ell \in \Obs}},
\end{equation}
because when $\boundM$ is design compatible, $\Pr{k, \ell \in \Obs}$ is zero only when $\boundMkl$ is zero.
Taking expectation yields
\begin{equation}
	\frac{1}{n^2} \sum_{k = 1}^{K} \sum_{\ell = 1}^{K} \boundMkl \parak \paral = \frac{1}{n^2} \parav^\tran \boundM \parav,
\end{equation}
which completes the proof of the first direction.

We will show the other direction by proving the contrapositive: a quadratic form that is not design compatible implies that no unbiased estimator exists.
For sake of contradiction, suppose that there exists an unbiased estimator $Q$ such that $n^2 \E{Q} = \parav^\tran \mat{A} \parav$.
Let $k, \ell \in \set{P}$ be such that $\Pr{k, \ell \in \set{S}} = 0$.
We know that such a pair exists because we have stipulated that the quadratic form is not design compatible.
Use the law of iterated expectation to write
\begin{equation}
	\E{Q} = \Pr{k \in \set{S}} \func{f}{\parav}  + \Pr{k \notin \set{S}} \func{g}{\parav},
\end{equation}
where $\func{f}{\parav} = \E{Q \given k \in \set{S}}$ and $\func{g}{\parav} = \E{Q \given k \notin \set{S}}$.
We know that $\func{g}{\parav}$ does not depend on $\parak$ because the coordinate is never observed when $k \notin \set{S}$.
Recall that $\Pr{k, \ell \in \set{S}} = 0$, so we know that $\func{f}{\parav}$ does not depend on $\paral$ because the coordinate is never observed when $k \in \set{S}$.
It is not possible to write a quadratic form $\parav^\tran \boundM \parav$ with $\boundMkl \neq 0$ as a sum of two functions where one does not depend on $\parak$ and the other does not depend on $\paral$.
\end{proof}

\subsection{Proposition~\mainref{prop:admissible-implies-sharp}: Admissibility Implies Sharpness}

\begin{lemma} \label{lemma:zero-diag-means-tight}
	If $\boundM \in \allbounds$ has a diagonal element which is zero, then $\boundM$ is sharp.
\end{lemma}
\begin{proof}
	Recall that the variance matrix $\varM$ is positive semi-definite, which means that each diagonal element is non-negative, i.e. $\varM(k,k) \geq 0$ for all $k \in [K]$.

	Suppose that $\boundM(k,k) = 0$ for some $k \in [K]$.
	Because $\boundM \succeq \varM$, this means that
	\[
	0
	= \boundM(k,k)
	\geq \varM(k,k)
	\geq 0
	\]
	so that $\varM(k,k) = 0$.
	Consider $\parav = \vec{e}_k$ which is 1 in coordinate $k$ and 0 in all other coordiantes.
	Then, $\parav \neq \vec{0}$, and
	\[
	\parav^\tran \boundM \parav
	= \boundM(k,k)
	= \varM(k,k)
	= \parav^\tran \varM \parav
	\enspace,
	\]
	so that $\boundM$ is sharp.
\end{proof}

\begin{refproposition}{\mainref{prop:admissible-implies-sharp}}
	\admissibleimpliessharp
\end{refproposition}

\begin{proof}
	Using proof by contradiction, suppose that $\boundM \in \allbounds$ is an admissible bound that is not sharp.
	Because $\boundM$ is admissible, there exists no other $\widetilde{\boundM} \in \allbounds$ that dominates $\boundM$.
	Because $\boundM$ is not sharp, we have that $\parav^\tran \boundM \parav > \parav^\tran \varM \parav$ for every $\parav \in \Reals^K$ such that $\parav \neq \vec{0}$.

	Let $\delta \triangleq \eigvalmin(\slackM)$ be the smallest eigenvalue of the slack matrix $\slackM = \boundM - \varM$.
	The fact that $\boundM$ is not sharp implies that $\delta > 0$.
	Define a new bound by the matrix $\widetilde{\boundM} = \boundM - \delta \cdot \unitM$.
	We will now show that $\widetilde{\boundM}$ dominates $\boundM$.

	First, we will show that the new bound is weakly larger than the variance matrix in the Loewner order: $\widetilde{\boundM} \succeq \varM$.
	To this end, observe that
	\begin{align*}
		\eigvalmin(\widetilde{\boundM} - \varM)
		&= \eigvalmin(\boundM - \delta \cdot \unitM- \varM) \\
		&= \eigvalmin(\boundM - \varM) - \delta \\
		&= \delta - \delta \\
		&= 0.
	\end{align*}
	Next, we will show that the new matrix $\widetilde{\boundM}$ is design compatible.
	Because the original bound matrix $\boundM$ is not sharp, we have by Lemma~\ref{lemma:zero-diag-means-tight} that its diagonal entries are all positive.
	Because $\boundM$ is design compatible, the corresponding diagonal entries are observed with positive probability: $(k,k) \notin \unobspairs$.
	By construction, the new matrix $\widetilde{\boundM}$ is different from $\boundM$ only in its diagonal entries.
	This means that $\widetilde{\boundM}(k,\ell) = \boundM(k,\ell)$ for all $(k,\ell) \in \unobspairs$ and thus design compatibility of $\widetilde{\boundM}$ follows from design compatibility of $\boundM$.

	A matrix that is greater than the variance matrix in the Loewner order and design compatibility is by definition a variance bound, so we have that $\widetilde{\boundM} \in \allbounds$.
	By construction, we have that $\widetilde{\boundM} \prec \boundM$ because $\delta > 0$.
	Thus, $\widetilde{\boundM}$ dominates $\boundM$.
	Because $\boundM$ is dominated by another bound, it cannot be admissible, so we have our contradiction.
\end{proof}

\subsection{Proposition~\mainref{prop:neyman-is-admissible}: Neyman Bound is Admissible}

In this section, we prove that the Neyman bound is admissible.
By Theorem~\mainref{thm:sdp-procedure-admissibility}, it suffices to show that Neyman's bound may be obtained through VB-OPT with a positive definite linear objective.

This style of proof has two challenges.
The first challenge is that of producing a candidate positive definite linear objective for which Neyman's bound is an optimal solution.
The second challenge is verifying that Neyman's bound is indeed optimal with respect to this objective.
For this second challenge, we present a proof which is derived from the duality theory of semi-definite programming.
Indeed, the matrix $\dualM$ in the proof below may be understood as a matrix-valued dual variable; however, no knowledge of duality theory is required to verify the proof.
While we find these proofs to be useful and relatively short, we remark that they are somewhat challenging to produce, due to the ``guess-and-check'' nature involved with identifying (1) a positive definite linear objective (2) dual certificates simultaneously.

\begin{refproposition}{\mainref{prop:neyman-is-admissible}}
	\Neymanisadmissible
\end{refproposition}

\begin{proof}
	Recall that, as discussed in Section~\mainref{sec:prev-examples}, that the variance of the difference-in-means estimator under a completely randomized experiment is given by the quadratic function
	\[
	n^2 \cdot \Varsub{\parav}{\hat{\tau}} =  \parav^\tran \varM \parav
	\enspace,
	\]
	where the potential outcomes are grouped as $\parav = \paren{ y_1(1) \dots y_n(1), y_1(0) \dots v_n(0) } \in \Reals^{2n}$
	and the $2n$-by-$2n$ matrix $\varM$ is
	\[
	\varM = \paren[\Big]{ \frac{n}{n-1}}
	\begin{bmatrix}
		\mat{H} & \mat{H} \\
		\mat{H} & \mat{H}
	\end{bmatrix}
	\qquadtext{and}
	\mat{H} = \unitM - \onevec \onevec^\tran / n
	\enspace.
	\]
	The unobservable pairs $\unobspairs$ are those that correspond to different potential outcomes for the same unit $y_i(1) y_i(0)$, i.e.
	\[
	\unobspairs = \setb[\Big]{ (i,i+n) \text{ and } (i+n, i) \text{ for } i \in [n]  }
	\enspace.
	\]
	Neyman's bound can be written as $n^2 \cdot \textrm{VB} = \parav^\tran \boundM \parav$, where
	\[
	\boundM =
	2 \paren[\Big]{ \frac{n}{n-1}}
	\begin{bmatrix}
		\mat{H} & \mathbf{0} \\
		\mathbf{0} & \mat{H}
	\end{bmatrix}
	\]
	which corresponds to the slack matrix
	\[
	\slackM =
	\paren[\Big]{ \frac{n}{n-1}}
	\begin{bmatrix}
		\mat{H} & - \mat{H} \\
		- \mat{H} & \mat{H}
	\end{bmatrix}
	\enspace.
	\]

	To show that Neyman's variance bound is admissible, it suffices to show (by Theorem~\mainref{thm:sdp-procedure-admissibility}) that the slack matrix $\slackM$ is a solution to the VB-OPT program for a linear objective $\objfS = \iprod{\slackM , \objM}$ where $\objM$ is positive definite.
	We claim that $\objM = \unitM$ is one such objective where Neyman's bound is the optimal solution.

	To begin, we claim that the slack matrix $\slackM$ corresponding to Neyman's bound attains the objective value $\iprod{\slackM , \objM} = 2n$.
	This is verified by the following calculation:
	\begin{align*}
		\iprod{\slackM , \objM}
		&= \iprod{\slackM}
			&\text{($\objM = \unitM$)} \\
		&= \tr[\Bigg]{
			\paren[\Big]{ \frac{n}{n-1}}
			\begin{bmatrix}
				\mat{H} & - \mat{H} \\
				- \mat{H} & \mat{H}
			\end{bmatrix}
		}
		&\text{(def of Neyman's $\slackM$)} \\
		&= 2 \paren[\Big]{ \frac{n}{n-1}} \tr{ \mat{H} }
			&\text{(linearity of $\trsym$)} \\
		&= 2 \paren[\Big]{ \frac{n}{n-1}} \sum_{i=1}^n \mat{H}_{i,i}
			&\text{(definition of $\trsym$)} \\
		&= 2 \paren[\Big]{ \frac{n}{n-1}}  \cdot n \cdot \paren[\Big]{ 1 - \frac{1}{n}}
			&\text{(definition of $\mat{H}$)} \\
		&= 2n \enspace.
	\end{align*}

	Next, we will show that any feasible slack matrix $\slackM \in \allslacks$ has an objective value which is bounded below as $\iprod{\slackM , \objM} \geq 2n$.
	This would immediately imply optimality of Neyman's bound for this linear objective and render the proof complete.
	To this end, we define a new $2n$-by-$2n$ matrix $\dualM$ which is given by
	\[
	\dualM =
	\begin{bmatrix}
		\mathbf{0} & \unitM\\
		\unitM & \mathbf{0}
	\end{bmatrix}
	\enspace.
	\]
	It is straightforward to verify that all of the eigenvalues of $\dualM$ are $\pm 1$, so that $\dualM \preceq \unitM$.
	Let $\slackM \in \allslacks$ be a valid variance bound.
	Given the pattern of unobservability associated with the completely randomized design, this means that the valid slack matrix $\slackM$ must satisfy that for all $i \in [n]$,
	\begin{equation} \label{eq:slack-constrained-neyman}
	\slackM(i,i+n)
	= - \varM(i,i+n)
	= \paren[\Big]{ \frac{n}{n-1} } \paren[\Big]{ 1 - \frac{1}{n} } = 1
	\end{equation}
	and likewise, $\slackM(i+n,i) = 1$.

	This implies that the objective value of any feasible solution may be bounded below as
	\begin{align*}
		\iprod{\slackM , \objM}
		&\geq \iprod{ \slackM, \dualM }
			&\text{($\dualM \preceq \unitM$ and $\mathbf{0} \preceq \slackM$)} \\
		&= \tr[\Big]{\slackM
			\begin{bmatrix}
				\mathbf{0} & \unitM\\
				\unitM & \mathbf{0}
			\end{bmatrix}
		 	}
 			&\text{(def of $\dualM$)} \\
 		&= 2 \sum_{i \in [n]} \slackM(i,i+n)
 			&\text{(def of $\trsym$)} \\
 		&= 2n \enspace,
	\end{align*}
	where the last line followed from equation~\eqref{eq:slack-constrained-neyman}.
\end{proof}

\subsection{Proposition~\mainref{prop:AS-inadmissible}: Inadmissibility of the Aronow--Samii Bound}

\begin{refproposition}{\mainref{prop:AS-inadmissible}}
	\ASinadmissible{}
\end{refproposition}

\begin{proof}
We prove the proposition with an example.
Indeed, the illustration in Section~\mainref{sec:illustration} in the main paper is an example where the Aronow--Samii bound is inadmissible.
With $\parav = \paren{a_1, a_2, b_1, b_2}$, the variance of the estimator is given by the matrix
\begin{equation}
	\varM =
	\begin{bmatrix}
		\phantom{-}1 & -1 & \phantom{-}1 & -1 \\
		-1 & \phantom{-}1 & -1 & \phantom{-}1 \\
		\phantom{-}1 & -1 & \phantom{-}1 & -1 \\
		-1 & \phantom{-}1 & -1 & \phantom{-}1 \\
	\end{bmatrix}.
\end{equation}
The two bounds in the illustration, $B_1$ and $B_2$, are given by
\begin{equation}
	\boundM_1 =
	\begin{bmatrix}
		\phantom{-}2 & \phantom{-}0 & \phantom{-}0 & -2 \\
		\phantom{-}0 & \phantom{-}2 & -2 & \phantom{-}0 \\
		\phantom{-}0 & -2 & \phantom{-}2 & \phantom{-}0 \\
		-2 & \phantom{-}0 & \phantom{-}0 & \phantom{-}2 \\
	\end{bmatrix}
	\qquadand
	\boundM_2 =
	\begin{bmatrix}
		\phantom{-}3 & \phantom{-}0 & \phantom{-}0 & -1 \\
		\phantom{-}0 & \phantom{-}3 & -1 & \phantom{-}0 \\
		\phantom{-}0 & -1 & \phantom{-}3 & \phantom{-}0 \\
		-1 & \phantom{-}0 & \phantom{-}0 & \phantom{-}3 \\
	\end{bmatrix},
\end{equation}
of which $\boundM_2$ corresponds to the Aronow--Samii bound.
Both these bounds are conservative and design compatible, so they valid according to Definition~\mainref{def:valid-bound}.
However, the difference
\begin{equation}
	\boundM_2 - \boundM_1 =
	\begin{bmatrix}
		1 & 0 & 0 & 1 \\
		0 & 1 & 1 & 0 \\
		0 & 1 & 1 & 0 \\
		1 & 0 & 0 & 1 \\
	\end{bmatrix}
\end{equation}
is positive semidefinite, so Definition~\mainref{prop:quadratic-admissible} tells us that $\boundM_2$ is inadmissible.
\end{proof}

\subsection{Proposition~\mainref{prop:composite-objective}: Monotonicity of Composite Objectives}

The following proposition guarantees that positive combinations of monotone functions is strictly monotone, provided that one of the functions is strictly monotone.

\begin{refproposition}{\mainref{prop:composite-objective}}
	\compositeobjective
\end{refproposition}

\begin{proof}
	Let $f = g + \gamma h$, and let $\varM$ be a positive semidefinite matrix and let $\slackM$ be a positive semidefinite, nonzero matrix.
	We have that
	\begin{align*}
		f(\varM)
		&= g(\varM) + \gamma h(\varM) \\
		&< g(\varM + \slackM) + \gamma h(\varM) \\
		&\leq g(\varM + \slackM) + \gamma h(\varM + \slackM) \\
		&= f(\varM + \slackM),
	\end{align*}
	where the equalities follow by definition of $f$, the strict inequality follows by strict monotonicity of $g$, and the next inequality follows by monotonicity of $h$ and non-negativity of $\gamma$.
\end{proof}

\subsection{Proposition~\mainref{prop:finite-sample-MSE-bound}: Precision of Variance Bound Estimator}\label{app:consistent-var-est}

In the main body, Proposition~\mainref{prop:finite-sample-MSE-bound} gave an upper bound on the mean squared error of the Horvitz--Thompson estimator of the variance bound.
This upper bound was the product of three terms, corresponding to the design, the variance bound, and the potential outcomes.
The bound depended on the largest magnitude of the potential outcomes.
In this section, we prove this bound, and demonstrate how the bound may be generalized so that only second moment conditions are required on the potential outcomes.

Before continuing, we introduce the entry-wise $L_{p,q}$ matrix norm.
Given an $n \times n$ matrix $\mat{A}$, the \emph{$L_{p,q}$ matrix norm} is defined as
\[
\norm{\mat{A}}_{p,q}
= \bracket[\Bigg]{ \sum_{i=1}^n
	\paren[\bigg]{
		\sum_{j=1}^n \abs{a_{i,j}}^p
	}^{q/p}
}^{1/q}
\enspace.
\]
When $p = q = 2$, then we recover the usual Frobenius norm.
The more general finite sample bound on the MSE of the variance bound estimator appears below as Proposition~\ref{prop:finite-sample-MSE-bound-general}.

\begin{proposition}\label{prop:finite-sample-MSE-bound-general}
Suppose that the variance bound $\boundM$ is design-compatible.
Then, for any $p,q\geq 1$ with $1/p + 1/q = 1$, the mean squared error of the Horvitz--Thompson estimator is bounded as
\begin{equation}
	\E[\Big]{ \paren[\big]{ n \varbf{\parav} - n \varbfest{\parav} }^2 }
	\leq
	\frac{1}{n^2}
	\pnorm{\infty}{\Cov{\vec{R}}} \times
	\pnorm{2p,2p}{\boundM}^2 \times
	\paren[\bigg]{\sum_{k=1}^{K} \parak^{2q} }^{2/q}.
\end{equation}
\end{proposition}
\begin{proof}
	Because $\Pr{i,j \in \Obs} > 0$ for all pairs $(i, j) \notin \unobspairs$, the Horvitz--Thompson estimator of $\varbf{\parav}$ is unbiased.
	Thus, the mean squared error is equal to the variance of the estimator, which can be computed as
	\begin{align*}
		\Var{n \varbfest{\parav}}
		&= \Var[\Bigg]{ \frac{1}{n} \sum_{k = 1}^{K} \sum_{\ell = 1}^{K} \frac{\indicator{k, \ell \in S} \times \indicator{\boundMkl > 0}}{\Pr{k, \ell \in S}} \boundMkl \parak \paral } \\
		&= \frac{1}{n^2} \sum_{k = 1}^{K} \sum_{\ell = 1}^{K} \sum_{s = 1}^{K} \sum_{t = 1}^{K}
		\Cov[\Big]{
			R_{k\ell} \boundMkl \parak \paral,
			R_{st} \boundMe{st} \para{s} \para{t}
		} \\
		&= \frac{1}{n^2} \sum_{k = 1}^{K} \sum_{\ell = 1}^{K} \sum_{s = 1}^{K} \sum_{t = 1}^{K}
		\Cov[\big]{ R_{k\ell}, R_{st} }
		\paren{\boundMkl \parak \paral} \paren{\boundMe{st} \para{s} \para{t}} \\
		&= \frac{1}{n^2} \tilde{\vec{b}}^\tran \Cov{\vec{R}} \tilde{\vec{b}} \enspace,
	\end{align*}
	where $\tilde{\vec{b}}$ is a vector of length $K^2$ whose coordinates are indexed by pairs $(i,j) \in \Pop^2$.
	The entries of $\tilde{\vec{b}}$ are the product of the variance bound and outcomes, $\boundMkl \parak \paral$.
	Using the operator norm bound on the above, we have that the mean squared error is bounded as
	\[
	\E[\Big]{ \paren[\big]{ n \varbf{\parav} - n \varbfest{\parav} }^2 }
	= \frac{1}{n^2} \tilde{\vec{b}}^\tran \Cov{\vec{R}} \tilde{\vec{b}}
	\leq \frac{1}{n^2} \pnorm{\infty}{\Cov{\vec{R}}} \times\enorm{\tilde{\vec{b}}}^2.
	\]
	Finally, we bound the squared $\ell_2$ norm of the vector $\tilde{\vec{b}}$.
	Using H{\"o}lder's inequality, we have that for any $p,q\geq 1$ with $1/p + 1/q = 1$,
	\[
	\enorm{\tilde{\vec{b}}}^2
	= \sum_{k = 1}^{K} \sum_{\ell = 1}^{K} \boundMkl^2 \paren{\parak \paral}^2
	\leq \paren[\bigg]{\sum_{k = 1}^{K} \sum_{\ell = 1}^{K} \boundMkl^{2p}}^{2/2p}
	\paren[\bigg]{\sum_{k = 1}^{K} \sum_{\ell = 1}^{K} \paren{\parak \paral}^{2q}}^{1/q}.
	\]
	The first factor on the right is $\pnorm{2p,2p}{\boundM}^2$.
	We can write the second factor as
	\[
	\paren[\bigg]{\sum_{k = 1}^{K} \sum_{\ell = 1}^{K} \paren{\parak \paral}^{2q}}^{1/q}
	= \paren[\bigg]{\sum_{k = 1}^{K} \parak^{2q}}^{2/q}.
	\qedhere
	\]
\end{proof}

Proposition~\mainref{prop:finite-sample-MSE-bound} is obtained by letting $p \to 1$ and $q \to \infty$.
Using Proposition~\ref{prop:finite-sample-MSE-bound-general}, we can now establish more general conditions under which consistent estimation of the variance bound is possible.

\begin{corollary}\label{cor:consistency-var-bound-general}
Suppose that the variance bound $\boundM$ is design compatible, $\pnorm{\infty}{\Cov{\vec{R}}}$ is asymptotically bounded by a constant, and there exists integers $p$ and $q$ with $1/p + 1/q = 1$ such that $\norm{\boundM}^2_{2p,2p} \paren{\sum_{k=1}^{K} \parak^{2q} }^{2/q} = \littleO{n^2}$, then the Horvitz--Thompson estimator is a consistent estimator of the variance bound:
	$\E{\paren{n\varbf{\parav} - n\varbfest{\parav} }^2} \rightarrow 0$.
\end{corollary}

Motivated by Corollary~\ref{cor:consistency-var-bound-general}, some experimenters may wish to modify the regularized objective by replacing the square of the Frobenius norm with the square of the entry-wise $L_{2p,2p}$ norm.

\subsection{Proposition~\ref*{prop:formal-limiting-argument}: Regularization of Operator Norm Objective}\label{sec:proof-formal-limiting-argument}

We now prove Proposition~\ref{prop:formal-limiting-argument}, which describes the variance bound obtained as the regularization parameter goes to zero.
We prove a more general proposition about solutions to regularized convex optimization problems in the limit where the regularization goes to zero.

\begin{proposition}\label{prop:general-limiting-argument}
	Let $\mathcal{D}$ be a closed convex subset of $\Reals^d$ and let $f: \Reals^d \rightarrow \Reals$ and $g: \Reals^d \rightarrow \Reals$ be continuous convex functions with compact sub-level sets.
	Furthermore, suppose that $g$ is strictly convex.
	Let $\setb{\delta_k}_{k=1}^\infty$ be a sequence of positive values converging to zero.
	Let $\setb{\vx_k}_{k=1}^\infty$ be a sequence of solutions to the following regularized program:
	\begin{equation}
		\vx_k = \argmin_{\vx \in \mathcal{D}} f(\vx) + \delta_k g(\vx) \enspace.
	\end{equation}
	The sequence of solutions converges to $\vx_k \rightarrow \vx^*$, where $\vx^*$ is the solution minimizing $g$ among all minimizers of $f$:
	\begin{equation}
		\begin{aligned}
			\vx^* = \argmin_{\vx \in \mathcal{D}} \quad & g(\vx)\\
			\textrm{s.t.} \quad & f(\vx) = \inf \braces{ f(\vy) : \vy \in \mathcal{D}}.
		\end{aligned}
	\end{equation}
\end{proposition}

To prove Proposition~\ref{prop:general-limiting-argument}, we will need the following elementary fact from real analysis,
which gives a characterization of when a sequence converges in terms of the behavior of all convergent subsequences.

\begin{lemma}\label{lemma:analysis-fact}
	Let $\setb{\vx_k}_{k=1}^\infty$ be a bounded sequence in a compact metric space $\mathcal{X}$. If every convergent subsequence has the same limit point $\vx^* \in \mathcal{X}$, then the sequence $\setb{\vx_k}_{k=1}^\infty$ converges to $\vx^*$.
\end{lemma}
\begin{proof}
	We give a sketch of the proof here.
	For sake of contradiction, suppose that $\setb{\vx_k}_{k=1}^\infty$ does not converge to $\vx^*$.
	Then, there exists a subsequence $\setb{\vx_{k_\ell}}_{\ell=1}^\infty$ which is bounded away from $\vx^*$.
	Because the subsequence $\setb{\vx_{k_\ell}}_{\ell=1}^\infty$ is contained in a compact metric space, it has, itself, a subsequenece $\setb{\vx_{k_{\ell_r}}}_{r=1}^\infty$ which converges to a point $\vx \in \mathcal{X}$.
	Because the subsequence $\setb{\vx_{k_\ell}}_{\ell=1}^\infty$ was bounded away from $\vx^*$, we have that $\vx \neq \vx^*$.
	However, $\setb{\vx_{k_{\ell_r}}}_{r=1}^\infty$  is a convergent subsequence of $\setb{\vx_k}_{k=1}^\infty$ and so by the hypothesis of the lemma, its limit point must be $\vx^*$; and so, we have arrived at a contradiction.
\end{proof}

With this lemma in hand, we are ready to prove Proposition~\ref{prop:general-limiting-argument}.

\begin{proof}[Proof of Proposition~\ref{prop:general-limiting-argument}]
	Our approach will be to apply Lemma~\ref{lemma:analysis-fact}, first by showing that the sequence $\setb{\vx_k}_{k=1}^\infty$ is contained in a compact set and then to show that every convergent subsequence has $\vx^*$ as its limit point.

	To this end, let $k \in \Naturals$ be given.
	By optimality of $\vx_k$ for the regularized program, we have that
	\[
	f(\vx_k) + \delta_k g(\vx_k) \leq f(\vx^*) + \delta_k g(\vx^*)
	\]
	Similarly, because $\vx^*$ minimizes $f$ over the set $\mathcal{D}$, we have that
	\[ f(\vx^*) \leq f(\vx_k) \enspace. \]
	Rearranging terms and putting these inequalities together, we obtain the bound
	\begin{equation}\label{eq:fg_inequality}
	0
	\leq f(\vx_k) - f(\vx^*)
	\leq \delta_k \paren[\big]{g(\vx^*) - g(\vx_k)}
	\enspace.
	\end{equation}

	Because $g$ has compact sub-level sets, it is bounded from below on its domain.
	More precisely, there exists a real number $B$ such that $g(\vx) \geq B$ for all $\vx \in \Reals^d$.
	Additionally, because the sequence of regularization parameters $\setb{\delta_k}_{k=1}^\infty$ converges to zero, there exists a bound $b$ such that $\delta_k \leq b$ for all $k \in \Naturals.$
	Thus, we have the inequality
	\[
	f(\vx_k)
	\leq f(\vx^*) + \delta_k \paren[\big]{g(\vx^*) - g(\vx_k)}
	\leq f(\vx^*) + b \paren[\big]{g(\vx^*) - B}
	\]
	so that the sequence $\setb{\vx_k}_{k=1}^\infty$ is contained in the sub-level set $\setb{\vx \in \Reals^d : f(\vx) \leq C}$, where $C = f(\vx^*) + b \paren[\big]{g(\vx^*) - B}$.
	By assumption, the sub-level sets of $f$ are compact so that the sequence $\setb{\vx_k}_{k=1}^\infty$  is contained in a compact metric space.
	In other words, we have established that $\setb{\vx_k}_{k=1}^\infty$ satisfies the assumption of Lemma~\ref{lemma:analysis-fact}.

	We now seek to show that every convergent subsequence of $\setb{\vx_k}_{k=1}^\infty$ has $\vx^*$ as its limit point.
	To this end, suppose that $\setb{\vx_{k_\ell}}_{\ell=1}^\infty$ is a convergent subsequence of $\setb{\vx_k}_{k=1}^\infty$ which has limit point $\vx^\dagger$.
	In the next step of the proof, we will establish that
	\begin{equation}\label{eq:limit-facts}
		f(\vx^\dagger) = f(\vx^*)
		\quadand
		g(\vx^\dagger) \leq g(\vx^*)
	\end{equation}

	We begin by verifying the equality $f(\vx^\dagger) = f(\vx^*)$.
	Observe that
	\begin{align*}
		f(\vx^\dagger) - f(\vx^*)
		&=\lim_{k \rightarrow \infty} f(\vx_k) - f(\vx^*)
			&\text{(continuity of $f$)} \\
		&\leq \lim_{k \rightarrow \infty} \delta_k \paren[\big]{g(\vx^*) - g(\vx_k)}
			&\text{(Inequality \eqref{eq:fg_inequality})} \\
		&= \lim_{k \rightarrow \infty} \delta_k \paren[\big]{g(\vx^*) - B}
			&\text{(boundedness of $g$)} \\
		&= 0 \enspace,
			&\text{($\delta_k \rightarrow 0$)}
	\end{align*}
	which verifies that $f(\vx^\dagger) \leq f(\vx^*)$.
	Because $\mathcal{D}$ is closed, we have that $\vx^\dagger \in \mathcal{D}$ and so by definition of $\vx^*$, we have that $f(\vx^\dagger) \geq f(\vx^*)$.
	Thus, $f(\vx^\dagger) = f(\vx^*)$, which establishes that $\vx^\dagger$ is a minimizer of $f$ over $\mathcal{D}$.

	We now verify the inequality $g(\vx^\dagger) \leq g(\vx^*)$.
	Recall that the regularization parameters are positive so that by dividing both sides of \eqref{eq:fg_inequality} and rearranging terms, we have that $g(\vx_k) \leq g(\vx^*)$.
	The result now follows from continuity of $g$ and taking the limit as $k \rightarrow \infty$.

	Thus, we have established that the limit point $\vx^\dagger$ of the convergent subsequence $\setb{\vx_{k_\ell}}_{\ell=1}^\infty$ satisfies the properties in equation \eqref{eq:limit-facts}.
	We now show that under the assumptions of the proposition, this implies that $\vx^\dagger = \vx^*$.
	For sake of contradiction, suppose that $\vx^\dagger \neq \vx^*$, and consider the midpoint between them, $\vx = 1/2 \cdot \vx^\dagger + 1/2 \cdot \vx^*$.
	First, observe that $\vx \in \mathcal{D}$ because it is the convex combination of the points $\vx^\dagger$ and $ \vx^*$ in the convex set $\mathcal{D}$.
	Second, observe that $\vx$ is a minimizer of $f$ over $\mathcal{D}$, as convexity of $f$ yields that
	\[
	f(\vx) \leq 1/2 \cdot f(\vx^\dagger) + 1/2 \cdot f(\vx^*) = f(\vx^*) = \inf \setb{ f(\vy) \mid \vy \in \mathcal{D} }
	\enspace.
	\]
	Finally, observe that $g(\vx) < g(\vx^*)$, as strict convexity of $g$ yields that
	\[
	g(\vx)
	< 1/2 \cdot g(\vx^\dagger) + 1/2 \cdot g(\vx^*)
	\leq g(\vx^*)
	\enspace.
	\]
	However, we have arrived at a contradiction, as $\vx^*$ as the minimizer of $g$ among all minimizers of $f$ over $\mathcal{D}$.
	Thus, $\vx^\dagger = \vx^*$.
	This establishes that every convergent subsequence of $\setb{\vx_k}_{k=1}^\infty$ has $\vx^*$ as its limit point.
	Thus, by Lemma~\ref{lemma:analysis-fact}, we have that the entire sequence $\setb{\vx_k}_{k=1}^\infty$ converges to $\vx^*$, as desired.
\end{proof}

Proposition~\ref{prop:formal-limiting-argument} now follows as a specific instance of Proposition~\ref{prop:general-limiting-argument}, where $f$ is the operator norm, $g$ is the Frobenius norm, and $\mathcal{D}$ is the set of slack matrices.
More generally, Proposition~\ref{prop:general-limiting-argument} may be used to analyze the limiting behavior of any regularized composite objectives for \optprocedure{}.

\subsection{Lemma~\mainref{lem:linear-est-variance}: Variance of Linear Estimators}

\begin{reflemma}{\mainref{lem:linear-est-variance}}
	\lemlinearestvariance
\end{reflemma}

\begin{proof}
Because $\parav$ is nonrandom, $\Var{\est} = \Var[\big]{n^{-1} \coefv^\tran \parav} = n^{-2} \parav^\tran \Cov{\coefv} \parav$.
\end{proof}

\newcommand{\maketable}[4]{
\begin{table}[ht]
\centering
\caption{#2}
\resizebox{.99\textwidth}{!}{%
	\begin{tabular}{lrrrrcrrrr}
	\toprule
	& \multicolumn{4}{c}{Panel A: #3} & & \multicolumn{4}{c}{Panel B: #4} \\ \cmidrule{2-5} \cmidrule{7-10}
	& Bias & Precision & Coverage & Width & & Bias & Precision & Coverage & Width \\
	\midrule
	\input{#1}\\ \bottomrule
\end{tabular}
}
\end{table}
}

\newcommand{\maketabletwopanels}[3]{
\begin{table}[ht]
\centering
\caption{#3}
\resizebox{.99\textwidth}{!}{%
	\begin{tabular}{lrrrrcrrrr}
	\toprule
	& \multicolumn{4}{c}{Panel A: Synthetic outcomes} & & \multicolumn{4}{c}{Panel B: Real data outcomes} \\ \cmidrule{2-5} \cmidrule{7-10}
	& Bias & Precision & Coverage & Width & & Bias & Precision & Coverage & Width \\
	\midrule
	\input{#1}\\ \midrule
	& \multicolumn{4}{c}{Panel C: Disciplinary outcome} & & \multicolumn{4}{c}{Panel D: Wristband outcome} \\ \cmidrule{2-5} \cmidrule{7-10}
	& Bias & Precision & Coverage & Width & & Bias & Precision & Coverage & Width \\
	\midrule
	\input{#2}\\ \bottomrule
\end{tabular}
}
\end{table}
}

\clearpage
\section{Additional Simulation Results}\label{sec:additional-sims}

We here describe additional results from our simulation investigation.
In addition to the variance estimators investigated in the main paper, we also include the variance estimator described by \citet{Leung2022Causal}, who consider a type of misspecification referred to as Approximate Neighborhood Interference (ANI).
This alternative estimator is in the class of the Newey--West or heteroskedasticity and autocorrelation consistent (HAC) estimators, which takes a simpler form than the estimators we consider in this paper.
While the comparison to the HAC-type estimator is of interest given that it was developed for a similar setting to the one we study here, we note that \citet{Leung2022Causal} never claims that the estimator is appropriate in our setting.
In particular, the assumptions used by \citet{Leung2022Causal} are different than the ones we consider, and he restricts his focus to Bernoulli designs.
For this reason, the poor performance of the HAC-type estimator displayed in the simulation here should not necessarily be interpreted as representative of the performance of the estimator when used in the settings it was created for.
The addition of the variance estimator by \citet{Leung2022Causal} to the simulation study was suggested during the review process.

In his simulations, \citet{Leung2022Causal} uses the same data from \citet{Paluck2016Changing} as we do, but \citet{Leung2022Causal} departs considerably from the original study in his setup, changing both the sample and the design, adapting it to align with the setting he studies.
Our simulation is more faithful to the design of the original study.
This explains why \citet{Leung2022Causal} finds that the HAC-type estimator performs well with the data from \citet{Paluck2016Changing}, when we find that the HAC-type estimator performs poorly.

We see the estimator described by \citet{Leung2022Causal} as complementary to the estimators we describe in this paper.
In particular, experimenters who use a Bernoulli design in a setting with ANI-type misspecification should prefer Leung's HAC estimator over the estimators described in this paper.

\clearpage
\subsection{Alternative outcomes}\label{sec:supp-alt-outcomes}

This simulation is identical to the one conducted in the main paper, but with the difference that we consider the wristband outcome and disciplinary action outcome separately.
In the main paper, we combined these outcomes to construct a mixed outcome, where the wristband outcome was one of the potential outcomes, and the disciplinary action outcome was the other.
Here, we set both potential outcomes to be the same outcome, being the wristband and disciplinary action outcomes, meaning that the treatment effect is constant at zero.
We include the potential outcomes investigated in the main paper as well for comparison, as well as the HAC-type estimator as described above.

The results are in line with those in the main paper.
The biases for the optimized variance estimators are considerably lower than for the Aronow--Samii estimator, which is reflected in the coverage rates and confidence interval widths.
The HAC-type variance estimator is anti-conservative with a large negative bias, leading to undercoverage and invalid inference.
The negative bias of the HAC-type estimator ranges from $22\%$ to $71\%$ of the true variance of the point estimator.

\begin{table}[ht]
\centering
\caption{Simulation results with alternative outcomes}
\resizebox{.99\textwidth}{!}{%
	\begin{tabular}{lrrrrcrrrr}
	\toprule
	& \multicolumn{4}{c}{Panel A: Synthetic outcomes} & & \multicolumn{4}{c}{Panel B: Real data outcomes} \\ \cmidrule{2-5} \cmidrule{7-10}
	& Bias & Precision & Coverage & Width & & Bias & Precision & Coverage & Width \\
	\midrule
	Aronow-Samii &  4.0461 & 0.771 & 1.000 & 1.000 &   &  2.781 & 0.707 & 1.000 & 1.000 \\
Trace &  0.1144 & 0.417 & 0.968 & 0.465 &   &  0.770 & 0.419 & 0.992 & 0.682 \\
Frobenius &  0.0947 & 0.391 & 0.961 & 0.461 &   &  0.764 & 0.392 & 0.992 & 0.682 \\
Targeted &  0.0452 & 0.370 & 0.960 & 0.450 &   &  0.743 & 0.436 & 0.991 & 0.677 \\
Composite &  0.0301 & 0.348 & 0.955 & 0.448 &   &  0.764 & 0.391 & 0.992 & 0.682 \\
HAC & -0.2198 & 0.378 & 0.911 & 0.387 &   & -0.425 & 0.320 & 0.844 & 0.382\\ \midrule
	& \multicolumn{4}{c}{Panel C: Disciplinary outcome} & & \multicolumn{4}{c}{Panel D: Wristband outcome} \\ \cmidrule{2-5} \cmidrule{7-10}
	& Bias & Precision & Coverage & Width & & Bias & Precision & Coverage & Width \\
	\midrule
	Aronow-Samii &  2.0431 & 0.633 & 1.000 & 1.000 &   &  0.8566 & 0.288 & 0.992 & 1.000 \\
Trace &  0.0935 & 0.335 & 0.970 & 0.596 &   &  0.0549 & 0.204 & 0.961 & 0.752 \\
Frobenius &  0.0494 & 0.340 & 0.956 & 0.583 &   &  0.0266 & 0.183 & 0.952 & 0.743 \\
Targeted &  0.0889 & 0.342 & 0.969 & 0.594 &   &  0.0555 & 0.231 & 0.959 & 0.752 \\
Composite &  0.0486 & 0.341 & 0.956 & 0.583 &   &  0.0268 & 0.183 & 0.952 & 0.743 \\
HAC & -0.4977 & 0.306 & 0.811 & 0.396 &   & -0.7111 & 0.132 & 0.692 & 0.388\\ \bottomrule
\end{tabular}
}
\end{table}

\clearpage
\subsection{Weak misspecification}

This simulation considers weak misspecification in the exposures.
We implement this by considering a setting where the social network that is the basis for the definition of the exposures is measured with error.
We remove, at random, $20\%$ of the reported edges in the network, and add the same number of edges, also at random, ensuring the number of edges within each school is the same.
This means that a removed edge could be added back again, but this will be rare since the networks are fairly sparse.
The two networks will therefore differ in approximately $20\%$ of their edges, leading to exposure misspecification.

Because the original network was used for the experimental design, we take the altered network to be the true (unobserved) network.
When an exposure based on the observed, misspecified network disagrees with the exposure based on the unobserved, true network, the corresponding potential outcome is doubled.
For example, consider some exposure $\exa \in \allexposures$, and the corresponding potential outcome $\yi{\exa}$.
If the realized exposure indeed was the exposure of interest, $\Exmi = \exa$, then we would observe $\Yi = \yi{\exa}$.
But if the exposures are misspecified, we might believe that $\Exmi = \exa$, even when this is not the case, meaning that we would not observe the corresponding potential outcome: $\Yi \neq \yi{\exa}$.
In our simulation, when the misspecified exposures say that $\Exmi = \exa$ but the true exposures say $\Exmi \neq \exa$, we set the observed outcome to twice the true potential outcome corresponding to the exposure of interest: $\Yi = 2 \yi{\exa}$.
This structure was chosen because it allows for a computationally efficient implementation of the simulations.

The results are largely in line with those in the main paper.
We present both the two outcomes considered in the main paper, and the two outcomes considered in Section~\ref{sec:supp-alt-outcomes} in the supplement.
The biases of the variance estimators are generally smaller than when the exposures are correctly specified, but they are all still conservative.
Coverage is still at or above the nominal rate for the confidence intervals, with the exception of the wristband outcome (Panel D), where the coverage is just below $95\%$.
The relative differences between the Aronow--Samii bound and the optimized bound are the same as in the main simulation.
Overall, the type of weak misspecification studied here does not affect the results.

The HAC-type estimator is anti-conservative with large negative bias, leading to invalid inferences.

\begin{table}[ht]
\centering
\caption{Simulation results under weak misspecification}
\resizebox{.99\textwidth}{!}{%
	\begin{tabular}{lrrrrcrrrr}
	\toprule
	& \multicolumn{4}{c}{Panel A: Synthetic outcomes} & & \multicolumn{4}{c}{Panel B: Real data outcomes} \\ \cmidrule{2-5} \cmidrule{7-10}
	& Bias & Precision & Coverage & Width & & Bias & Precision & Coverage & Width \\
	\midrule
	Aronow-Samii &  3.7692 & 0.999 & 1.000 & 1.000 &   &  2.799 & 1.295 & 1.000 & 1.000 \\
Trace &  0.0949 & 0.514 & 0.967 & 0.472 &   &  0.639 & 0.629 & 0.983 & 0.655 \\
Frobenius &  0.0892 & 0.497 & 0.961 & 0.471 &   &  0.672 & 0.649 & 0.986 & 0.661 \\
Targeted &  0.0269 & 0.471 & 0.959 & 0.457 &   &  0.622 & 0.638 & 0.983 & 0.651 \\
Composite &  0.0255 & 0.464 & 0.955 & 0.457 &   &  0.671 & 0.647 & 0.986 & 0.661 \\
HAC & -0.1879 & 0.486 & 0.917 & 0.403 &   & -0.345 & 0.618 & 0.850 & 0.398\\ \midrule
	& \multicolumn{4}{c}{Panel C: Disciplinary outcome} & & \multicolumn{4}{c}{Panel D: Wristband outcome} \\ \cmidrule{2-5} \cmidrule{7-10}
	& Bias & Precision & Coverage & Width & & Bias & Precision & Coverage & Width \\
	\midrule
	Aronow-Samii &  2.0512 & 1.199 & 1.000 & 1.000 &   &  0.9297 & 0.457 & 0.994 & 1.000 \\
Trace &  0.0637 & 0.590 & 0.961 & 0.584 &   &  0.0372 & 0.261 & 0.951 & 0.732 \\
Frobenius &  0.0491 & 0.613 & 0.954 & 0.578 &   &  0.0234 & 0.261 & 0.947 & 0.727 \\
Targeted &  0.0610 & 0.596 & 0.961 & 0.582 &   &  0.0373 & 0.279 & 0.949 & 0.732 \\
Composite &  0.0485 & 0.616 & 0.954 & 0.578 &   &  0.0234 & 0.260 & 0.947 & 0.728 \\
HAC & -0.4366 & 0.573 & 0.816 & 0.410 &   & -0.6701 & 0.213 & 0.693 & 0.403\\ \bottomrule
\end{tabular}
}
\end{table}

\clearpage
\subsection{Strong misspecification}

This simulation considers severe misspecification in the exposures.
The setting is the same as in the previous section with weak misspecification, but instead of randomly deleting and adding $20\%$ of the edges, we here delete and add $50\%$ of the edges.

We find that the biases of the variance estimators are similar to under weak misspecification, indicating that the variance bounds are relatively robust to misspecification.
While we expect there are situations where the bias of the variance estimators would be negative under misspecification, the fact that the variance bounds dominate the true variance for all (correctly specified) potential outcomes provide a fairly large safety margin for misspecification without introducing excessive conservativeness.
However, unlike under weak misspecification, the coverage rates are clearly below the nominal level for some of the outcomes here.
In particular, coverage is between $91$ and $92\%$ for the disciplinary action and wristband outcomes.
The reason for the poor coverage despite positive bias of the variance estimators is bias in the point estimator in this setting.

The HAC-type estimator is anti-conservative with large negative bias, leading to invalid inferences.

\begin{table}[ht]
\centering
\caption{Simulation results under strong misspecification}
\resizebox{.99\textwidth}{!}{%
	\begin{tabular}{lrrrrcrrrr}
	\toprule
	& \multicolumn{4}{c}{Panel A: Synthetic outcomes} & & \multicolumn{4}{c}{Panel B: Real data outcomes} \\ \cmidrule{2-5} \cmidrule{7-10}
	& Bias & Precision & Coverage & Width & & Bias & Precision & Coverage & Width \\
	\midrule
	Aronow-Samii &  3.6087 & 1.470 & 1.000 & 1.000 &   &  2.888 & 1.098 & 1.000 & 1.000 \\
Trace &  0.0885 & 0.747 & 0.966 & 0.475 &   &  0.564 & 0.536 & 0.955 & 0.632 \\
Frobenius &  0.0918 & 0.733 & 0.961 & 0.476 &   &  0.620 & 0.558 & 0.964 & 0.643 \\
Targeted &  0.0244 & 0.691 & 0.959 & 0.461 &   &  0.553 & 0.551 & 0.954 & 0.629 \\
Composite &  0.0296 & 0.689 & 0.955 & 0.463 &   &  0.620 & 0.557 & 0.964 & 0.643 \\
HAC & -0.1459 & 0.733 & 0.924 & 0.417 &   & -0.322 & 0.517 & 0.784 & 0.405\\ \midrule
	& \multicolumn{4}{c}{Panel C: Disciplinary outcome} & & \multicolumn{4}{c}{Panel D: Wristband outcome} \\ \cmidrule{2-5} \cmidrule{7-10}
	& Bias & Precision & Coverage & Width & & Bias & Precision & Coverage & Width \\
	\midrule
	Aronow-Samii &  2.1742 & 1.006 & 1.000 & 1.000 &   &  1.0020 & 0.488 & 0.990 & 1.000 \\
Trace &  0.0626 & 0.499 & 0.920 & 0.573 &   &  0.0388 & 0.285 & 0.914 & 0.719 \\
Frobenius &  0.0610 & 0.521 & 0.918 & 0.571 &   &  0.0353 & 0.291 & 0.912 & 0.718 \\
Targeted &  0.0600 & 0.505 & 0.919 & 0.571 &   &  0.0376 & 0.300 & 0.911 & 0.718 \\
Composite &  0.0602 & 0.521 & 0.917 & 0.571 &   &  0.0353 & 0.290 & 0.912 & 0.718 \\
HAC & -0.4157 & 0.475 & 0.754 & 0.415 &   & -0.6481 & 0.211 & 0.624 & 0.410\\ \bottomrule
\end{tabular}
}
\end{table}

\clearpage
\subsection{Moderately dependent exposures}

This simulation considers exposures with less dependence than those in the main simulation.
In the main paper, we consider the treatment effect of exposure $e_1 = (1,1,0)$ relative to $e_0 = (0,0,0)$.
As we discussed there, we focus on these exposures because they are highly dependent with many second-order positivity violations, which makes it challenging to find a well-performing variance bound.
In this section, we consider two exposures with slightly less dependence than the exposures in the main paper.

We here consider the treatment effect of exposure $(1,0,1)$ relative to $(0,0,0)$.
That is, the indirect effect, contrasting when one of a student's peers are treated versus when no one in the student's school is treated.
A way to measure the overall dependence of exposures in an experiment is with the Frobenius norm of the correlation matrix of the exposures normalized with the size of the matrix (to account for that different number of units will satisfy first-order positivity for different exposures).
For the exposures in the main paper, the normalized Frobenius norm was $6.9952$.
For the exposures considered here, the normalized Frobenius norm was $6.3039$, which is a reduction of approximately $10\%$.
The dependence is thus less than in the main paper, but still fairly high.

The results mirror those in the main paper.
The biases are considerably smaller with the optimized bounds, resulting in considerably narrower confidence intervals.
All variance estimators are still conservative, and coverage is at or above the nominal rate.

The HAC-type estimator is here approximately unbiased for the synthetic outcomes, but anti-conservative with large negative bias for the real data outcomes.

\begin{table}[ht]
\centering
\caption{Simulation results with exposures $(1,0,1)$ and $(0,0,0)$}
\resizebox{.99\textwidth}{!}{%
	\begin{tabular}{lrrrrcrrrr}
	\toprule
	& \multicolumn{4}{c}{Panel A: Synthetic outcomes} & & \multicolumn{4}{c}{Panel B: Real data outcomes} \\ \cmidrule{2-5} \cmidrule{7-10}
	& Bias & Precision & Coverage & Width & & Bias & Precision & Coverage & Width \\
	\midrule
	Aronow-Samii & 5.108235 & 0.2275 & 1.000 & 1.000 &   &  3.187 & 0.4110 & 1.000 & 1.000 \\
Trace & 0.029873 & 0.1449 & 0.951 & 0.410 &   &  0.950 & 0.3130 & 0.992 & 0.681 \\
Frobenius & 0.030953 & 0.1450 & 0.951 & 0.410 &   &  0.953 & 0.3134 & 0.992 & 0.682 \\
Targeted & 0.033216 & 0.1836 & 0.956 & 0.410 &   &  0.970 & 0.3241 & 0.992 & 0.684 \\
Composite & 0.029862 & 0.1448 & 0.951 & 0.410 &   &  0.952 & 0.3133 & 0.992 & 0.681 \\
HAC & 0.000841 & 0.0476 & 0.950 & 0.405 &   & -0.309 & 0.0896 & 0.894 & 0.406\\ \bottomrule
\end{tabular}
}
\end{table}

\clearpage
\subsection{Weakly dependent exposures}

This simulation considers exposures that are only weakly dependent.
We consider the treatment effect of exposure $(1,1,0)$ relative to $(1,0,1)$.
Unlike the previously considered exposure pairs, a student can be assigned the first exposure when one of their peers is assigned the second exposure, meaning that there is less dependence in the assignment of these exposures.
This is reflected in the normalized Frobenius norm, discussed in the previous section.
For the previous exposure pairs, this norm was $6.9952$ and $6.3039$, respectively.
Here, the normalized Frobenius norm is $2.0335$, showing that the dependence is less than a third compared to before.

Because of the lower dependence, most of the potential outcome pairs are observable, so less bounding is needed.
For this reason, the differences between the various bounds are expected to be less stark compared to the exposures with high dependence.
In settings where there is no dependence between the exposures, no bounding is needed other than to address the fundamental problem of causal inference.

The results are in line with those in the main paper, but the differences between the optimized variance estimator and the Aronow--Samii estimator are less pronounced.
This is because less bounding is required when the dependence is weaker.
Nevertheless, the bias for the Aronow--Samii estimator is more than 6 times as large for the synthetic outcomes and 3 times as large for the real data outcomes, indicating that considerable improvements are to be expected also for weakly dependent exposures.

The HAC-type estimator is here approximately unbiased for the synthetic outcomes, but anti-conservative with large negative bias for the real data outcomes.

\begin{table}[ht]
\centering
\caption{Simulation results with exposures $(1,1,0)$ and $(1,0,1)$}
\resizebox{.99\textwidth}{!}{%
	\begin{tabular}{lrrrrcrrrr}
	\toprule
	& \multicolumn{4}{c}{Panel A: Synthetic outcomes} & & \multicolumn{4}{c}{Panel B: Real data outcomes} \\ \cmidrule{2-5} \cmidrule{7-10}
	& Bias & Precision & Coverage & Width & & Bias & Precision & Coverage & Width \\
	\midrule
	Aronow-Samii & 1.3759 & 1.507 & 0.998 & 1.000 &   &  0.654 & 0.788 & 0.982 & 1.000 \\
Trace & 0.0892 & 0.735 & 0.962 & 0.674 &   &  0.200 & 0.441 & 0.970 & 0.858 \\
Frobenius & 0.1946 & 0.733 & 0.971 & 0.709 &   &  0.185 & 0.446 & 0.963 & 0.852 \\
Targeted & 0.0854 & 0.645 & 0.962 & 0.678 &   &  0.221 & 0.460 & 0.969 & 0.865 \\
Composite & 0.0398 & 0.648 & 0.958 & 0.663 &   &  0.187 & 0.447 & 0.963 & 0.853 \\
HAC & 0.0154 & 0.669 & 0.954 & 0.653 &   & -0.365 & 0.361 & 0.868 & 0.617\\ \bottomrule
\end{tabular}
}
\end{table}

\clearpage
\subsection{Different second order cutoff}

The final set of supplementary simulation results investigate how different cutoffs on the second-order assignment probabilities affect the performance of the estimators.
As discussed in the main paper, second-order probabilities close to zero behaves similar to probabilities at zero, even if the corresponding pairs of potential outcomes in principle are observable.
Hence, it is advisable to construct the bound so that second-order probabilities that are close to zero are considered unobservable, and thereby not estimated.
In the main paper, we set the cutoffs for second-order probabilities at $0.002$.
This choice was somewhat arbitrary, and here we investigate the cutoffs $0.001$ and $0.004$.

The results when the cutoff is at $0.001$ is close to identical to those in the main paper.
The biases are overall higher when the cutoff is at $0.004$, reflecting the fact that a higher cutoff makes for a more challenging bounding problem.
However, the relative width of the confidence intervals are similar.
The main difference is that the targeted objective performs worse here relative to the other optimized bounds.
We have not found an explanation for this behavior, and we suspect that it might be an artifact of the SCS solver of the underlying semidefinite program.

The HAC-type estimator is anti-conservative with large negative bias.

\begin{table}[ht]
\centering
\caption{Simulation results when cutoff is 0.001}
\resizebox{.99\textwidth}{!}{%
	\begin{tabular}{lrrrrcrrrr}
	\toprule
	& \multicolumn{4}{c}{Panel A: Synthetic outcomes} & & \multicolumn{4}{c}{Panel B: Real data outcomes} \\ \cmidrule{2-5} \cmidrule{7-10}
	& Bias & Precision & Coverage & Width & & Bias & Precision & Coverage & Width \\
	\midrule
	Aronow-Samii &  3.8453 & 0.688 & 1.000 & 1.000 &   &  2.613 & 0.644 & 1.000 & 1.000 \\
Trace &  0.0677 & 0.405 & 0.964 & 0.464 &   &  0.670 & 0.407 & 0.991 & 0.678 \\
Frobenius &  0.0623 & 0.384 & 0.957 & 0.463 &   &  0.728 & 0.399 & 0.991 & 0.690 \\
Targeted &  0.0544 & 0.443 & 0.953 & 0.459 &   &  0.571 & 0.452 & 0.987 & 0.656 \\
Composite &  0.0249 & 0.359 & 0.953 & 0.455 &   &  0.728 & 0.399 & 0.991 & 0.690 \\
HAC & -0.2198 & 0.378 & 0.911 & 0.395 &   & -0.425 & 0.320 & 0.844 & 0.390\\ \bottomrule
\end{tabular}
}
\end{table}

\begin{table}[ht]
\centering
\caption{Simulation results when cutoff is 0.004}
\resizebox{.99\textwidth}{!}{%
	\begin{tabular}{lrrrrcrrrr}
	\toprule
	& \multicolumn{4}{c}{Panel A: Synthetic outcomes} & & \multicolumn{4}{c}{Panel B: Real data outcomes} \\ \cmidrule{2-5} \cmidrule{7-10}
	& Bias & Precision & Coverage & Width & & Bias & Precision & Coverage & Width \\
	\midrule
	Aronow-Samii &  4.4119 & 0.850 & 1.000 & 1.000 &   &  3.092 & 0.774 & 1.000 & 1.000 \\
Trace &  0.2093 & 0.452 & 0.975 & 0.467 &   &  0.953 & 0.467 & 0.993 & 0.689 \\
Frobenius &  0.1486 & 0.414 & 0.966 & 0.456 &   &  0.813 & 0.377 & 0.993 & 0.665 \\
Targeted &  1.4290 & 0.710 & 0.995 & 0.671 &   &  2.020 & 0.558 & 0.999 & 0.859 \\
Composite &  0.0469 & 0.355 & 0.957 & 0.436 &   &  0.815 & 0.378 & 0.993 & 0.665 \\
HAC & -0.2198 & 0.378 & 0.911 & 0.374 &   & -0.425 & 0.320 & 0.844 & 0.367\\ \bottomrule
\end{tabular}
}
\end{table}

\end{document}